\documentclass[a4paper,fleqn,usenatbib]{mnras}

\usepackage[T1]{fontenc}
\usepackage{ae,aecompl}
\usepackage{graphicx}
\usepackage{amsfonts}
\usepackage{amssymb}
\usepackage{amsmath}

\def\apgt{\ {\raise-.5ex\hbox{$\buildrel>\over\sim$}}\,}
\def\aplt{\ {\raise-.5ex\hbox{$\buildrel<\over\sim$}}\,}
\newcommand{\ve}[1]{\boldsymbol{#1}}

\title[Transfer of debris during stellar encounters]{Mass transfer between debris discs during close stellar encounters}

\author[J\'{i}lkov\'{a} et al.]
{Lucie J\'{i}lkov\'{a},$^{1}$\thanks{E-mail: jilkova@strw.leidenuniv.nl} 
Adrian~S. Hamers,$^{1}$
Michael Hammer,$^{2,3}$ 
\newauthor and Simon Portegies Zwart$^{1}$ 
\vspace*{0.2cm} \\
$^{1}$Leiden Observatory, Niels Bohrweg 2, Leiden, 2333\,CA, The Netherlands \\
$^{2}$Cornell University, 614 Space Sciences Building, Ithaca, NY\,14853, USA \\
$^{3}$Department of Astronomy and Steward Observatory, University of Arizona, Tuscon, AZ\,85721, USA}

\pubyear{2016}

\begin{document}
\label{firstpage}

% These dates will be filled out by the publisher
\date{Accepted 2016 January 29. Received 2016 January 21; in original form 2015 November 17}

% Enter the current year, for the copyright statements etc.

\pagerange{\pageref{firstpage}--\pageref{lastpage}} \pubyear{2002}
\maketitle

\begin{abstract}
We study mass transfers between debris discs during stellar encounters.
We carried out numerical simulations of close flybys of two stars, one of which has a disc of planetesimals represented by test particles.
We explored the parameter space of the encounters, varying the mass ratio of the two stars, their pericentre and eccentricity of the encounter, and its geometry.
We find that particles are transferred to the other star from a restricted radial range in the disc and the limiting radii of this transfer region depend on the parameters of the encounter.
We derive an approximate analytic description of the inner radius of the region.
The efficiency of the mass transfer generally decreases with increasing encounter pericentre and increasing mass of the star initially possessing the disc.
Depending on the parameters of the encounter, the transfer particles have a specific distributions in the space of orbital elements (semimajor axis, eccentricity, inclination, and argument of pericentre) around their new host star.
The population of the transferred particles can be used to constrain the encounter through which it was delivered.
We expect that many stars experienced transfer among their debris discs and planetary systems in their birth environment. 
This mechanism presents a formation channel for objects on wide orbits of arbitrary inclinations, typically having high eccentricity but possibly also close-to-circular (eccentricities of about 0.1).
Depending on the geometry, such orbital elements can be distinct from those of the objects formed around the star.
\end{abstract}


%%%%%%%%%%%%%%%%%%%%%

\section{Introduction}
\label{sec:intro}

Stars form in giant molecular clouds and in most cases, a group containing 10 to $10^4$ stars form at a similar time from the same cloud \citep{2003ARA&A..41...57L}. 
Depending on the number of stars and their spatial and velocity distributions, these groups, with stellar densities relatively high compared to those of the field stars, are classified as stellar associations or star clusters \citep[e.g.][]{2010MNRAS.409L..54B,2011MNRAS.410L...6G}.
The gravitational interactions in these crowded environments result in close stellar encounters \citep[for example][]{1987gady.book.....B,2001MNRAS.322..859B,2006A&A...454..811P,2006ApJ...642.1140O,2008A&A...488..191O,2012ApJ...756..123O} that can strongly influence the properties of protoplanetary discs around the still young stars, and eventually the planetary systems formed from these \citep{1993MNRAS.261..190C,1994ApJ...424..292O}.
Several authors presented $N$-body simulations of planetary systems (for example \citealp{2002ApJ...565.1251H}; \citealp{2009ApJ...697..458S}; \citealp{2012MNRAS.419.2448P}; \citealp{2015MNRAS.453.2759Z}) and hydro-dynamical simulations of protoplanetary discs in star clusters \citep{2014MNRAS.441.2094R}.
These works confirmed the importance of the star clusters and close stellar encounters for the dynamics of planetary systems and circumstellar discs.
The rate and parameters of the encounters depend on the characteristics of the star cluster, such as its mass, density, initial spatial and velocity distributions (for example \citealt{1987gady.book.....B}; \citealp{2001MNRAS.322..859B}; \citealt{2006ApJ...641..504A}; \citealp{2007MNRAS.378.1207M}; \citealp{2010A&A...509A..63O}; \citealt{2013ApJ...769..150C}).
Using $N$-body simulations, \citet{2013ApJ...769..150C} tabulated the number and properties of encounters as a function of the cluster characteristics.
For example,  they measured the encounter rate (counting flybys closer than 1000\,AU) for a solar-like star (0.8--1.2\,M$_{\sun}$) of $\sim1.9\times10^{-6}$\,yr$^{-1}$ experienced in a cluster with mass of 1000\,M$_{\sun}$, typical radius of about 2.5\,pc, virial ratio of 0.75, and a moderate degree of substructure (fractal dimension $D=2.2$).

After stars form, gas and dust are present in their circumstellar discs, where planets and debris form later.
The current knowledge of the debris discs has been summarized by \citet{2014prpl.conf..521M}, \citet{2010RAA....10..383K} and\citet{2008ARA&A..46..339W}.
Debris discs are result of planet formation around main sequence stars and they consist of dust and larges bodies (such as comets, asteroids, or planetesimals) which determine the dynamics of the discs.
The dust grains are heated by the central star and they re-radiate in the infrared (IR), producing so called IR excess in the spectral energy distribution of their host star, or (sub)millimeter wavelengths.
It is the radiation of the dust that is observed.
The dust has relatively short lifetime and is constantly replenished by collisions between the larger bodies \citep[for example][]{2002MNRAS.334..589W}.
Debris discs have been observed around hundreds of stars.
The detection rates vary depending on the wavelength, stellar type and age, and the relative sensitivity of the surveys \citep[see][for the summary and references]{2014prpl.conf..521M}.
Cold dust (at $\geq60\micron$) has been detected around $24 \pm 5$ and $32 \pm 5$\,per\,cent of A stars (\citealp{2014MNRAS.445.2558T} at $100\,\micron$ and \citealp{2006ApJ...653..675S} at $70\,\micron$, respectively) and around $20 \pm 2$\,per\,cent of solar type FGK stars \citep{2013A&A...555A..11E}.
Dust in the mid-IR wavelengths ($\leq60\,\micron$) was detected around $\sim11$\,per\,cent of solar type stars \citep{2011AJ....141...11D}.

The debris discs are observed to decay with time, as the planetesimals are depleted by collisions which grind them into dust \citep{2003ApJ...598..626D}.
The planetesimals in debris discs must be stirred so that their relative velocities are sufficient for grinding.
The origin of the stirring is still under discussion and several mechanisms have been suggested \citep{2014prpl.conf..521M}: 
pre-stirring as a result of the protoplanetary phase \citep[see][and references therein]{2008ARA&A..46..339W}; 
stirring by planets in the same system (\citealp{2004AJ....127..513K}, \citealp{2009MNRAS.399.1403M}); 
self-stirring by sufficiently massive planetesimals (for example \citealp{2008ApJS..179..451K}, \citealp{2010MNRAS.405.1253K}, and references therein); 
or stirring by external process, such as stellar flybys \citep{2002AJ....123.1757K}.

\subsection{Simulations of discs during stellar encounters}

In the early gas-rich stages, the viscosity and pressure of the gas in a circumstellar disc are important for the general disc dynamics when considering the effects of external perturbers and close stellar encounters.
For the more distant ones, where the periastron is larger than the size of the disc, gas and dust free simulations are commonly used to study the perturbations due the encounter and the debris disc is often modeled by test particles 
(\citealp{1993MNRAS.261..190C} were among the first to use this approach, while they considered dissipation in the disc through pseudo-viscosity; discussion on the role of self-gravity, pressure and viscosity of the disk was carried for example by \citealp{2005ApJ...629..526P}).

The dynamics of a planetesimal during a stellar encounter can be approximately described as a general restricted three-body problem.
The planetesimals are much less massive than the two stars and their gravitational influence\,---\,mutual as well as on the stars\,---\,can be neglected.
Planetesimals are then represented as (zero mass) test particles that live in the time-dependent gravitational potential of the two stars that move on a conic section orbit.
A similar approach was used already in the seminal work of \citet{1972ApJ...178..623T}, who pioneered the simulations of mergers of disc galaxies.
The particles of the disc are perturbed during the encounter and can, in general, stay bound to their parent star, become bound to the perturbing star, or become unbound from the system.
Previous work showed that depending on the parameters of the encounter and the initial size of the disc, the fate of the particles depends on their initial location in the disc around the parent star.
For example, \citet{1993MNRAS.261..190C} noticed that in the prograde coplanar parabolic encounter of equal mass stars, all the particles located closer than $\sim0.3$ of the pericentre of the encounter to the parent star stay bound.
\citet{2001Icar..153..416K} confirmed this result and gave a more detailed description of the perturbation in eccentricity and semimajor axis of the disc particle orbits.
A natural result of the encounter is a truncation of the disc and the dependence of the resulting disc size on the mass ratio between the encountering stars was recently described in detail by \citet{2014A&A...565A.130B}.
The encounter can also induce various structures in the disc, such as rings or spiral patterns, or cause 
the disc to have an elliptical shape \citep{2001MNRAS.323..402L,2003ApJ...592..986P}.
Simulations of the influence of a stellar flyby on a debris disc were also motivated to explain individual observed systems, for example in case of $\beta$\,Pictoris by \citet{2001MNRAS.323..402L} and HD\,141569 by \citet{2009A&A...493..661R}, or to understand the scattered Kuiper belt in the Solar system \citep[e.g.][]{2005Icar..173..559M,2005Icar..177..246K,2014MNRAS.444.2808P}.
Results on the change of the disc mass or size due to the flybys were often applied in the simulations of star clusters, where the encounters occur (for example \citealp{2006ApJ...642.1140O,2006A&A...454..811P,2011A&A...532A.120L,2015A&A...577A.115V}; \citealp{2016MNRAS.457..313P}).

\subsection{Previous studies of captured planetesimals}
\label{sec:intro_previous_work}

Most of the studies that model a debris disc during stellar encounters focused on the evolution of the disc itself (its mass loss, change of its size, morphology, energy or angular momentum).
Depending on the parameters of the encounter, a portion of the disc can also be transferred from the parent star to the perturber.
However a systematic study of the mass transfer among debris discs during a stellar flyby is still missing.
As we describe below, captured bodies present an important output of stellar encounters and material originating from other stars might be present in many debris discs\,---\,including in the Solar system.

\citet{1993MNRAS.261..190C} investigated the response of an accretion disc to a stellar flyby.
They represented the disc by test particles with pseudo-viscosity and carried out simulations of parabolic encounters of equal mass stars considering different initial geometries of the encounter orbit and the disc\,---\,coplanar prograde and retrograde, and orthogonal.
They found that out of these three configurations, the mass transfer occurs only in the coplanar prograde encounter and that only the particles located initially $\apgt 0.35$ of the pericentre distance from the parent star can be captured by the star initially without disc.
These results were further extended by \citet{1996MNRAS.278..303H} who focused on energy and angular momentum exchanged during the encounter.
They considered a more extended disc, up to four times the pericentre distance, and showed that particles are transferred from larger initial radii (beyond the pericentre distance) also in configurations when the disc is inclined with respect to the orbital plane (inclinations of 45, 90, 135 and 180$\degr$, where the latter corresponds to the coplanar retrograde encounter).

\citet{2001MNRAS.323..402L} studied prograde coplanar encounters and geometries with inclinations of the encounter orbit with respect to the disc of $30$ and $60\degr$.
They varied the eccentricity of the encounter and found that planetesimals can be captured for eccentricities $\leq2$ while none are captured for eccentricities $\geq5$.
The transferred material appeared to form an asymmetrical disc around their new host and showed clustering in eccentricities and semimajor axes.

From their parameter space study of stellar encounters, \citet{2005A&A...437..967P} concluded that the mass transfer occurs nearly exclusively in prograde encounters.
They investigated the dependence of the relative mass captured during prograde parabolic encounters on the mass ratio of the stars and the pericentre of the encounter.
They found that the captured mass increases with the stellar mass ratio, up to value of about 0.8 when the captured mass becomes almost independent of the stellar masses (they measured the mass ratio as the mass of the star initially without disc relative to the one initially with disc).
They concluded that this might indicate the existence of an upper limit to the mass transfer for a given pericentre distance.
They also found that the captured mass decreases almost linearly with increasing pericentre of the encounter.

\citet{2005ApJ...629..526P} presented the most detailed study of the mass transfers among circumstellar discs.
They compared captures in parabolic and hyperbolic stellar orbits and confirmed that the transfer is smaller in the latter case (negligible for periastron larger than three times the disc size).
They also found that the majority of the captured mass moves on highly eccentric orbits\,---\,in the equal mass parabolic encounter about 80\% of the particles have eccentricities above 0.8 and almost none below 0.4.

Planetesimals transferred during a stellar encounter were also suggested to explain the origin of some Solar system planetesimals in peculiar orbits \citep{2004AJ....128.2564M,2004Natur.432..598K,2010Sci...329..187L,sednitos2015}.
The formation of the population of a so called inner Oort cloud\,---\,which includes objects with pericentre $\apgt50$\,au and semimajor axis in range of 150--1500\,au \citep{2014Natur.507..471T}\,---\,is still not well understood, because these orbits are too far away from the Sun to be influenced by planetary perturbations and too close to be substantially perturbed by the Galactic potential and encounters with the field stars \citep{2015MNRAS.451..144P}.
\citet{sednitos2015} constrained the parameter space of encounters that could result in a population of planetesimals transferred to the Solar system that is consistent with the observed orbits of Sedna-like objects \citep{2004ApJ...617..645B,2014Natur.507..471T}. 
Constraining the encounter is possible due to specific characteristics of the captured orbits.

Here we present a systematic study of the mass captured from a debris disc during a stellar encounter. 
We carried out simulations (Sect.~\ref{sec:simulations}) and measured how much material is transferred depending on the parameters of the encounter and what are the orbital characteristics of the particles before and after the transfer (Sect.~\ref{sec:results}).
We derive an analytic approximation for the minimal radial distance of the particles to be captured by the encountering star and compare it with the results from the simulations (Sect.~\ref{sect:analytic}).
We summarize and conclude in Sect.~\ref{sec:conclusions}.

\section{Simulations}
\label{sec:simulations}

We carried out simulations of stellar encounters where one of the stars has a planetesimal disc and we followed the planetesimals transferred to the star initially without a disc. 
Because we approximate the disc by zero-mass particles (see below), the results will not change in case both stars have a disc.

\subsection{Numerical method}
\label{sec:num}

We assumed that the masses of planetesimals are small compared to the stars \citep[debris discs are typically less massive than 1\,M$_{\oplus}$, for example][]{2008ARA&A..46..339W}.
Under this assumption we represented the planetesimals by zero-mass points.
Such particles move in the gravitational potential of the two stars, they do not interact with each other and neither do they influence the motion of the two stars.

We integrate the equations of motions using a combination of $N$-body and hybrid methods (the same as in \citealp{sednitos2015} and similarly as in \citealp{hd666_2015}).
The orbit of the two stars is integrated using the symplectic $N$-body code \textsc{huayno} \citep{2012NewA...17..711P}.
As long as the stars are well separated (at least three times the disc size), the orbits of the planetesimals around their parent star at the beginning of the encounter and around the encountering star in the later times, are calculated by solving Kepler's equations using universal variables \citep[here we adopted the solver from the \textsc{sakura} code,][]{2014MNRAS.440..719G}.
In this hybrid approach the gravitational influence of the other star is considered as a perturbation and is coupled to the planetesimals using \textsc{bridge} \citep{2007PASJ...59.1095F}.
All calculations and the coupling of codes are realized using the Astronomical Multi-purpose Software Environment or \textsc{AMUSE} \citep{2013CoPhC.184..456P,2013A&A...557A..84P}.\footnote{\url{http://amusecode.org}.}
During the initial stages of the flybys (when the distance between the stars is large and decreasing) the influence of the star initially without the disc is considered as a perturbation, while during the later stages (when the distance between the stars is large and increasing) the influence of the disc parent star is considered as a perturbation of the particles captured by the other star.
By comparing with self-consistent $N$-body simulations in which all the particles are integrated directly, we tested that the hybrid approach treats the captured particles correctly.
However during the later stages of the flyby, larger inaccuracies can be introduced in the orbits of the particles that are not captured by the other star, that is those particles that are still bound to the parent star or completely unbound from the system.
In these cases the influence of the parent star is comparable to or stronger that of the other star and the assumption of our hybrid method, that it can be considered as a perturbation, is not fulfilled.
Our approach is appropriate only for the particles whose dynamics during the later stages of the encounter are dominated by the star initially without disc, the transferred particles, on which we focus here.
This approximate approach allows us to carry out many fast simulations and map parameter space of the encounters systematically.

\subsection{Initial conditions}
\label{sec:ic}

\begin{figure}
\center
\includegraphics[width=0.49\textwidth]{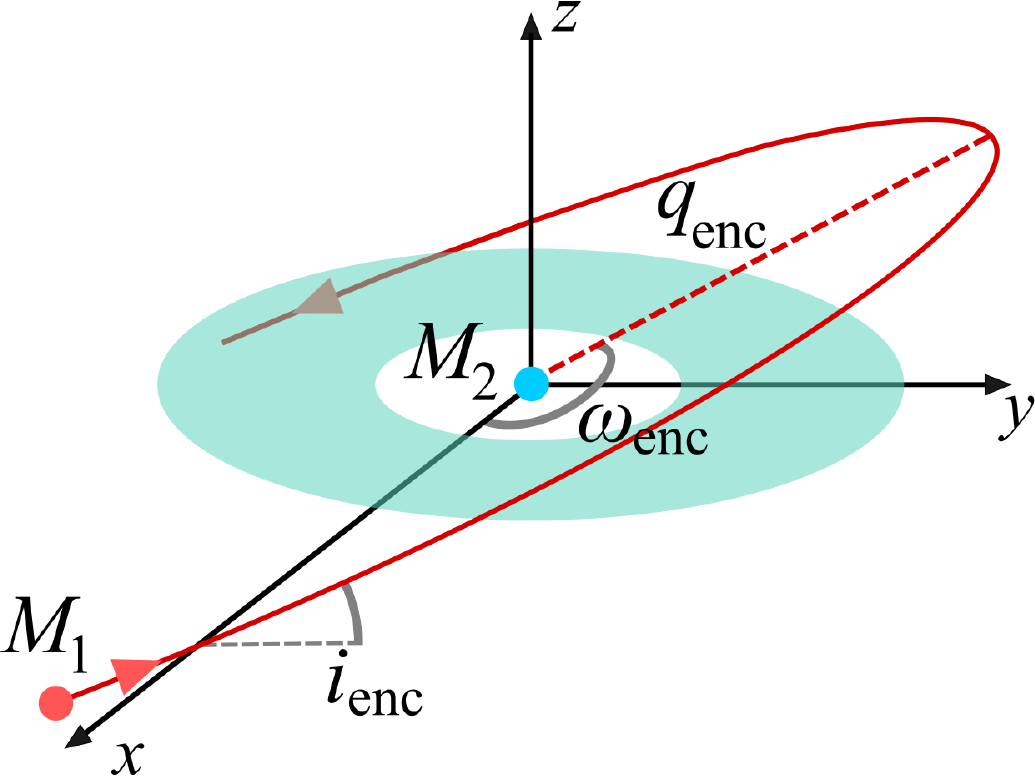} \\
\caption{A doodle of the initial conditions of the encounter simulation.
The red bullet indicates star $M_1$ which is initially without the disc, while the blue bullet indicates star $M_2$ and its disc is shown by the light green annulus.
The coordinate system is centered on star $M_2$ and the reference plane $xy$ is defined by the disc.
The stars are moving on a parabolic orbit which is indicated by the full red line.
The pericenter of the encounter $q_{\rm{enc}}$ is indicated by the dashed red line.
The plane of the encounter is inclined by the inclination angle $i_{\rm{enc}}$ around the $x$-axis.
The argument of pericenter $\omega_{\rm{enc}}$ is measured in the encounter plane between the $x$-axis and the direction to the pericenter.
The doodle is not scaled.}
\label{fig:encounter_ic}
\end{figure}

We set the mass of the star without the disc, $M_1$, to 1\,${\rm M}_{\sun}$ in all our simulations.
We systematically varied the mass of the star initially with disc $M_2$, the pericentre of the encounter $q_{\rm{enc}}$, and the eccentricity of the encounter $e_{\rm{enc}}$, in the ranges of 0.1--2\,${\rm M}_{\sun}$, 200--500\,au, and 1.001--4.5, respectively.
These values correspond to encounters that occur in star clusters and associations \citep[for example][and references therein]{2011A&A...532A.120L}.
Here we restrict to mass ratios $M_2/M_1$ from 0.5 to 10. 
We note that encounters with higher mass ratios (up to few hundred) are also expected in clustered environment \citep[for example][]{2010A&A...509A..63O}.

We further run simulations with different mutual inclination of the encounter plane and the plane of the disc (which is the plane of reference, see Fig.~\ref{fig:encounter_ic}), $i_{\rm{enc}}$, and different argument of periastron of the encounter, $\omega_{\rm{enc}}$.
Because the disc is axisymmetric (see below), the third angle defining the mutual geometry of the orbit of the two stars and the disc\,---\,the longitude of ascending node\,---\,does not play a role.
The initial conditions are illustrated in Fig.~\ref{fig:encounter_ic} and all the considered parameters values are specified in Table~\ref{tab:grid_par}.

The initial separation of the encountering stars is set up in such way that the amplitude of the gravitational force from the parent star $M_2$ at the distance of outer edge of the disc (200\,au from the star of mass $M_2$, see below) is ten times larger then the amplitude of the gravitational force from the other star with mass $M_1$.
Such an initial separation is sufficiently large and the influence of the star initially without the disc on the planetesimals is negligible.
We tested that increasing the initial separation up to where the force at the outer disc edge from star $M_1$ is 1\% of the one from star $M_2$ \citep[as used by][]{2014A&A...565A.130B} does not change the results \citep[in agreement with][]{1993MNRAS.261..190C,1996MNRAS.278..303H}.
We integrate the encounter until the separation between the two stars again reaches the initial value.
\citet{1996MNRAS.278..303H} pointed out that most of the interactions between the disc particles and the perturber takes place shortly after the closest approach.
We tested and confirmed that our integration time is sufficient for the captured particles to settle to their final state.

Planetesimals initially orbit star $M_2$ in a flat disc (that is we do not consider any vertical profile).
Unless specified otherwise, we use $10^4$ disc particles.
By increasing the number of particles (up to $5\cdot10^4$) in several simulations, we tested that the results do not change for higher resolutions.
For several specific encounters, we run simulations with six different values of the random seed. 
We estimate the error of our results by the standard deviations of the quantities we study below (such as the radii of the transfer region and the transfer efficiency, Sect.~\ref{sec:results}), which are smaller than 2 per cent for all the studied cases.

The initial radius and phase of the planetesimals in the disc are uniformly distributed, which corresponds to the surface number density $\propto 1/r$, where $r$ is the radial distance of the particles from the parent star measured in the plane of the disc.
Such a profile is often used to model protoplanetary discs \citep[e.g.][and references therein]{2013MNRAS.429..895P,2012A&A...538A..10S} and is supported by observations \citep[e.g.][]{2007ApJ...659..705A}.
Since the disc is represented by zero-mass particles, it is possible to re-scale the surface number density profile in post-processing of the simulations to represent different initial surface mass density radial profiles.
Each particle can be considered with a weight given by its initial radius so that the surface radial profile is a specific function of $r$ \citep[similarly as in][]{2012A&A...538A..10S}.
We discuss the role of the initial disc surface density in Sect.~\ref{sec:sigma}.
We set the radial extent of the disc to 30--200\,au. 
Such choice is consistent with disc sizes typically observed in clustered environments \citep[see for example][and references therein]{2015A&A...577A.115V}.
Unless specified otherwise, the planetesimals are initialized on circular orbits.

For most of our initial conditions, the disc and the orbit of the two stars are in the same plane and the $z$-components of their angular momentum have the same direction.
Such a coplanar prograde case results in the most violent encounters and with the highest number of transferred particles.
We carried out 1000 of such simulations and the grid of encounter parameters is given in the 1st section of Table~\ref{tab:grid_par}.
To estimate the effect of the general geometry, we varied the relative inclination of the plane of the encounter with respect to the disc, $i_{\rm{enc}}$, and also the argument of periastron of the orbit of the two stars, $\omega_{\rm{enc}}$.
We specify the encounter parameters in the 2nd and 3rd sections of the Table~\ref{tab:grid_par}.

To estimate the effect of eccentricity of the planetesimals orbits, $e_{{\rm disc}}$, we run simulations where the eccentricities are randomly selected from a uniform distribution from 0 to $e_{{\rm disc,max}}=0.05$ or $0.1$ (see last section of Table~\ref{tab:grid_par}).

\begin{table}
  \caption{Initial conditions for encounter simulations.
  The values of $q_{\rm{enc}}$, $M_2$, and $e_{\rm{enc}}$ on first section define a grid of 1000 runs.
  Further, the runs with $i_{\rm{enc}}$ and $\omega_{\rm{enc}}$ varied are listed.
  The last section describes the cases when the eccentricity of the disc particles, $e_{\rm{disc}}$, was varied.}
  \label{tab:grid_par}
  \begin{center}
  \begin{tabular}{lll}
  \hline
%   parameter & unit & values considered \\
  \multicolumn{3}{l}{grid parameters:} \\
%   \hline
  $q_{\rm{enc}}$ & [au] & 200, 220, 240, 260, 280, 300, 350, 400, 450, 500 \\
  $M_2$          & [M$_{\sun}$] & 0.1, 0.2, 0.3, 0.4, 0.5, 0.75, 1.0, 1.25, 1.5, 2.0 \\
  $e_{\rm{enc}}$ & & 1.001, 1.25, 1.5, 1.75, 2.0, 2.5, 3.0, 3.5, 4.0, 4.5 \\
  $i_{\rm{enc}}$ & [$\degr$] & 0 \\
  $\omega_{\rm{enc}}$ & [$\degr$] & 90 \\
  $e_{\rm{disc}}$ & & 0 \\
  \hline
  \multicolumn{3}{l}{varying $i_{\rm{enc}}$:} \\
  $q_{\rm{enc}}$ & [au] & 200, 300 \\
  $M_2$          & [M$_{\sun}$] & 0.1, 0.5, 1.0 \\
  $e_{\rm{enc}}$ & & 1.001, 1.5, 3.0 \\
  $i_{\rm{enc}}$ & [$\degr$] & 0--180, in steps of 15 \\
  $\omega_{\rm{enc}}$ & [$\degr$] & 90 \\
  $e_{\rm{disc}}$ & & 0 \\
  \hline
  \multicolumn{3}{l}{varying $\omega_{\rm{enc}}$:} \\
  $q_{\rm{enc}}$ & [au] & 200 \\
  $M_2$          & [M$_{\sun}$] & 0.1 \\
  $e_{\rm{enc}}$ & & 1.0 \\
  $i_{\rm{enc}}$ & [$\degr$] & 30, 60, 90, 120 \\
  $\omega_{\rm{enc}}$ & [$\degr$] & 0--180, in steps of 15 \\
  $e_{\rm{disc}}$ & & 0 \\
  \hline
  \multicolumn{3}{l}{varying $e_{\rm{disc}}$:} \\
  $q_{\rm{enc}}$ & [au] & 200, 300, 400 \\
  $M_2$          & [M$_{\sun}$] & 0.1, 0.5, 1.0, 1.5 \\
  $e_{\rm{enc}}$ & & 1.001, 1.5, 3.0 \\
  $i_{\rm{enc}}$ & [$\degr$] & 0 \\
  $\omega_{\rm{enc}}$ & [$\degr$] & 90 \\
  $e_{\rm{disc}}$ & & 0--0.05, 0--0.1 \\
  \hline
  \end{tabular}
  \end{center}
\end{table}
  
\section{Results}
\label{sec:results}

In each encounter experiment, we follow the disc particles that are transferred from star $M_2$ to star $M_1$ (initially with and without disc, respectively).
We calculate the orbital elements of the disc particles with respect to both stars\,---\,the semimajor axis, $a_s$, and eccentricity, $e_s$, inclination, $i_s$, and argument of periastron, $\omega_s$ (see Sect.~\ref{sec:orientation} for the discussion on the choice of the reference plane);
where the index $s$ identifies the star, that is $s=1,2$.
We identify the captured particles as those bound only to the star $M_1$ at the end of the simulations.
At the end of most of our simulations, a small fraction of the particles (typically less than 5\%) are bound to both stars.
By increasing the integration time (more than five times) and using full $N$-body simulations, we tested that these particles generally become bound only to star $M_2$ or unbound from the system and do not change the characteristics of the captured population.
To avoid low number statistics we consider only the simulations in which at least 100 particles are transferred (48 out of our 1000 coplanar prograde simulations result in 1--99 transferred particles).

We first focus on a description of the results of the systematic grid parameter space study (encounter parameters listed in the first part of Table~\ref{tab:grid_par}, Sects.~\ref{sec:transfer_rad}--\ref{sec:transferred_orbits}) and later on the cases with a more general geometry (Sects.~\ref{sec:inc_enc} and \ref{sec:omega_enc}), the case with an eccentric disc (Sect.~\ref{sec:ecc_disc}), and we also consider the role of the surface density of the disc (Sect.~\ref{sec:sigma})

\subsection{Transfer region}
\label{sec:transfer_rad}

The fate of a particle after the encounter is determined by its orbit in the initial disc around the parent star $M_2$.
We identify a minimal disc radius from where the particles can be transferred to $M_1$.
We call this radius $r_{\rm{tr,min}}$ and we show that it generally corresponds to the radius from where the particles can be unbound from the parent star $M_2$, $r_{\rm{un,min}}$.
It was already pointed out by \citet{2001Icar..153..416K} that $r_{\rm{un,min}}$ depends on the parameters of the encounter\,---\,the mass ratio $M_1/M_2$, pericentre $q_{\rm{enc}}$, and the eccentricity $e_{\rm{enc}}$\,---\,as 
\begin{align}
\label{eq:r_truncation_Kob2001}
r_{\rm{un,min}} \approx \alpha \left[(1+M_1/M_2)(1+e_{\rm{enc}})\right]^{(-1/3)} q_{\rm{enc}}, 
\end{align}
where $\alpha \approx 0.3$ or 0.5 for a prograde or retrograde encounter, respectively. 
Our results show similar trends and in Sect.~\ref{sect:analytic}, we provide a detailed description of a derivation of an approximate analytic formula for $r_{\rm{tr,min}}(M_1/M_2,q_{\rm{enc}},e_{\rm{enc}})$ and compare it with the results from our simulations in Sect.~\ref{sec:comparison}.

\begin{figure}
\center
\includegraphics[width=0.49\textwidth]{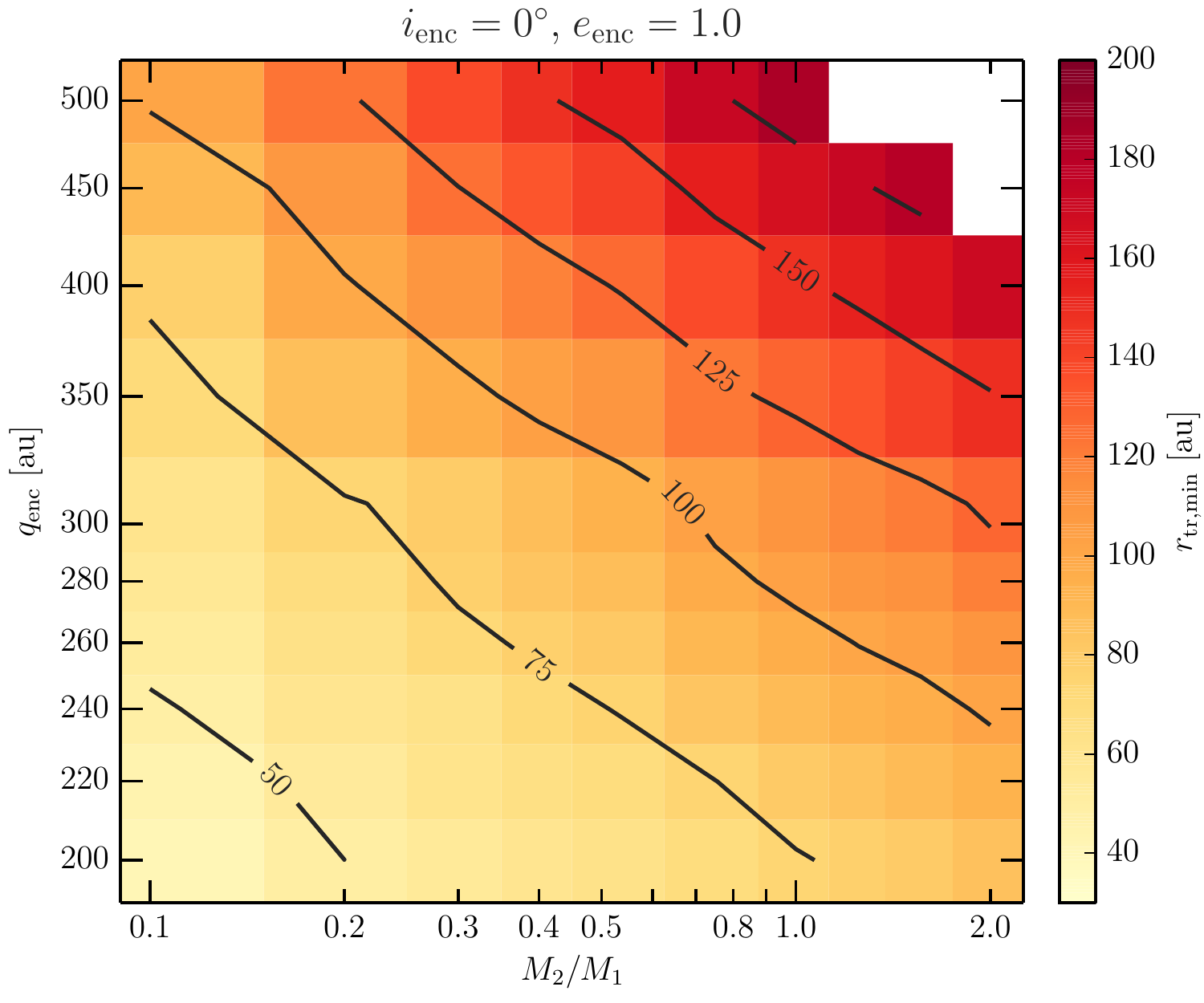}
\caption{Minimal disc radius from where the particles can be transferred $r_{\rm{tr,min}}$ for coplanar prograde  ($i_{\rm{enc}}=0\degr$) parabolic ($e_{\rm{enc}}=1.0$) encounters.
The horizontal axis shows the mass ratio of star initially with to without disc $M_2/M_1$,
the vertical axis shows the pericentre of the encounter $q_{\rm{enc}}$.
Note that both horizontal and vertical axes are logarithmic.
The color scale maps the minimal transfer radius $r_{\rm{tr,min}}$.
The contour levels are in au.}
\label{fig:rmin_map_q_m}
\end{figure}

\begin{figure}
\center
\includegraphics[width=0.49\textwidth]{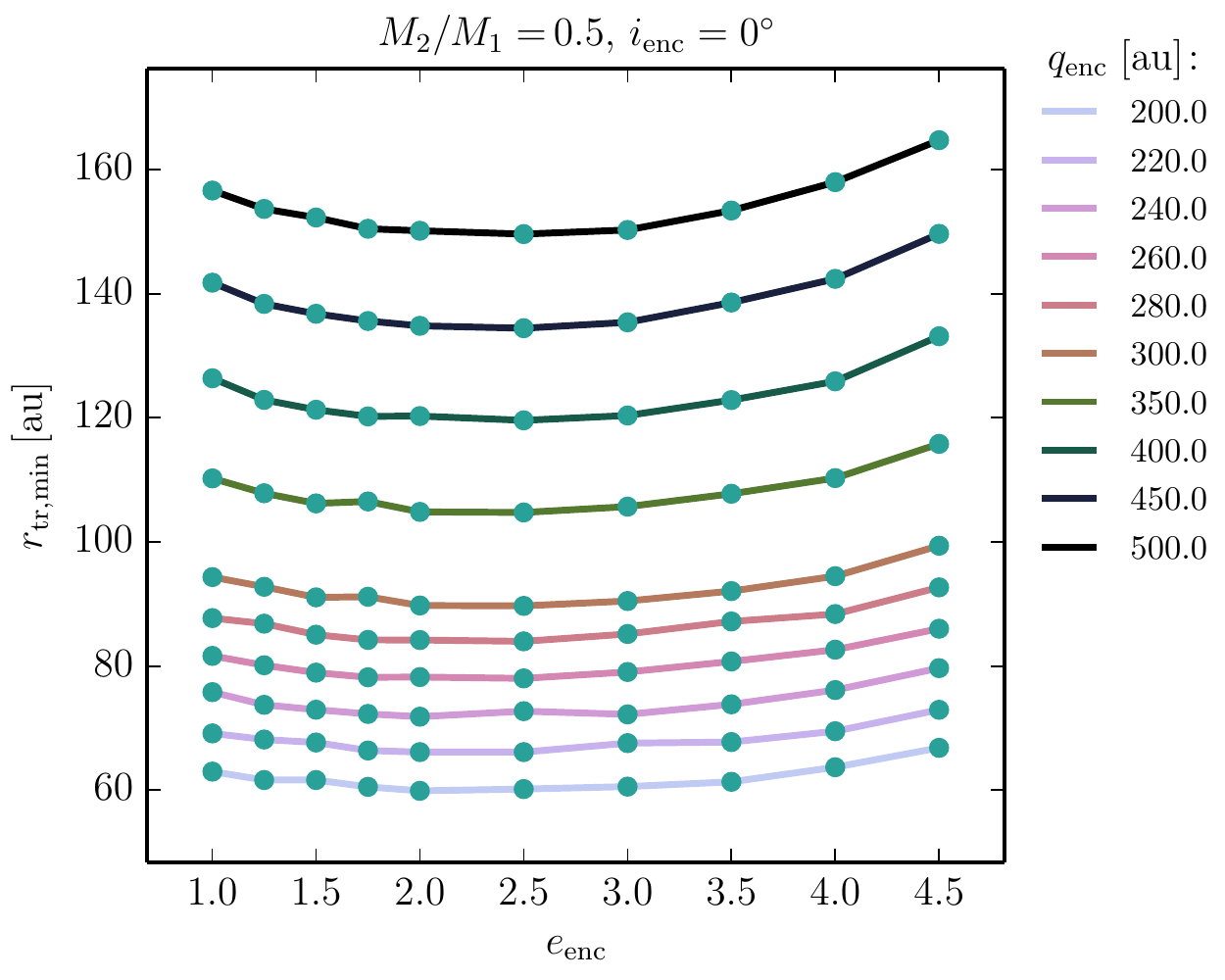}
\caption{Minimal transfer radius, $r_{\rm{tr,min}}$, for encounters with mass ratio $M_2/M_1=0.5$ and different pericentres $q_{\rm{enc}}$ as a function of eccentricity, $e_{\rm{enc}}$.
Here, $r_{\rm{tr,min}}$ for the parabolic orbits ($e_{\rm{enc}}=1.0$) correspond to the values mapped in Fig.~\ref{fig:rmin_map_q_m} for the mass ratio $M_2/M_1=0.5$ fixed on the horizontal axis.}
\label{fig:rmin_ecc_m05}
\end{figure}

In Fig.~\ref{fig:rmin_map_q_m}, we show the dependence of $r_{\rm{tr,min}}$ on the mass ratio of the encounter stars and the pericentre of the encounter for the coplanar prograde parabolic encounters (i.e. for which the highest number of particles is transferred).
The initial radial extent of the disc is 30--200\,au (Sect.~\ref{sec:ic}) which also sets the limits on $r_{\rm{tr,min}}$.
For the cases with large pericentres ($q_{\rm{enc}}\apgt450$\,au) and high mass ratios ($M_2/M_1\apgt1.25$), $r_{\rm{tr,min}}$ is close to or larger than the initial disc extent of 200\,au;
for smaller pericentres ($q_{\rm{enc}}\aplt200$\,au) and low mass ratios ($M_2/M_1\aplt0.1$), $r_{\rm{tr,min}}$ is decreasing up to the lower disc size limit of 30\,au.
Fig.~\ref{fig:rmin_map_q_m} demonstrates the dependency of $r_{\rm{tr,min}}$ on the pericentre of the encounter $q_{\rm{enc}}$ and the mass ratio $M_2/M_1$ and in Fig.~\ref{fig:rmin_ecc_m05}, we show that the minimal transfer radius $r_{\rm{tr,min}}$ depends only weakly on the encounter eccentricity $e_{\rm{enc}}$ \citep[in agreement with][]{2001Icar..153..416K}.

\begin{figure}
\center
\includegraphics[width=0.49\textwidth]{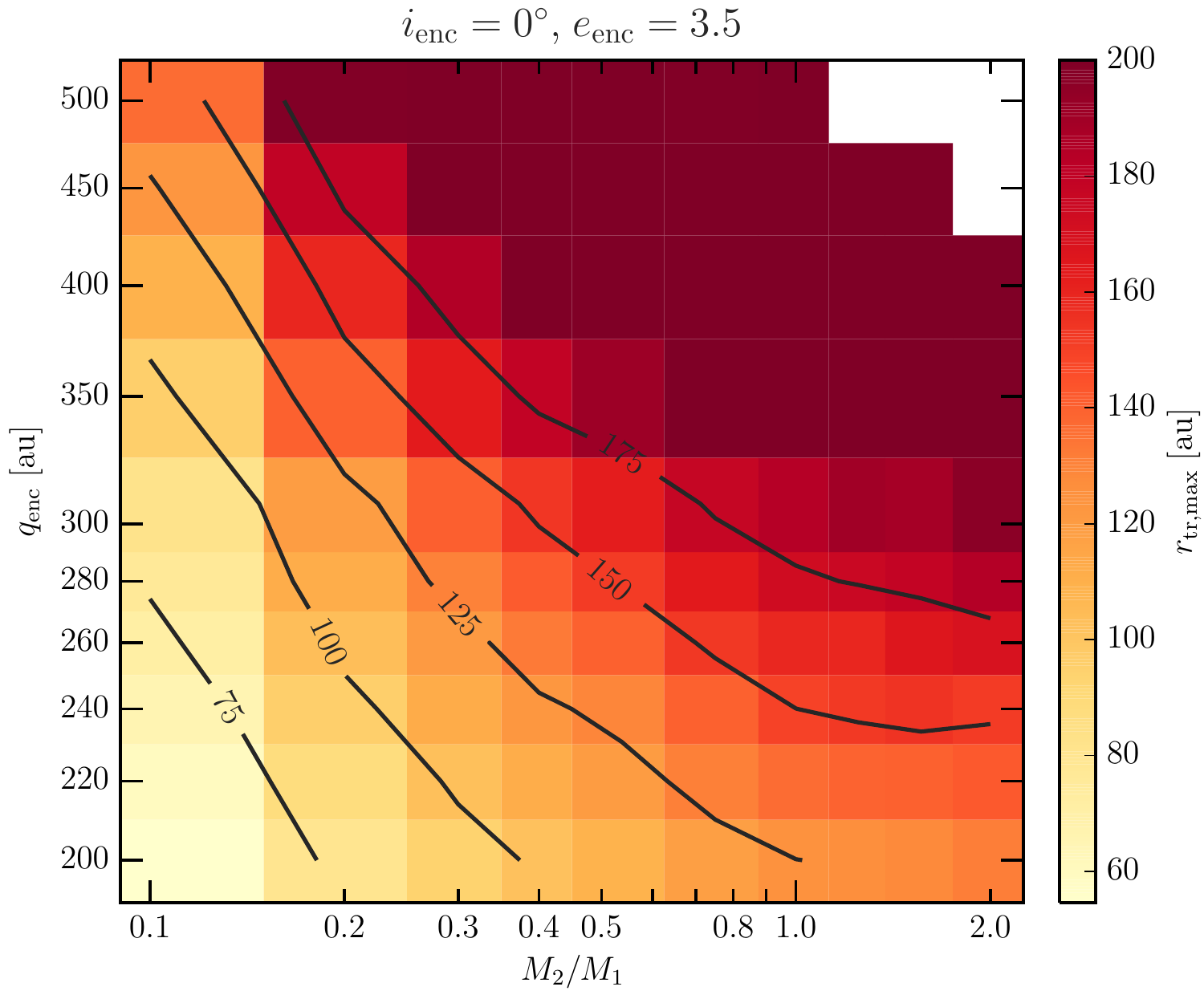}
\caption{Maximal disc radius from where the particles can be transferred, $r_{\rm{tr,max}}$, for coplanar prograde  ($i_{\rm{enc}}=0\degr$) encounters with $e_{\rm{enc}}=3.5$.
The horizontal axis shows the mass ratio of the stars $M_2/M_1$,
the vertical axis shows the pericentre of the encounter.
Note that both horizontal and vertical axes are logarithmic.
The color scale maps the $r_{\rm{tr,max}}$.
Here, 200\,au is the initial outer edge of the disc and therefore a lower limit of $r_{\rm{tr,max}}$.
The contour levels are in au.}
\label{fig:rmax_map_q_m}
\end{figure}

For faster hyperbolic encounters (with eccentricities $e_{\rm{enc}}\apgt 2.5$), we also identify a maximal radius from up to where the particles are transferred to star $M_1$, $r_{\rm{tr,max}}$.
For encounters with $e_{\rm{enc}}\aplt 2.5$, the particles are transferred from up to the outer edge of the disc of 200\,au which corresponds to the lower limit on $r_{\rm{tr,max}}$.
In Fig.~\ref{fig:rmax_map_q_m}, we show $r_{\rm{tr,max}}$ for coplanar prograde encounters with $e_{\rm{enc}}=3.5$.
As expected and similarly to $r_{\rm{tr,min}}$, $r_{\rm{tr,max}}$ also increases with larger mass ratios and larger pericentres.

\begin{figure}
\center
\includegraphics[width=0.49\textwidth]{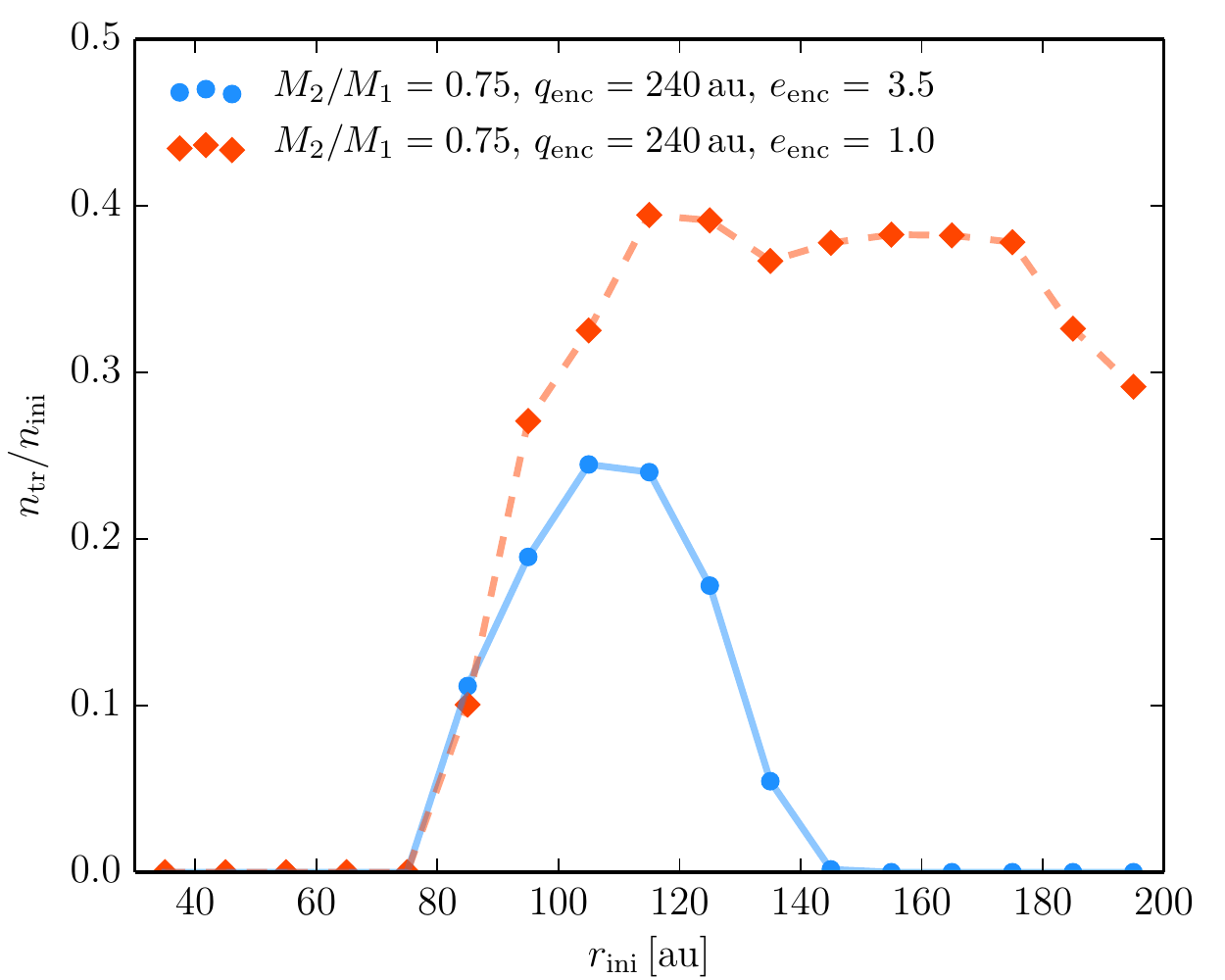}
\caption{Radial distribution of the relative number of transferred particles for two coplanar prograde encounters with the same mass ratio ($M_2/M_1=0.75$) and pericentre ($q_{\rm{enc}}=240$\,au) and different eccentricities.
The horizontal axis shows the initial disc radius of the particles $r_{\mathrm{ini}}$.
The distribution is calculated in equidistant radial bins of 10\,au.
The ratio of the transferred ($n_{\mathrm{tr}}$) to the initial ($n_{\mathrm{ini}}$) number of particles in each bin is shown on the vertical axis.
Blue bullets connected by solid line and red diamonds connected by dashed line correspond to the encounter eccentricity of 1.0 and 3.5, respectively.}
\label{fig:rdist_multi}
\end{figure}

\begin{figure*}
\center
\includegraphics[height=0.25\textheight]{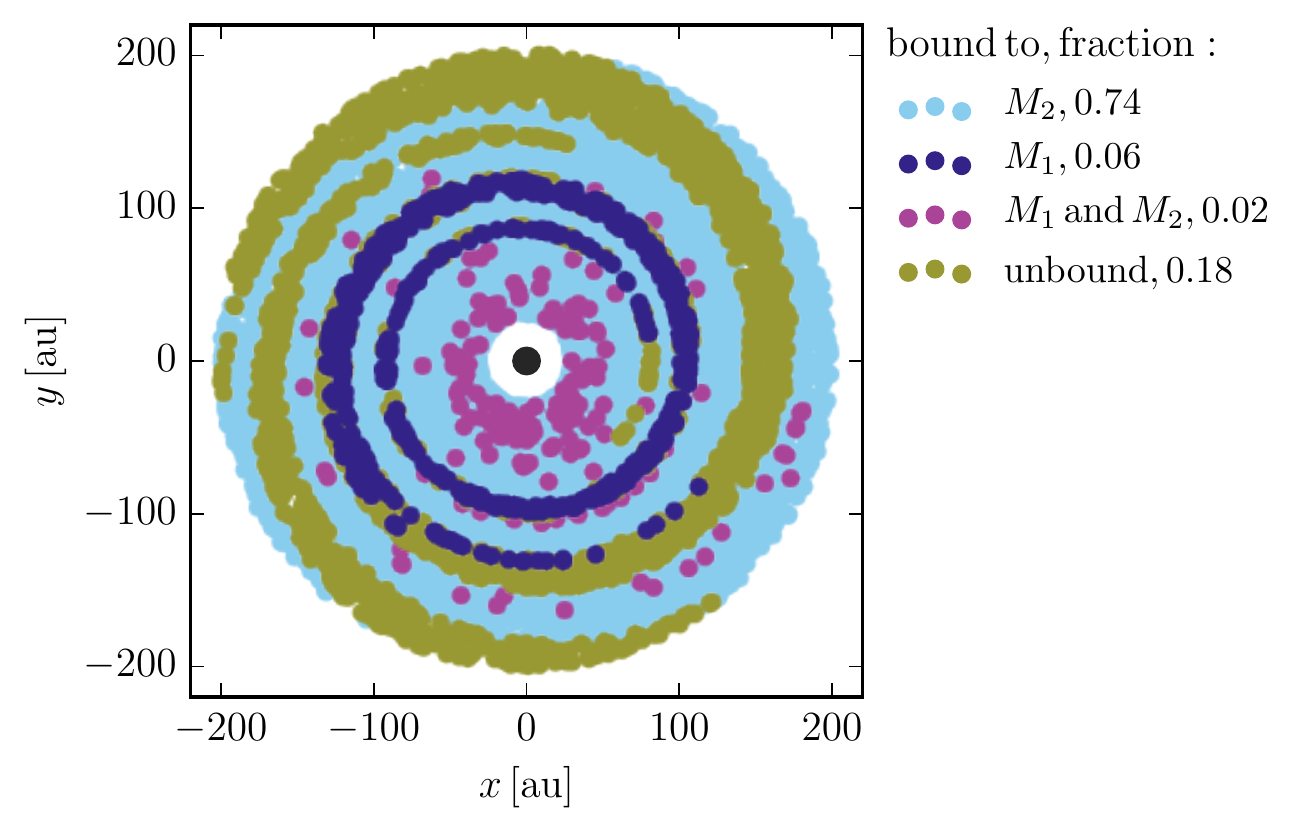}
\includegraphics[height=0.25\textheight]{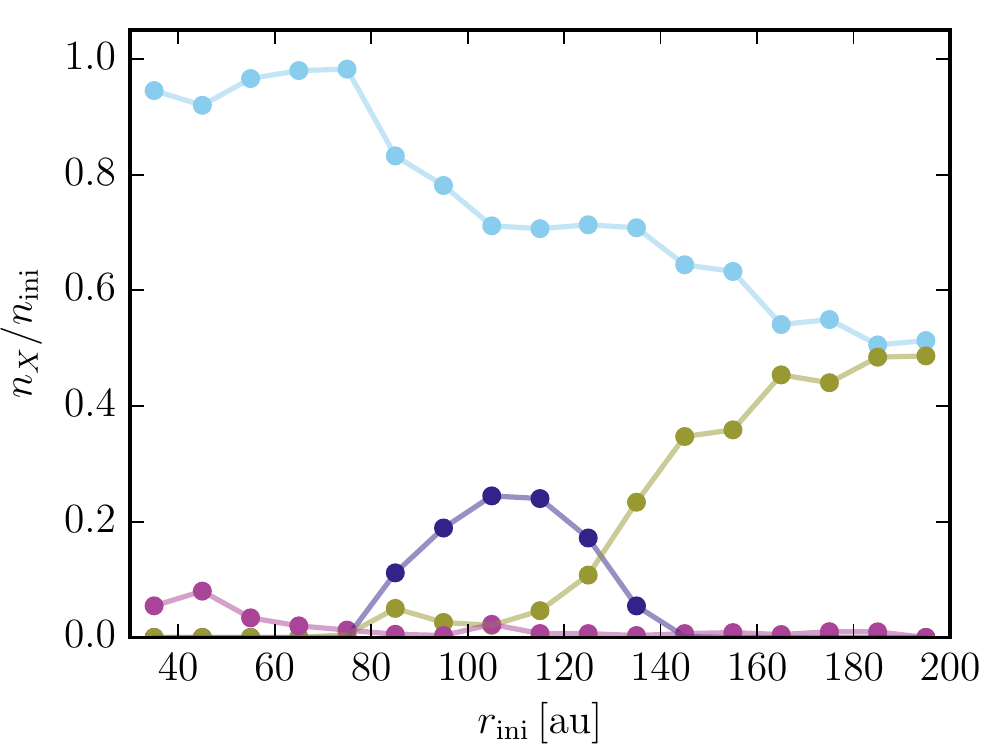}
\caption{Initial distribution of the disc for different fate of the particles for the encounter with $M_2/M_1=0.75$, $q_{\rm{enc}}=240$\,au, and $e_{\rm{enc}}= 3.5$.
{\it Left:} The plane of the disc with the coordinates in an non-inertial reference frame centered on the star $M_2$ (marked by black bullet in the centre). 
Disk particles are color-coded according to their final fate after the encounter as indicated in the legend where the fraction for each option is also given.
{\it Right:} Relative distributions of the initial disc radius of the particles of different final status.
The distributions are calculated in equidistant radial bins of 10\,au.
The ratio of the number of particles ($n_{X}$, where $X$ stands for bound to $M_2$ or/and $M_1$ or unbound from the system) to the initial ($n_{\mathrm{ini}}$) number of particles in each bin is shown on the vertical axis.}
\label{fig:disc_ini_fin}
\end{figure*}

\begin{figure}
\center
\includegraphics[width=0.49\textwidth]{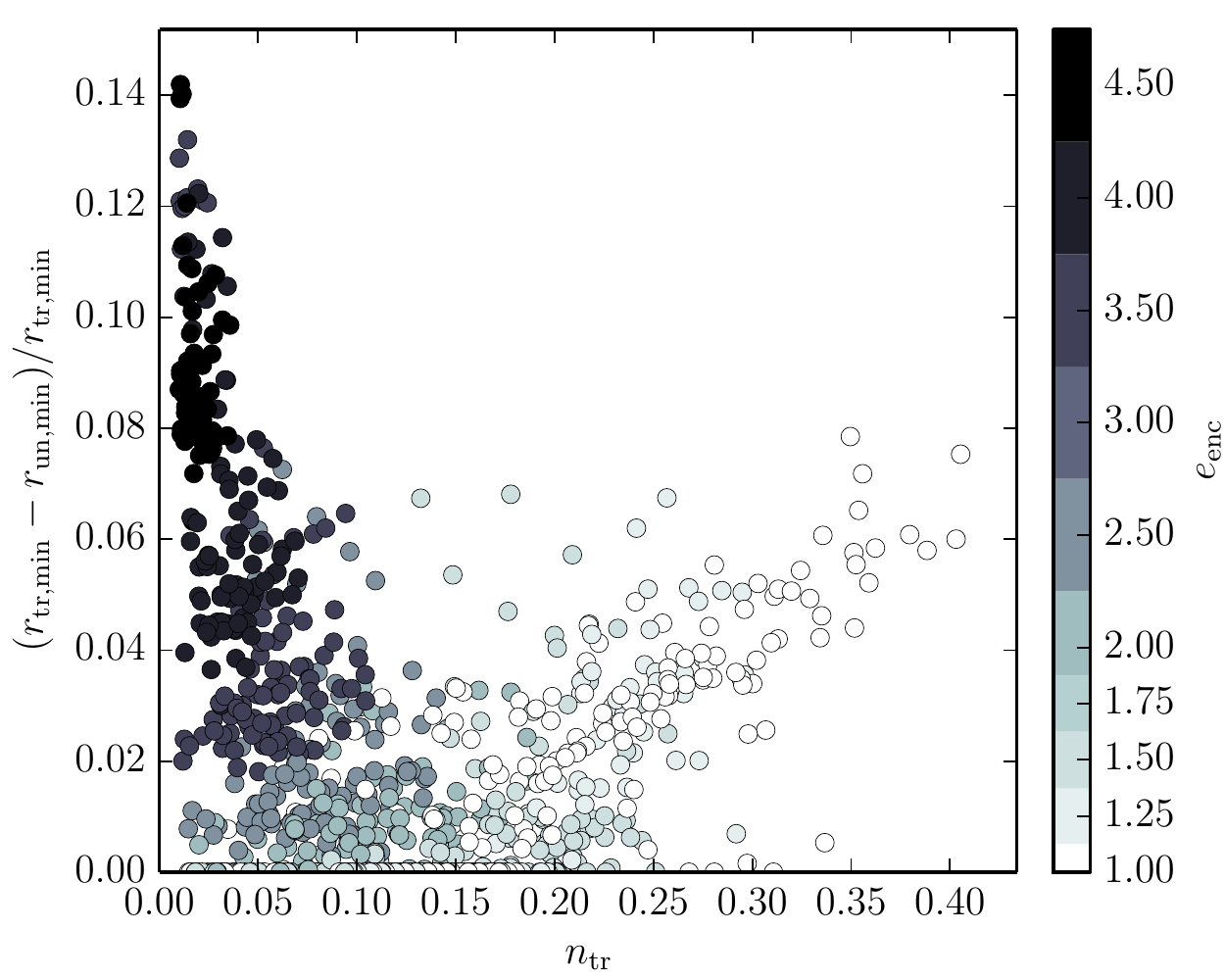}
\caption{Relative difference between the minimal transfer radius $r_{\rm{tr,min}}$, and the minimal radius of the unbound particles $r_{\rm{un,min}}$, as a function of the relative number of transferred particles $n_{\rm{tr}}$ for the 1000 encounters of the parameters grid.
Coplanar encounters are plotted and color-coded by the eccentricity $e_{\rm{enc}}$ (regardless the pericenter $q_{\rm{enc}}$ and mass $M_2$).}
\label{fig:rmin_tr_unbound}
\end{figure}

In Fig.~\ref{fig:rdist_multi}, we show the radial distribution of the relative number of transferred particles for the encounters with $M_2/M_1=0.75$ and $q_{\rm{enc}}=240$\,au and for two different encounter eccentricities $e_{\rm{enc}}=1.0$ and 3.5, which are also plotted in Figs.~\ref{fig:rmin_map_q_m} and \ref{fig:rmax_map_q_m}.
In the case of the slower, parabolic encounter, more particles are transferred from a wider radial range, where the outer edge is outside the initial extent of the disc (of 200\,au, see red diamonds and dashed line in Fig.~\ref{fig:rdist_multi})
The faster encounter results in fewer transferred particles and $r_{\rm{tr,max}}<200$\,au.

We note that in about 25 simulations (out of the total 1000) a small number of {\it outlier} particles (this is always \aplt 5 particles, which is never more than 5\% of the total number of the captured particles) are transferred from outside the initial disc region limited by $r_{\rm{tr,min}}$ and $r_{\rm{tr,max}}$. 
This is a result of the approximate hybrid approach when integrating the orbits of the disc particles (see Sect.~\ref{sec:num}) and we tested the method correcting for the outliers by comparing with the $N$-body simulations.

The initial position in the disc for the particles of different final fate is showed in Fig.~\ref{fig:disc_ini_fin}, for the encounter with $M_2/M_1=0.75$, $q_{\rm{enc}}=240$\,au, and $e_{\rm{enc}}=3.5$ (same encounter as already shown in Fig.~\ref{fig:rdist_multi} in the red distribution).
We show the initial disc around star $M_2$ color coded by the fate of the particles after the encounter in the left panel of Fig.~\ref{fig:disc_ini_fin}; and the radial distributions of the relative number of the particles in the right panel.
Most of the particles, 74\% of the initial $10^4$, stay bound to the parent star; 
18\% are unbound from the system and lost into interstellar space;
6\% are transferred to star $M_1$; 
and 2\% are left bound to both of the stars.
The small number of particles that are bound to both stars (shown in violet) is initially spread over the whole radial extent of the disc; these particles will typically end-up unbound from the system or bound only to star $M_2$.
As can be seen from the blue radial distribution in Fig.~\ref{fig:rdist_multi}, the transferred particles (shown in dark blue) are initially enclosed in between two radii.
Similarly, the particles that are eventually unbound from the system (shown in green) are limited by a minimal radius, called $r_{\rm{un,min}}$, that is very similar to the minimal radius of the transferred particles $r_{\rm{tr,min}}$.
In Fig.~\ref{fig:rmin_tr_unbound}, we show the relative difference between $r_{\rm{tr,min}}$ and $r_{\rm{un,min}}$ (the later is always smaller than the former).
The difference between the two minimal radii is less than 8\% of $r_{\rm{tr,min}}$ for all the encounter with $e_{\rm{enc}}\aplt 3.5$, and less than 15\% for the cases with higher eccentricities.
This might result from the faster encounters generally producing a smaller number of transferred particles.

\subsection{Transfer efficiency}
\label{sec:transfer_eff}

\begin{figure}
\center
\includegraphics[width=0.49\textwidth]{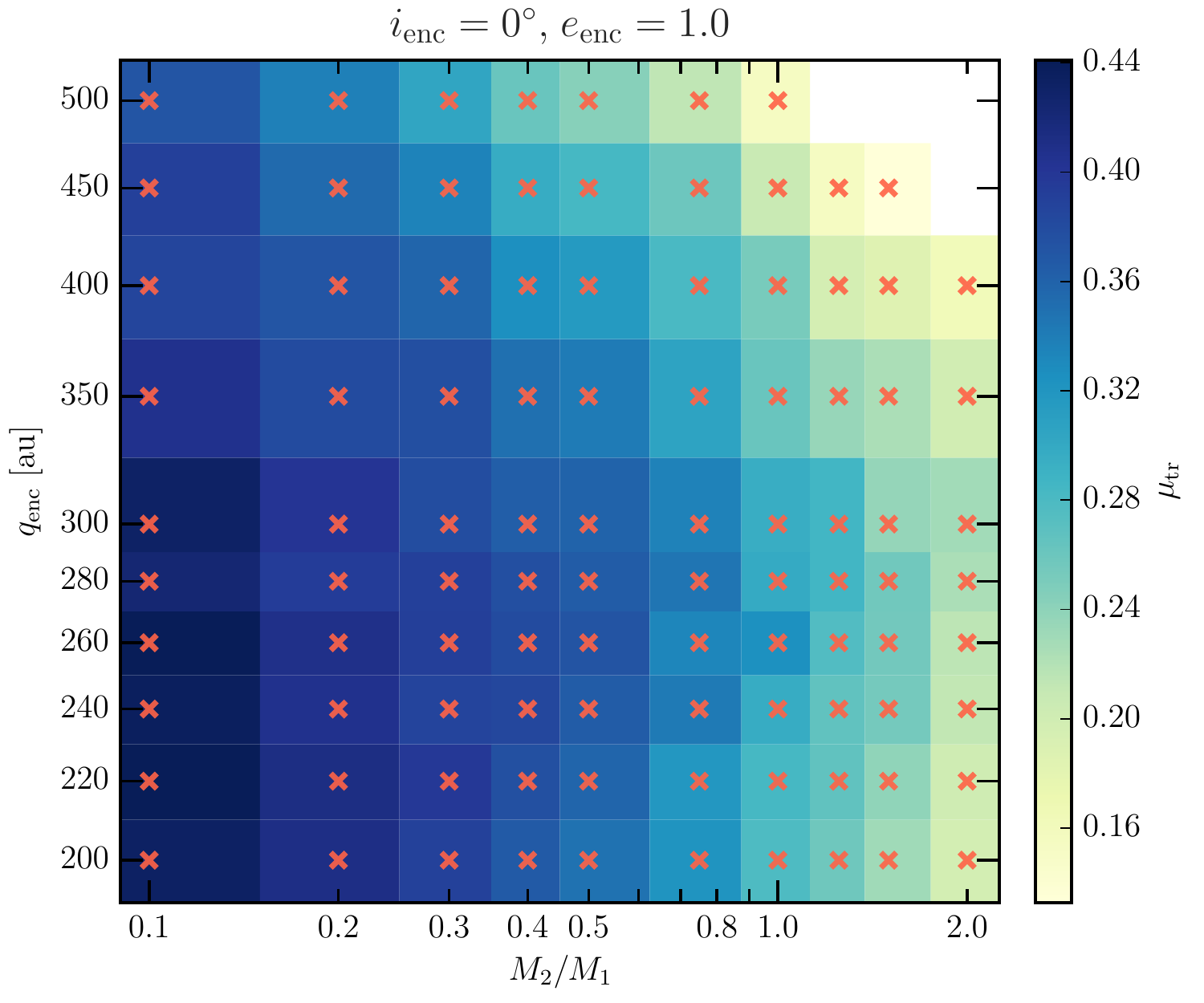}
\caption{Efficiency of mass transfer $\mu_{\rm{tr}}$ for coplanar prograde parabolic encounters.
The horizontal axis shows the mass ratio $M_2/M_1$, the vertical axis shows the pericentre of the encounter $q_{\rm{enc}}$.
Note that both horizontal and vertical axes are logarithmic.
The color scale maps the $\mu_{\rm{tr}}$.
The red crosses mark the bins where $r_{\rm{tr,max}}>200$\,au and the transfer region is not completely covered\,---\,which is the case for all encounters here.
}
\label{fig:mu_tr_e1}
\end{figure}

\begin{figure}
\center
\includegraphics[width=0.45\textwidth]{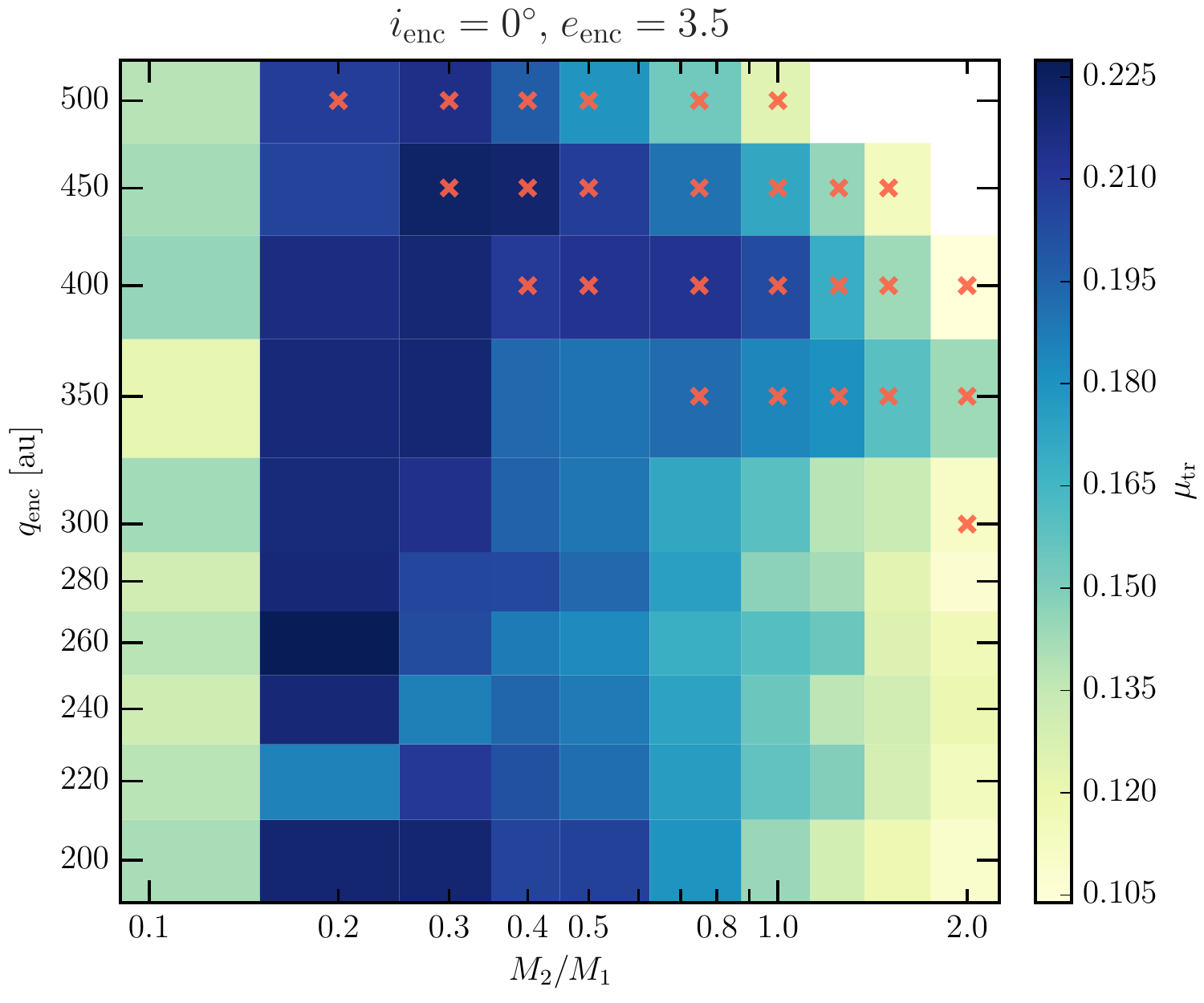}
\caption{Efficiency of mass transfer $\mu_{\rm{tr}}$ for coplanar prograde encounters with an eccentricity of 3.5.
See Fig.~\ref{fig:mu_tr_e1} for a detailed description.}
\label{fig:mu_tr_e3.5}
\end{figure}

\begin{figure}
\center
\includegraphics[width=0.45\textwidth]{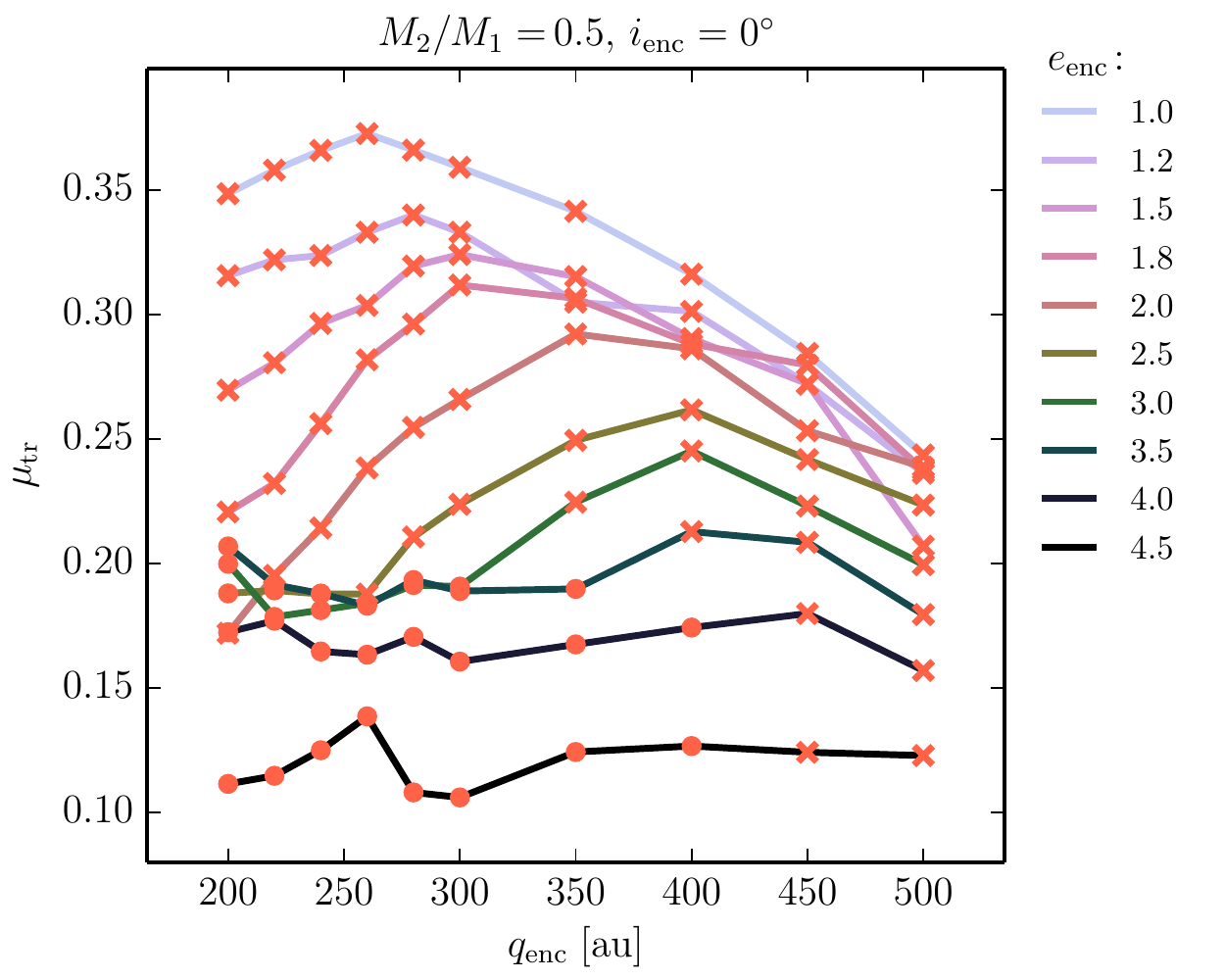}
\caption{Dependency of the efficiency of mass transfer $\mu_{\rm{tr}}$ on the pericentre of the encounter $q_{\rm{enc}}$.
The encounters are coplanar and prograde encounters with mass ratio $M_2/M_1=0.5$.
Lines of different colors correspond to different encounter eccentricities, $e_{\rm{enc}}$, as indicated to the right.
Bullets depict the encounters with completely covered transfer region, while crosses the encounters with $r_{\rm{tr,max}}>200$\,au.
}
\label{fig:mu_tr_mass05}
\end{figure}

To measure the efficiency of the transfer we follow the ratio of the number of transferred particles and the number of particles initially located in the original disc within the range $r_{\rm{tr,min}}$--$r_{\rm{tr,max}}$.
We call this quantity the transfer efficiency, $\mu_{\rm{tr}}$.
It is important to keep in mind that as showed in Sect.~\ref{sec:transfer_rad}, for a substantial part of the studied parameter space, the maximal disc radius of the transferred particles $r_{\rm{tr,max}}$ is larger than the considered outer edge of the disc of 200\,au.
The complete transfer region of transfers is not covered for these cases.

\begin{figure}
\center
\includegraphics[width=0.45\textwidth]{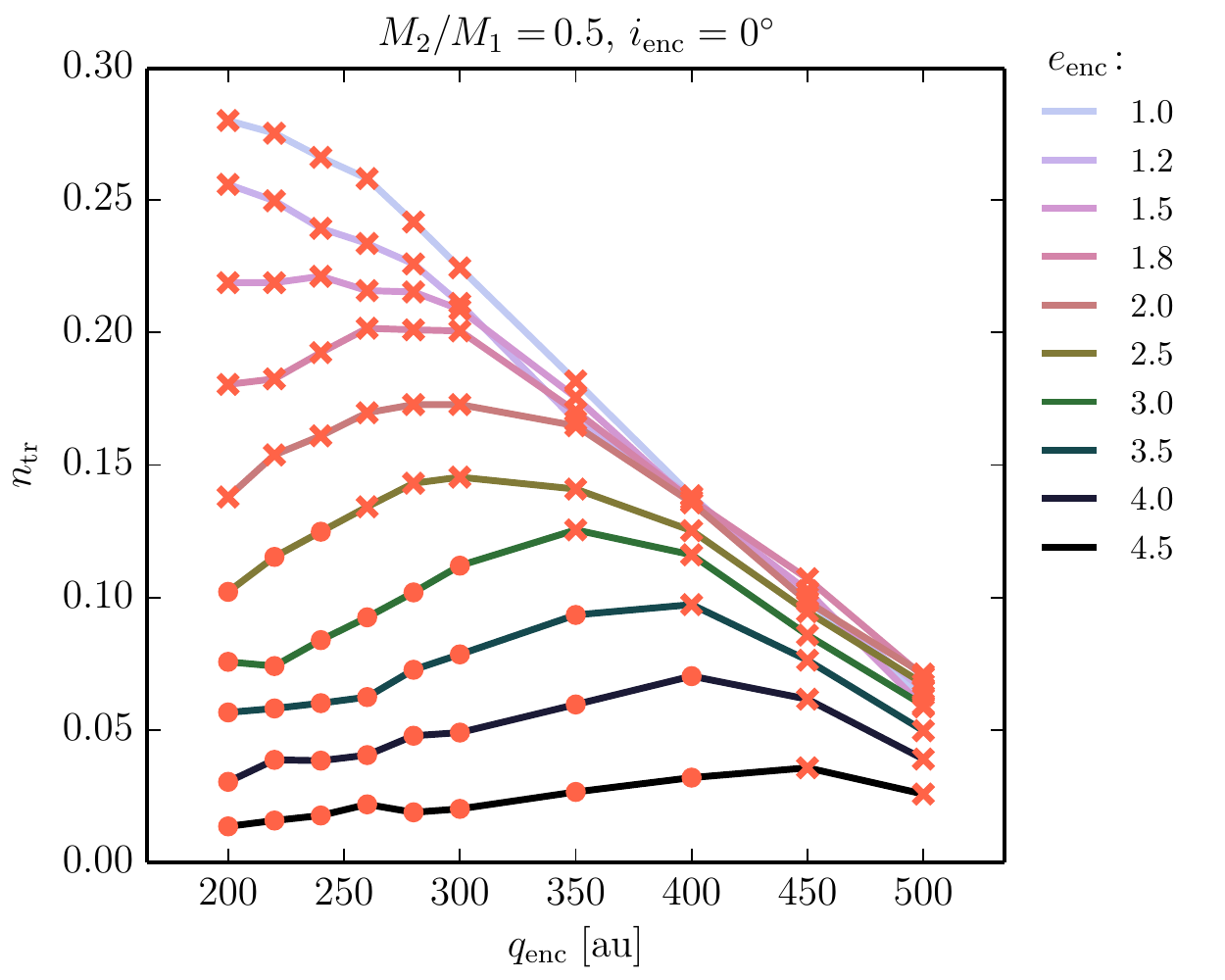}
\caption{Dependency of the relative number of the transferred particles $n_{\rm{tr}}$ on the pericentre of the encounter  $q_{\rm{enc}}$ for mass ratio $M_2/M_1=0.5$, and different eccentricities $e_{\rm{enc}}$.
See Fig.~\ref{fig:mu_tr_mass05} for a detailed description.}
\label{fig:n_tr_mass05}
\end{figure}

\begin{figure}
\center
\includegraphics[width=0.45\textwidth]{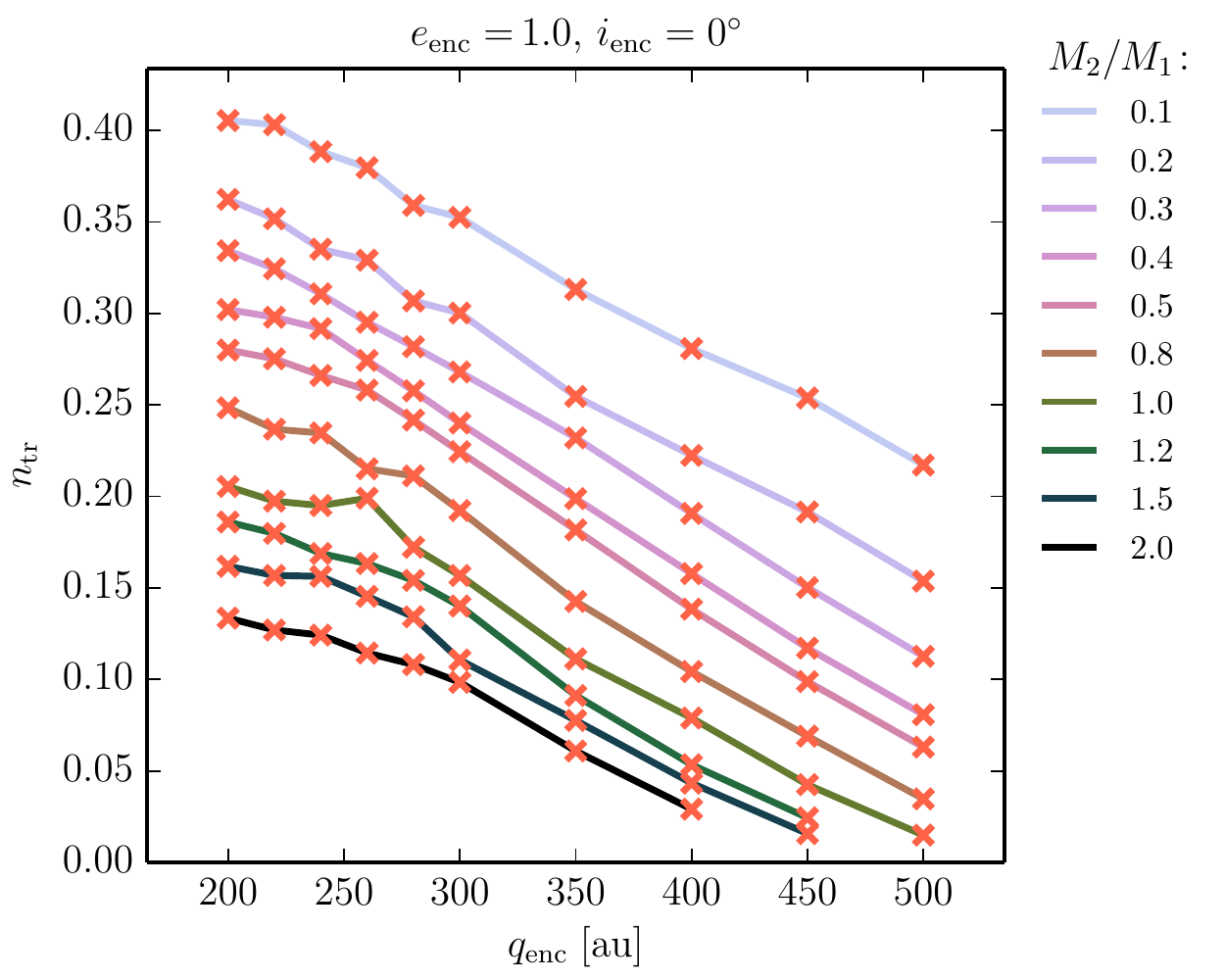}
\caption{Dependency of the relative number of the transferred particles $n_{\rm{tr}}$ on the pericentre of the encounter $q_{\rm{enc}}$ for coplanar prograde parabolic encounters for different mass ratio $M_2/M_1$.
Similarly to Fig.~\ref{fig:mu_tr_mass05}, lines of different colors correspond to different mass ratios $M_2/M_1$, as indicated to the right.
Here, points are indicated by crosses because the transfer region is not completely covered for any of the displayed simulations (i.e. $r_{\rm{tr,max}}>200$\,au; see Fig.~\ref{fig:mu_tr_mass05}).}
\label{fig:n_tr_e1}
\end{figure}

In Figs.~\ref{fig:mu_tr_e1} and ~\ref{fig:mu_tr_e3.5}, we present $\mu_{\rm{tr}}$ for the pericentres $q_{\rm{enc}}$, and mass ratios $M_2/M_1$, of the encounter for fixed eccentricities $e_{\rm{enc}}$ of 1.0 and 3.5, respectively.
We mark the encounters for which the transfer region is not completely covered (i.e., $r_{\rm{tr,max}}>200$\,au) by the red cross.
Regardless of the incompleteness of the data, lower mass ratios result in higher transfer efficiencies.
This is consistent with the conclusion of \citet[][note that they defined the mass ratio of the encountering star inverse to the one used here]{2005A&A...437..967P}.
However, for eccentric orbits (Fig.~\ref{fig:mu_tr_e3.5}), the encounters with the lowest considered mass ratio $M_2/M_1=0.1$, have a lower transfer efficiency than encounters with $M_2/M_1=0.2$.
The same feature is present for all higher encounter eccentricities $e_{\rm{enc}}\apgt3.0$.
By increasing the number of the disc particles (to $5\cdot10^4$) and by decreasing the initial disc extent (so that the inner and outer disc edges are closer to the values of $r_{\rm{tr,min}}$ and $r_{\rm{tr,max}}$, respectively), we tested that this result is not resolution dependent.
To increase the resolution in the mass ratio $M_2/M_1$ of our grid, we also run additional simulations with $M_2/M_1=0.125$, 0.15, and 0.175\,M$_{\sun}$ for the encounters with $e_{\rm{enc}}=3.5$, and we find that the transfer efficiency $\mu_{\rm{tr}}$ is indeed lowest for the $M_2/M_1=0.1$ and smoothly increasing up to $M_2/M_1=0.2$.
The transfer efficiency is generally higher for the parabolic encounters ($\mu_{\rm{tr}}=0.1$--0.23 for $e_{\rm{enc}}=3.5$ in Fig.~\ref{fig:mu_tr_e3.5} while $\mu_{\rm{tr}}=0.15$--0.45 for $e_{\rm{enc}}=3.5$ in Fig.~\ref{fig:mu_tr_e1}) as was already noted by \citet{2005ApJ...629..526P}.

Regarding the dependence of $\mu_{\rm{tr}}$ on the pericentre of the encounter $q_{\rm{enc}}$,
Figs.~\ref{fig:mu_tr_e1} and Fig.~\ref{fig:mu_tr_e3.5} indicate that the transfer ratio is maximal for a particular value of $q_{\rm{enc}}$ which is the same for different mass ratios\,---\,260\,au for the parabolic encounters (Fig.~\ref{fig:mu_tr_e1}) and $\sim350$--400\,au for $e_{\rm{enc}}=3.5$ (Fig.~\ref{fig:mu_tr_e1}).
In Fig.~\ref{fig:mu_tr_mass05}, we show the dependency of $\mu_{\rm{tr}}$ on $q_{\rm{enc}}$ for encounters with a fixed mass ratio but a range in eccentricities.
The transfer efficiency is higher at larger pericentre for higher values of $e_{\rm{enc}}$.
Regardless of the pericentre, the transfer efficiency is generally smaller for more eccentric encounters.

In previous studies, the number of transferred particles was followed rather than its ratio to the initial population in the transfer region, that is than what we call the transfer efficiency $\mu_{\rm{tr}}$.
In Fig.~\ref{fig:n_tr_mass05} we plot the number of transferred particles as the fraction of the total number of disc particles, $n_{\rm{tr}}$.
Fig.~\ref{fig:n_tr_mass05} has the same setup as Fig.~\ref{fig:mu_tr_mass05}, where the mass ratio is fixed and the dependence on the encounter pericentre is shown for different eccentricities.
Most of the previous work focused on the case of a coplanar prograde parabolic orbit which leads the most efficient transfer for any given mass ratio and pericentre.
The number of transferred particles $n_{\rm{tr}}$ decreases almost linearly with $q_{\rm{enc}}$ in this case in agreement with \citet{2005A&A...437..967P}.
This is further confirmed in Fig.~\ref{fig:n_tr_e1}, where we plot $n_{\rm{tr}}$ for all the coplanar prograde parabolic encounters for different $M_2/M_1$.
The linear decrease has an approximately constant slope irrespective of the masses of the stars, while is generally lower for higher mass ratios (that is for large masses of the star initially with the disc, $M_2$).

The number of transferred particles as well as the transfer efficiency can in principle depend on the surface number density profile of the disc particles.
As mentioned in Sect.~\ref{sec:ic}, we adopted an initial uniform distribution in $r$ which corresponds to the surface density $\propto 1/r$.
To address this effect, we consider different surface density profiles in Sect.~\ref{sec:sigma}.

\subsection{Orbits of the transferred planetesimals}
\label{sec:transferred_orbits}

\begin{figure}
\center
\includegraphics[width=0.45\textwidth]{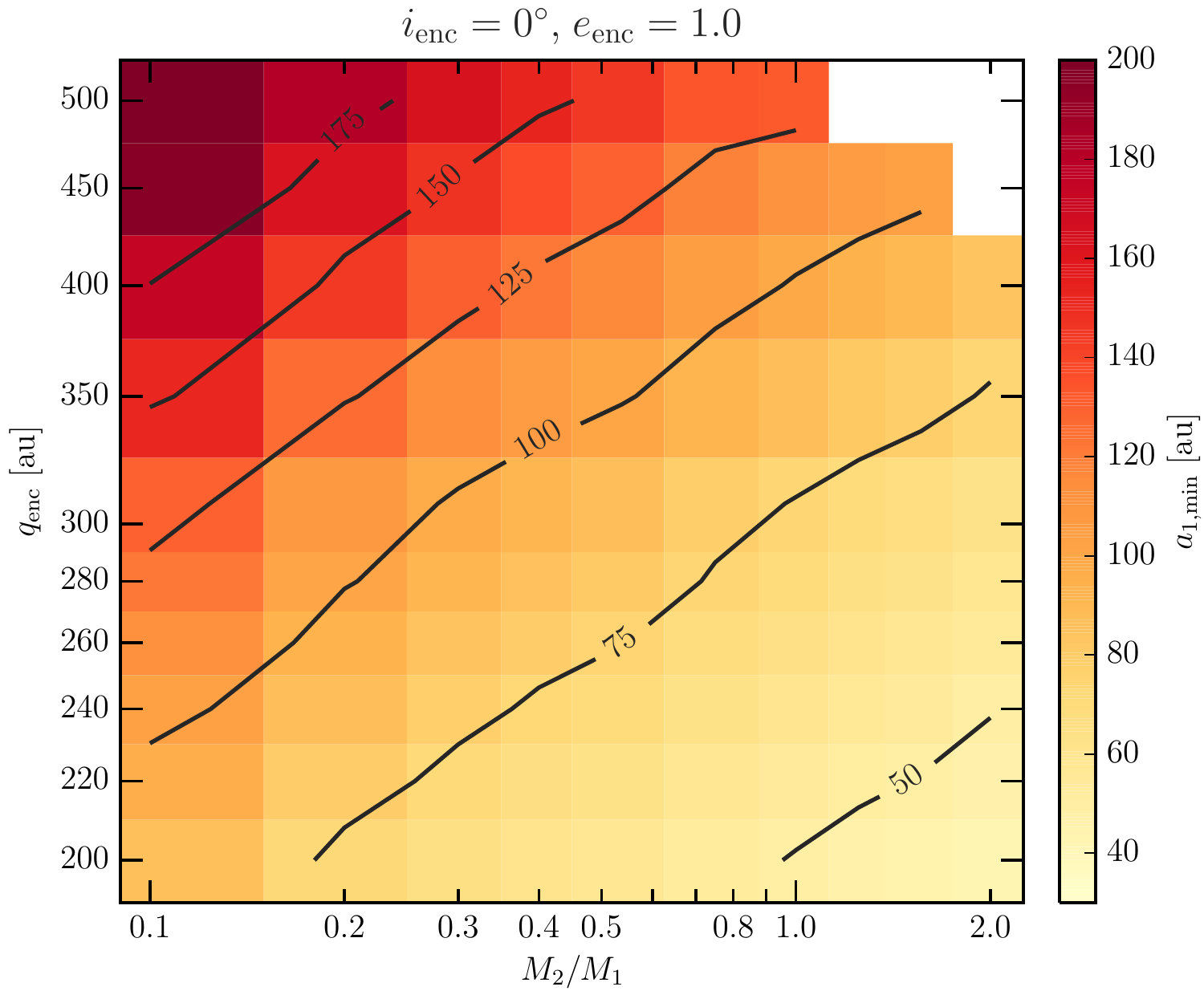}
\caption{Minimal semimajor axis of the transferred particles for the coplanar prograde parabolic encounters ($e_{\rm{enc}}=1.0$, $i_{\rm{enc}}=0\degr$).
The mass ratio $M_2/M_1$ and the pericentre of the encounter $q_{\rm{enc}}$ is increasing along the horizontal axis and vertical axis, respectively.
Note that both horizontal and vertical axes are logarithmic.
The color scale maps the minimal semimajor axis of the orbits transferred around the star $M_1$.
The contour levels are in au.}
\label{fig:amin_map_q_m}
\end{figure}

\begin{figure}
\center
\includegraphics[width=0.45\textwidth]{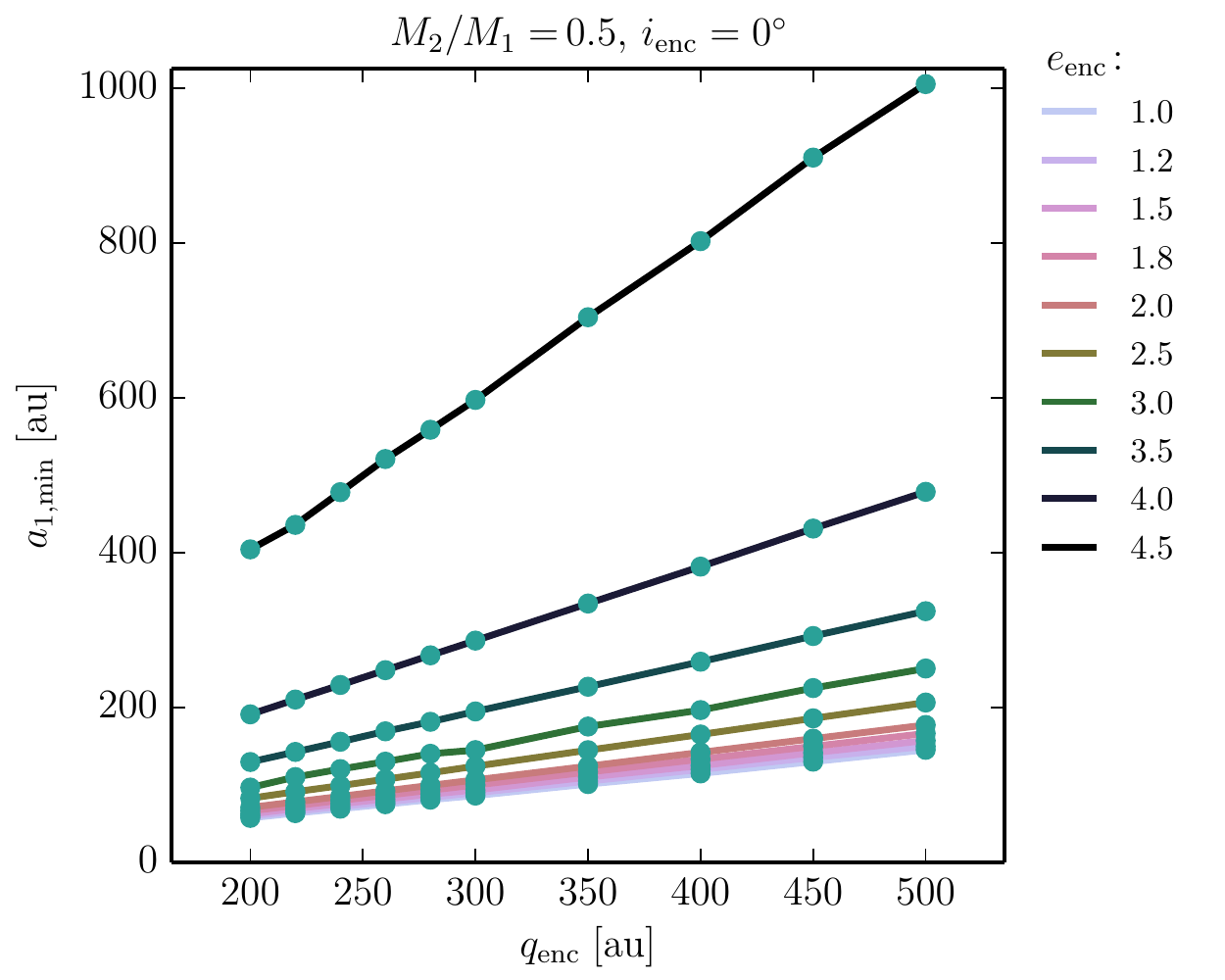}
\caption{Minimal semi-major axis of the transferred particles $a_{1,\mathrm{min}}$ as a function of the pericentre of the encounter $q_{\rm{enc}}$ for different eccentricities $e_{\rm{enc}}$ of coplanar prograde encounters.
The mass ratio $M_2/M_1$ is 0.5 for all the depicted encounters.
Lines of different colors show $a_{1,\mathrm{min}}(q_{\rm{enc}}, M_2/M_1=0.5, e_{\rm{enc}})$ for fixed $e_{\rm{enc}}$ as indicated on the right side of the plot.}
\label{fig:amin_mass_0.5}
\end{figure}

\begin{figure*}
\center
\includegraphics[width=0.325\textwidth]{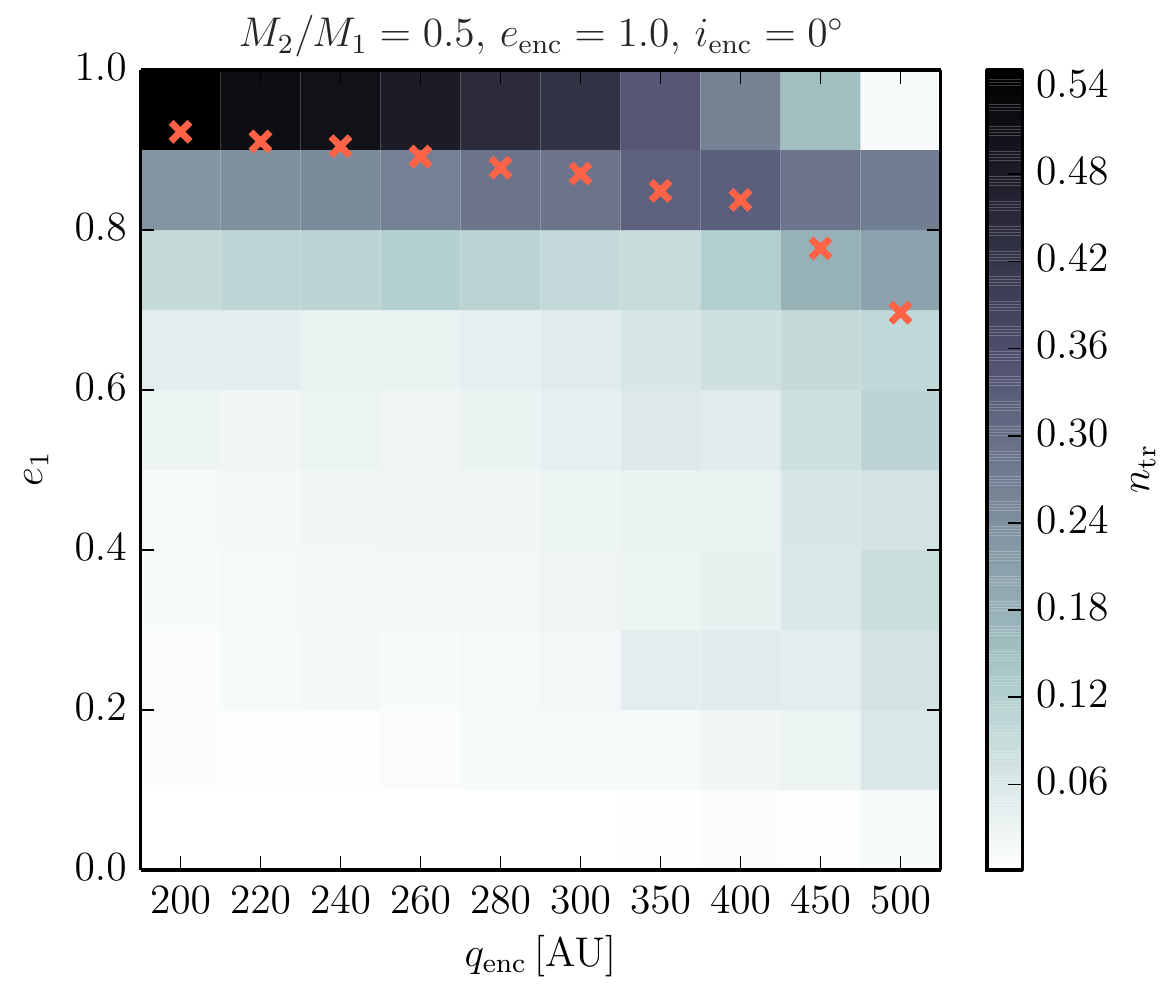}~
\includegraphics[width=0.325\textwidth]{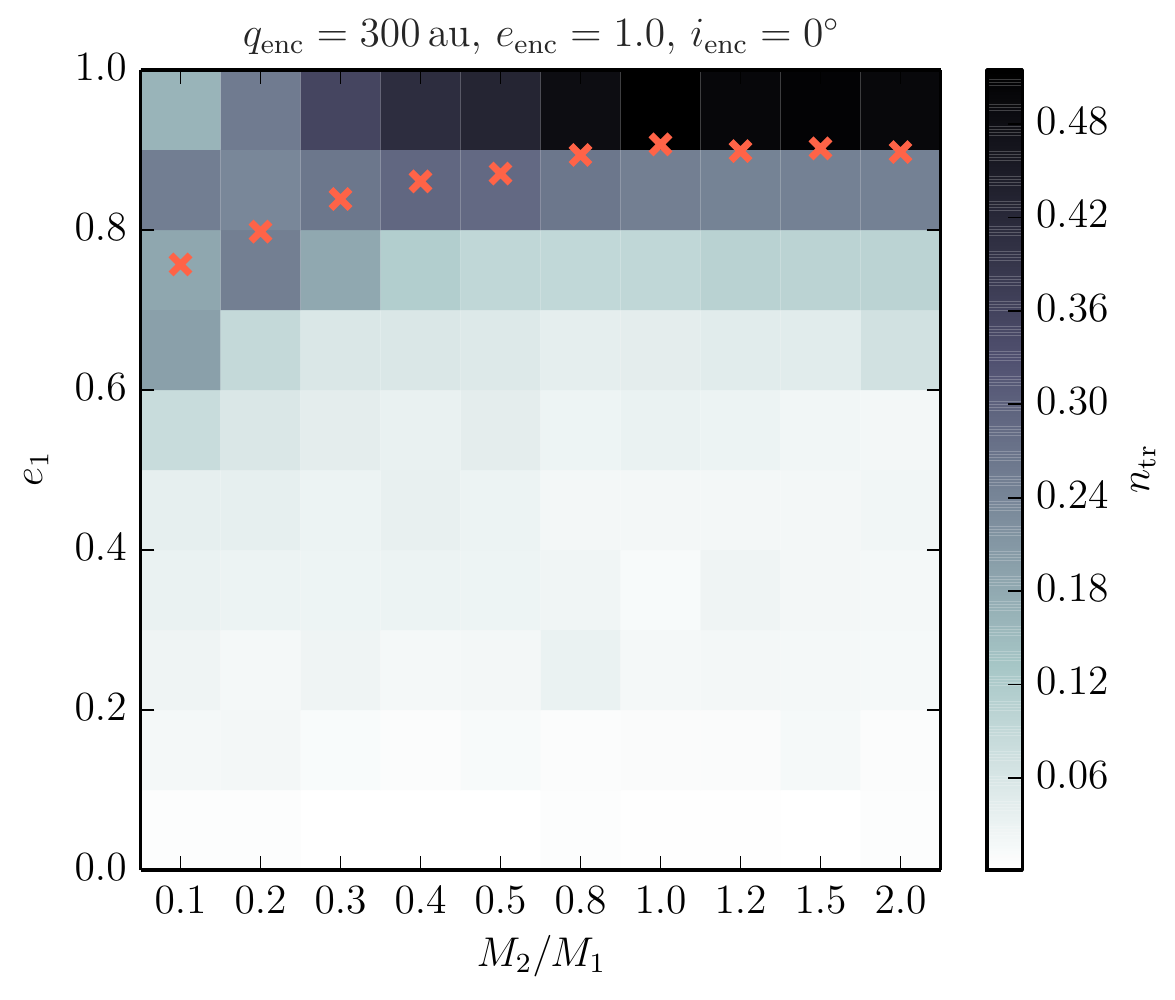}~
\includegraphics[width=0.325\textwidth]{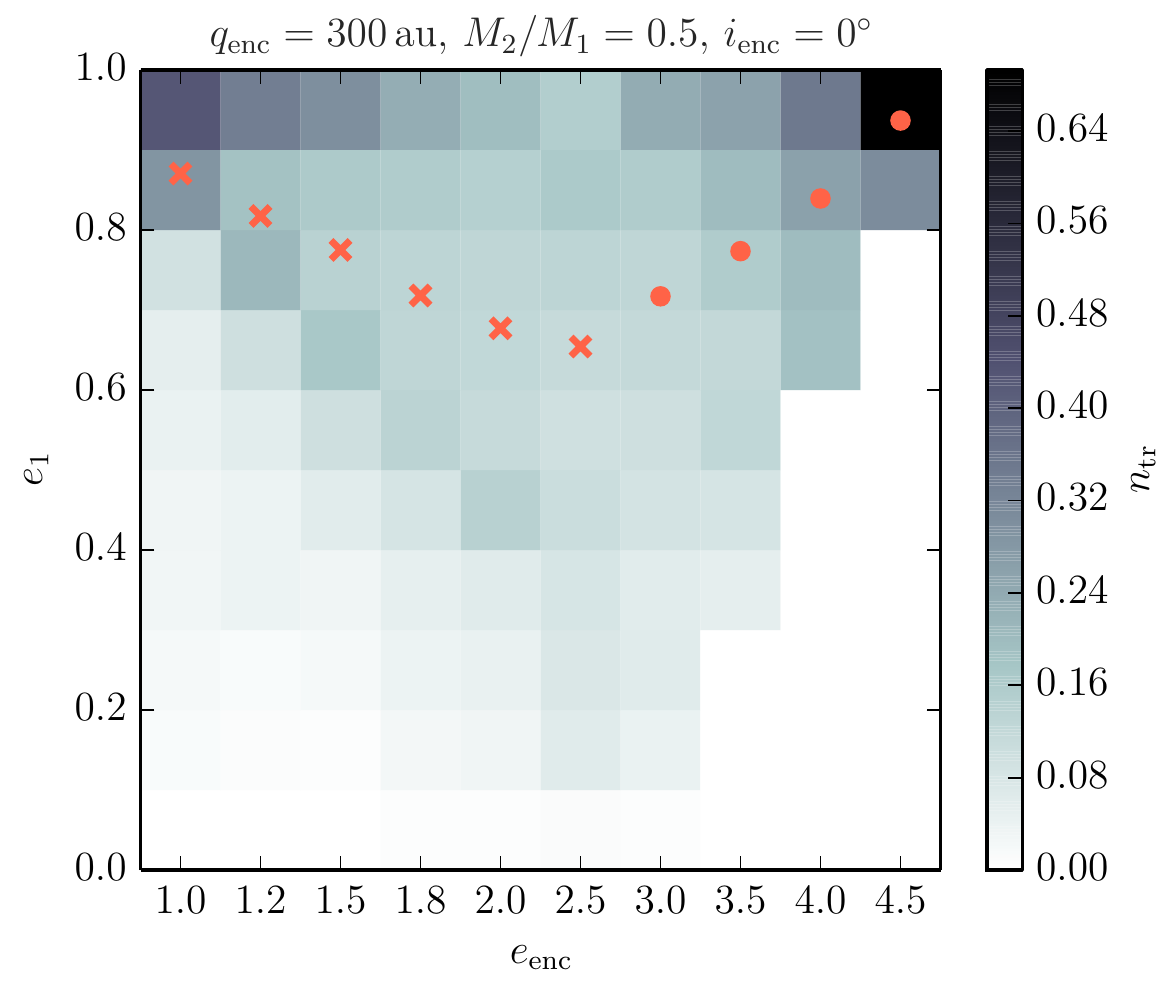}
\caption{Distributions of the eccentricity of the transferred particles $e_1$ for coplanar prograde encounters ($i_{\rm{enc}}=0\degr$).
The three plots show eccentricity distributions for different encounter pericentre $q_{\rm{enc}}$ ({\it Left}), mass ratio $M_2/M_1$ ({\it Middle}), and encounter eccentricity $e_{\rm{enc}}$ ({\it Right}).
{\it Left}: Eccentricity distributions for encounters with $M_2/M_1=0.5$, $e_{\rm{enc}}=1.0$ (parabolic encounters), and $q_{\rm{enc}}$ varying along the horizontal axis.
For clarity, the scale of the horizontal axis is arbitrary\,---\,values of pericentre are equidistantly distributed over the horizontal axis.
The vertical axis shows the eccentricity of the transferred particles $e_1$ in 10 equidistant bins.
The color scale maps the relative number of transferred particles in each eccentricity bin as measured for individual encounters with different $q_{\rm{enc}}$.
The red symbols correspond to the median value of $e_1$. 
Bullets depict the encounters with completely covered transfer region, while crosses the encounters with $r_{\rm{tr,max}}>200$\,au (see Sect.~\ref{sec:transfer_rad} and, e.g., Fig.~\ref{fig:mu_tr_mass05}).
{\it Middle}: Eccentricity distributions for encounters with $q_{\rm{enc}}=300$\,au, $e_{\rm{enc}}=1.0$ (parabolic encounters), and different $M_2/M_1$ along the horizontal axis.
{\it Right}: Eccentricity distributions for encounters with $q_{\rm{enc}}=300$\,au, $M_2/M_1=0.5$, and different $e_{\rm{enc}}$ along the horizontal axis.
The distribution at the $q_{\rm{enc}}=300$ on the horizontal axis of the {\it Left} plot, $M_2/M_1=0.5$ of the {\it Middle} plot, and $e_{\rm{enc}}=1.0$ of the {\it Right} one, is the same.}
\label{fig:de_m05_e10}
\end{figure*}

The transferred particles represent a specific population in orbit around their new host $M_1$.
In this section, we analyze the orbits of particles transferred in coplanar prograde encounters ($i_{\rm{enc}}=0\degr$), these are described in Sects.~\ref{sec:transfer_rad} and \ref{sec:transfer_eff}.
Orbits of particles transferred during encounters with non-zero inclination of the disc and the plane of the encounter, $i_{\rm{enc}}$, are described in Sects.~\ref{sec:inc_enc} and \ref{sec:omega_enc}.

In Fig.~\ref{fig:amin_map_q_m}, we show the minimal semimajor axis of the transferred particles $a_{1,\mathrm{min}}$ (which corresponds to the transferred orbit with the minimal energy) as a function of the mass ratio $M_2/M_1$, and the pericentre of the encounter $q_{\rm{enc}}$ for the parabolic prograde coplanar encounters ($e_{\rm{enc}}=1.0$, $i_{\rm{enc}}=0\degr$).
There is a clear trend\,---\,as expected, for larger pericentre and smaller mass ratio, the larger $a_{1,\mathrm{min}}$.
In Fig.~\ref{fig:amin_mass_0.5}, we show $a_{1,\mathrm{min}}$ as a function of $q_{\rm{enc}}$ for different $e_{\rm{enc}}$ and fixed $M_2/M_1$.
The minimal semimajor axis of the transferred orbits $a_{1,\mathrm{min}}$ is a linear function of the pericentre of the encounter $q_{\rm{enc}}$ and the coefficient of the proportionality depends on the eccentricity of the encounter $e_{\rm{enc}}$\,---\,
larger values of $e_{\rm{enc}}$ result in steeper increase of $a_{1,\mathrm{min}}$ with $q_{\rm{enc}}$.

\begin{figure}
\center
\includegraphics[width=0.45\textwidth]{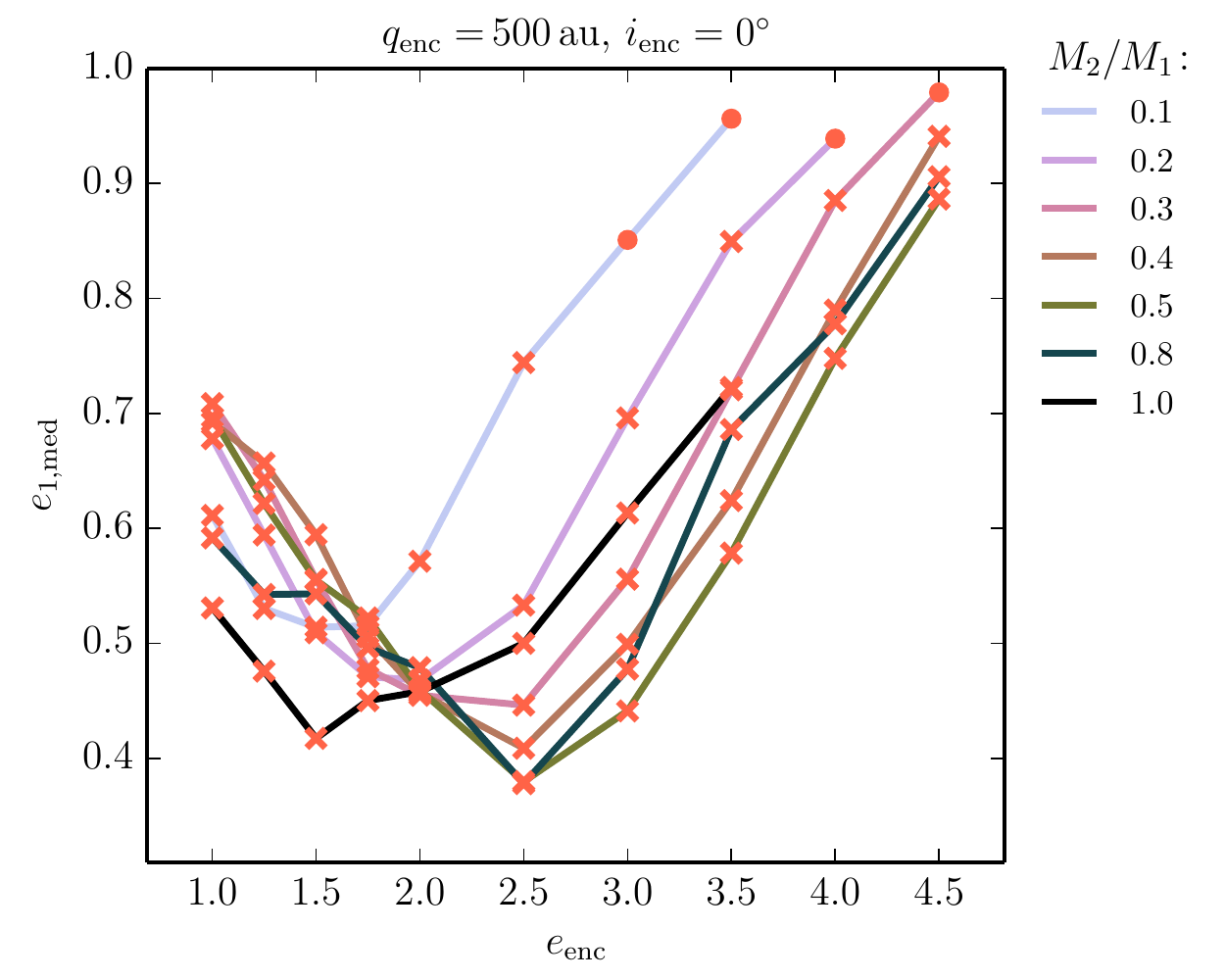}
\caption{Median value of the eccentricity of the transferred particles $e_{1,\mathrm{med}}$ as a function of the encounter eccentricity $e_{\rm{enc}}$ for coplanar prograde encounters with pericentre $q_{\rm{enc}}=500$\,au.
The lines of different colors correspond to encounters with stars of different mass ratios $M_2/M_1$, as indicated on the right.
Bullets depict the encounters with completely covered transfer region, while crosses the encounters with $r_{\rm{tr,max}}>200$\,au (see Sect.~\ref{sec:transfer_rad}).}
\label{fig:e_tr_med_p500}
\end{figure}

Most of the transferred particles are on eccentric orbits \citep{2005ApJ...629..526P}.
In Fig.~\ref{fig:de_m05_e10}, we show the eccentricity distributions of the transferred particles for coplanar prograde encounters.
Similarly to \citet[][their Figure\,7, {\it Bottom}]{2005ApJ...629..526P}, for the parabolic encounters the captured particles move on eccentric orbits with $e_1\apgt0.8$ irrespective of the pericentre $q_{\rm{enc}}$ and the mass ratio $M_2/M_1$ (Fig.~\ref{fig:de_m05_e10}, {\it Left} and {\it Middle}).
The median eccentricity is generally decreasing with $q_{\rm{enc}}$ (red crosses in Fig.~\ref{fig:de_m05_e10}, {\it Left}) and increasing with $M_2/M_1$ (red crosses in Fig.~\ref{fig:de_m05_e10}, {\it Middle}).
For the parabolic encounters ($e_{\rm{enc}}=1.0$, Fig.~\ref{fig:de_m05_e10}, {\it Left} and {\it Middle}), a small fraction of the captured particles (about 5\% or less) also have low eccentricities, $e_1\aplt0.2$.
For hyperbolic encounters ($e_{\rm{enc}}>1.0$) shown in Fig.~\ref{fig:de_m05_e10}, {\it Right}, some of the transferred particles have relatively low eccentricity. 
For fixed $q_{\rm{enc}}$ and $M_2/M_1$, the median value of $e_1$ is decreasing with the encounter eccentricity up to $e_{\rm{enc}}\sim2.5$, and increasing again for more eccentric encounters.
We find a similar dependence irrespective of the encounter pericentre $q_{\rm{enc}}$ and mass ratio $M_2/M_1$.
In Fig.~\ref{fig:e_tr_med_p500}, we show the median value of the transferred particles' eccentricity $e_{1,\rm{med}}$ as a function of $e_{\rm{enc}}$ for encounters with pericentre $q_{\rm{enc}}=500$\,au and different mass ratio $M_2/M_1$.
We find that the median of $e_1$ is generally lower than 0.5 if the encounter pericentre is large, $q_{\rm{enc}}\apgt350$\,au.

\subsection{Inclination of the encounter plane}
\label{sec:inc_enc}

\begin{figure}
\center
\includegraphics[width=0.49\textwidth]{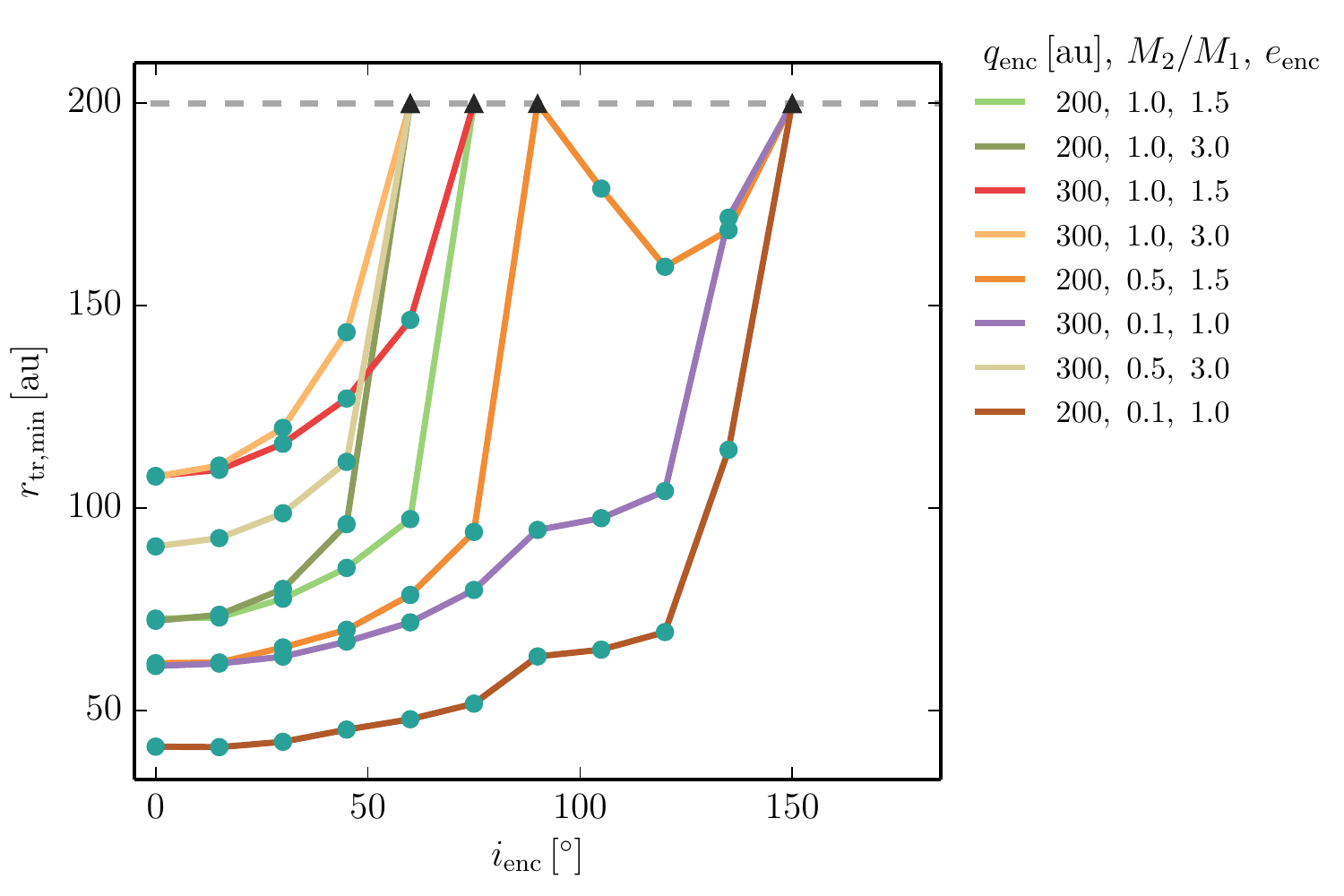}
\caption{Minimal transfer radius $r_{\rm{tr,min}}$ as a function of the encounter inclination $i_{\rm{enc}}$.
Eight encounters with different pericentre $q_{\rm{enc}}$, mass ratio $M_2/M_1$, and eccentricity $e_{\rm{enc}}$, are shown by lines of different colors, as indicated on the right.
Note that for some encounters with higher $i_{\rm{enc}}$, we used more particles (up to $5\cdot10^4$) and we used a larger value of the inner edge of the disc (while still smaller than $r_{\rm{tr,min}}$, that is from the range 30\,au--$r_{\rm{tr,min}}$) to increase the resolution.
The gray dashed line at 200\,au is the upper limit on $r_{\rm{tr,min}}$ given by the outer disc radius used in our simulations.
The black triangles indicate the encounters for which no particles were transferred in our high-resolution simulations and therefore we assume that $r_{\rm{tr,min}}>200$\,au for these cases.
The argument of periastron of the encounter $\omega_{\rm{enc}}$ is $90\degr$ for all encounters.}
\label{fig:rmin_incl}
\end{figure}

\begin{figure}
\center
\includegraphics[width=0.49\textwidth]{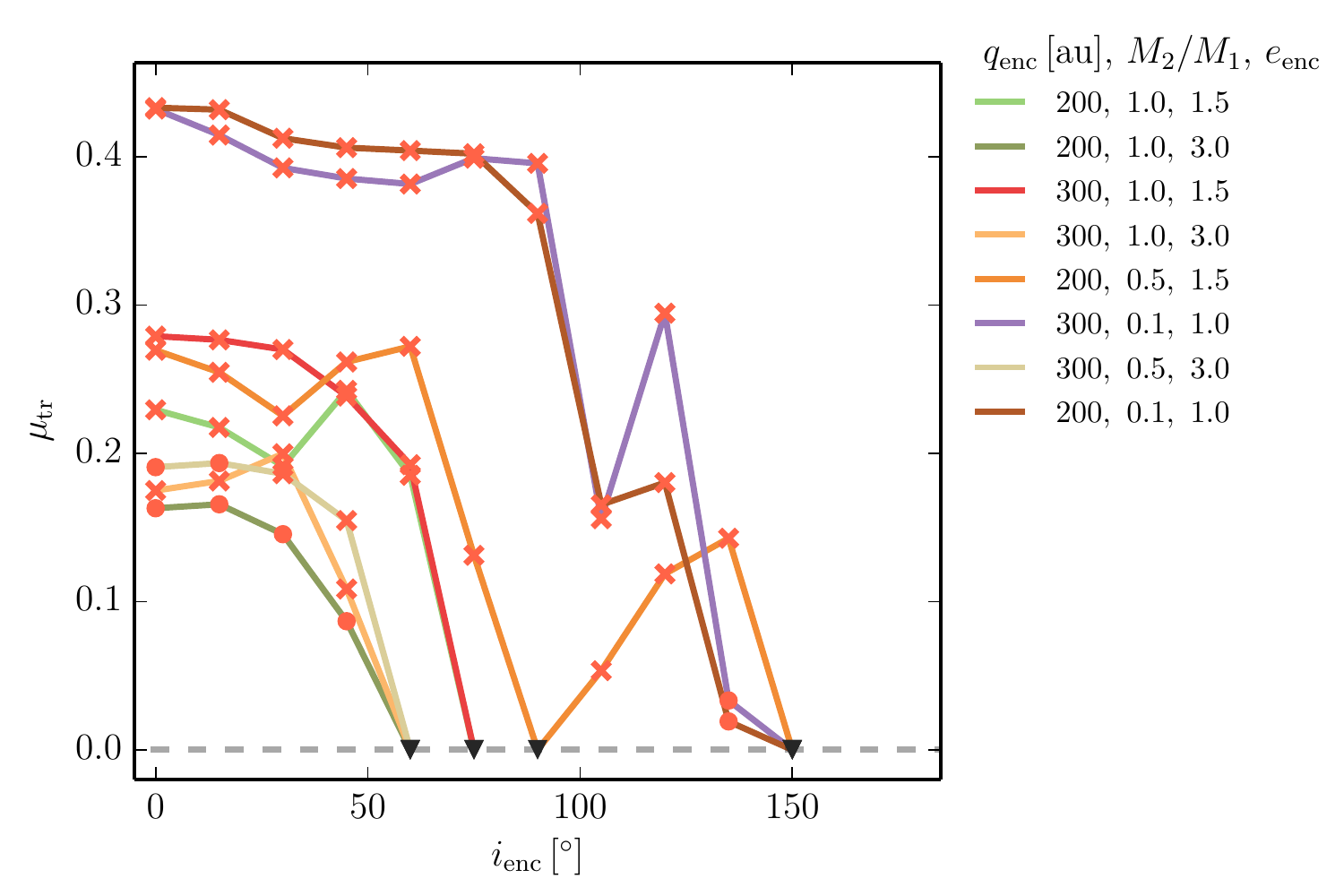}
\caption{Transfer efficiency $\mu_{\rm{tr}}$ as a function of encounter inclination $i_{\rm{enc}}$.
See Fig.~\ref{fig:rmin_incl} for a detailed description.
The gray dashed line corresponds to no transferred particles and the black triangles indicate the high-resolution simulations where no particles were transferred.
Bullets depict the encounters with completely covered transfer region, while crosses the encounters with $r_{\rm{tr,max}}>200$\,au (see Sect.~\ref{sec:transfer_rad}).}
\label{fig:mu_tr_incl}
\end{figure}

To explore how the transfer efficiency and the characteristics of the captured population depend on the inclination of the disc plane with respect to the plane of the encounter $i_{\rm{enc}}$ we carried out simulations with $i_{\rm{enc}}$ in the range 0--180$\degr$ (see Table~\ref{tab:grid_par}, section {\it varying} $i_{\rm{enc}}$).

In Fig.~\ref{fig:rmin_incl}, we show the dependency of the minimal transfer radius $r_{\rm{tr,min}}$ on the encounter inclination  $i_{\rm{enc}}$ for eight cases with different pericentre $q_{\rm{enc}}$, mass ratio $M_2/M_1$, and eccentricity $e_{\rm{enc}}$.
In general, the minimal transfer radius $r_{\rm{tr,min}}$ is increasing with inclination.
For most of the encounter parameters we explored here, $r_{\rm{tr,min}}$ increases beyond the outer edge of the disc of 200\,au for $i_{\rm{enc}}$ of about $60\degr$.
In the case of the encounter with $q_{\rm{enc}}=200$\,au, $M_2/M_1=0.5$, and $e_{\rm{enc}}=1.5$ (points connected by the orange line), $r_{\rm{tr,min}}>200$\,au for the orthogonal geometry ($i_{\rm{enc}}=90\degr$), while for $i_{\rm{enc}}\approx105$--$150\degr$, $r_{\rm{tr,min}}$ is smaller than 200\,au (decreasing for $i_{\rm{enc}}\approx105$--$120\degr$ and increasing again for $i_{\rm{enc}}\apgt120\degr$).

As we describe in Sect.~\ref{sec:transfer_rad} and shown in Fig.~\ref{fig:rmin_ecc_m05}, for given $q_{\rm{enc}}$ and $M_2/M_1$ the minimal transfer radius $r_{\rm{tr,min}}$ does not strongly depend on the eccentricity $e_{\rm{enc}}$.
This changes with increasing inclination\,---\,the encounters with higher eccentricity $e_{\rm{enc}}$ have also larger minimal transfer radius $r_{\rm{tr,min}}$.

Similarly as for the coplanar prograde encounters in Fig.~\ref{fig:rmin_tr_unbound}, the minimal radius of the transferred particles $r_{\rm{tr,min}}$ and the minimal radius or the unbound particles $r_{\rm{tr,min}}$ do not differ by more than about 10\% for the inclined encounters.

The dependency of the transfer efficiency $\mu_{\rm{tr}}$ on the encounter inclination $i_{\rm{enc}}$ is shown in Fig.~\ref{fig:mu_tr_incl}.
The relative number of transferred particles is expected to decrease with $i_{\rm{enc}}$ \citep{1993MNRAS.261..190C} and we find a similar trend\,---\,the transfer efficiency $\mu_{\rm{tr}}$ is generally also smaller for higher $i_{\rm{enc}}$.
Since $r_{\rm{tr,min}}$ increases steeply for encounters with $i_{\rm{enc}}>90\degr$, our simulations do not cover the transfer region for most of the considered encounters.

Realistic encounters have random inclination $i_{\rm{enc}}$ and the results presented in the previous sections, which assumed that the disc and the encounter are in the same plane (i.e., coplanar geometry), represent the lower limits for the minimum transfer radius $r_{\rm{tr,min}}$ (Fig.~\ref{fig:rmin_incl}) and the upper limits for the transfer efficiency $\mu_{\rm{tr}}$ (Fig.~\ref{fig:mu_tr_incl}).
Additionally, Figs.~\ref{fig:rmin_incl} and~\ref{fig:mu_tr_incl} show that $r_{\rm{tr,min}}$ and $\mu_{\rm{tr}}$ in the low inclination ($i_{\rm{enc}}\aplt 30\degr$) encounters hardly differ from those in their respective co-planar cases for most parameters.

\subsubsection{Orientation of the transferred orbits}
\label{sec:orientation}

For the coplanar encounters, all the transferred particles are orbiting their new host $M_1$ in the same plane\,---\,the plane of the disc and of the encounter orbit.
The situation is different for the encounters that are inclined with respect to the disc plane, that is $i_{\rm{enc}}>0\degr$.
Orbits are traditionally characterized by orbital elements and in Sect.~\ref{sec:transferred_orbits} we studied the distributions of the semimajor axes and eccentricities of the particles transferred in the coplanar prograde encounters.
The orientation of the orbital plane with respect to a given reference plane can be defined by the inclination and the longitude of ascending node, and the orientation of the orbit in the orbital plane is defined by the argument of periastron.
The plane of reference is in principle arbitrary and the orbital elements will differ based on the choice.
Therefore we study the orientation of the orbital planes of the transferred particles using their angular momentum vectors.

The orbital plane is defined by the plane perpendicular to the specific relative angular momentum vector $\mathbfit{h}$ of the orbiting body (cross product of the relative position and velocity vectors of the particles and star $M_1$).
To describe the alignment of the orbital planes of the transferred particles, we study the clustering of the directions of their $\mathbfit{h}$.
We calculate the mean relative angular momentum vector of the transferred population $\mathbfit{h}_{\rm{tr,mean}}$ as the mean vector of normalized $\mathbfit{h}$ of the transferred particles with respect to star $M_1$.
We calculate the angles between $\mathbfit{h}_{\rm{tr,mean}}$ and $\mathbfit{h}$, which we call $\phi_{\rm{tr}}$.
The distribution of $\phi_{\rm{tr}}$ then characterizes the clustering of the directions of $\mathbfit{h}$, and therefore also the alignment of the orbital planes of the transferred particles.
If the individual orbital planes have a similar orientation, we expect the angles $\phi_{\rm{tr}}$ to be small.

\begin{figure}
\center
\includegraphics[width=0.4\textwidth]{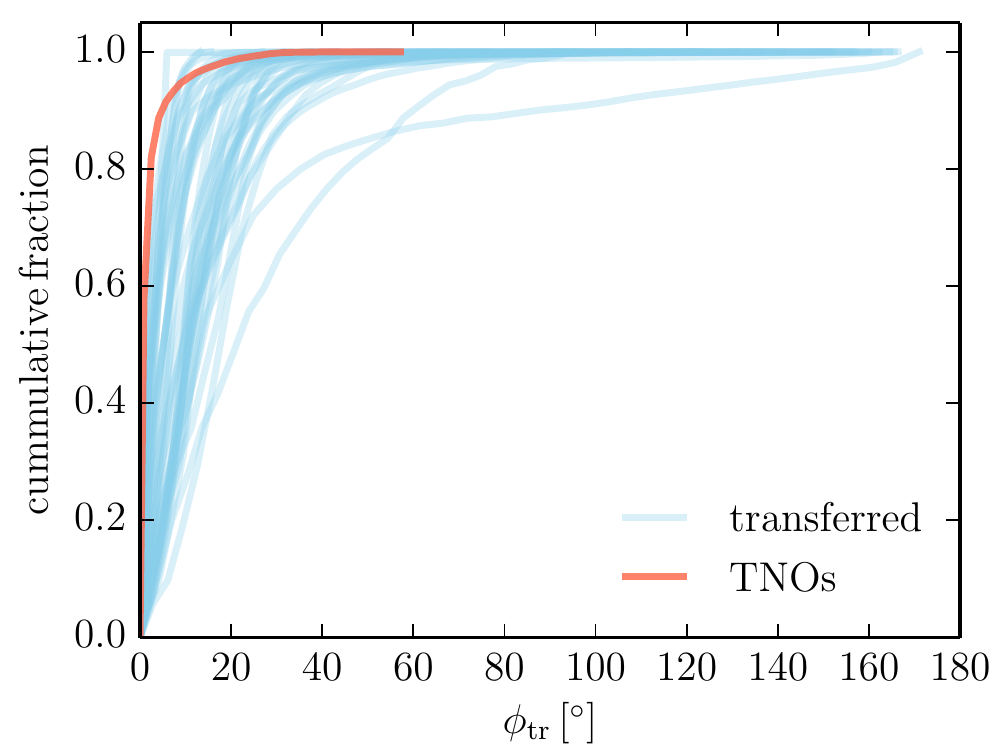}
\caption{Cumulative distributions of $\phi_{\rm{tr}}$ for the inclined encounters and for TNOs.
Blue lines show the distributions for simulated transferred particles for the 48 inclined encounters plotted in Figs.~\ref{fig:rmin_incl} and \ref{fig:mu_tr_incl}.
Red line shows the distribution for the TNOs of the Solar system.}
\label{fig:dmom_cum}
\end{figure}

In Fig.~\ref{fig:dmom_cum}, we show the cumulative distribution functions of $\phi_{\rm{tr}}$ for the particles transferred during our inclined encounters (see Table~\ref{tab:grid_par}, section {\it varying $i_{\rm{inc}}$}, and Figs.~\ref{fig:rmin_incl} and \ref{fig:mu_tr_incl}). 
For 95\% of the encounters (41 out of 43 simulations), half of the transferred particles orbit in a plane with $\phi_{\rm{tr}}\aplt15\degr$.
The same limit on $\phi_{\rm{tr}}$ also holds for 84\% of the particles in 36 out of 43 simulations. 

For comparison, we carried the analysis for 1351 Transneptunian Objects of the Solar system (TNOs, bodies that orbit the Sun with average distance larger than Neptune's semimajor axis of 30\,au).\footnote{We obtained the list of TNOs from the Minor Planet Center (MPC) database operated at the Smithsonian Astrophysical Observatory (SAO) under the auspices of the International Astronomical Union (IAU); \url{http://www.minorplanetcenter.net/iau/lists/TNOs.html}.}
For more than 90\% of the TNOs, $\phi_{\rm{tr}}<5\degr$.
The cumulative distribution of $\phi_{\rm{tr}}$ is also plotted in Fig.~\ref{fig:dmom_cum} (red line).
The orbital planes of the observed TNOs have a more similar orientation than the particles transferred in our simulations.

We calculated the orbital elements of the transferred orbits in the coordinate system with the plane of reference perpendicular to the mean vector of the normalized $\mathbfit{h}$ of the captured particles.
Regardless of the encounter parameters, the median values of the inclination of the captured population are typically $\aplt 30\degr$ with standard deviations $\aplt 35\degr$.
The median values of the argument of pericenter are within the range from $-30\degr$ to $30\degr$, with the standard deviations $\sim 100\degr$.
However, we observe that the argument of pericenter generally clusters in narrower distributions for the orbits captured at larger semimajor axes.

\begin{figure*}
\center
\includegraphics[width=0.7\textwidth]{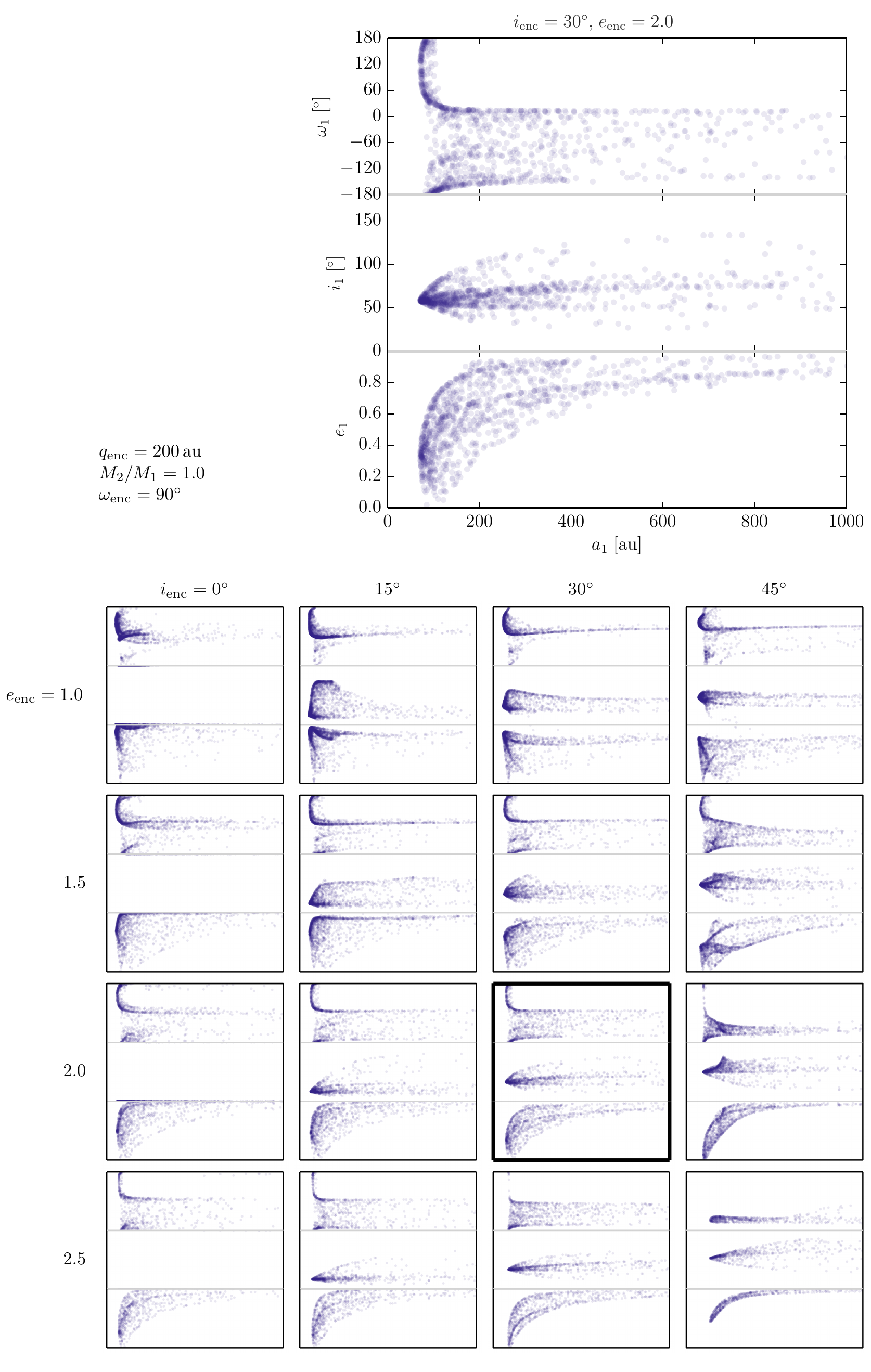}
\caption{Orbital elements of the transferred particles.
Each plot shows the argument of pericentre $\omega_1$, the inclination $i_1$, and the eccentricity $e_1$ versus the semimajor axis $a_1$ of the transferred particles which are shown in the upper, middle, and bottom panel, respectively.
All plots are for encounters with pericenter $q_{\rm{enc}}=200$\,au, mass ratio $M_2/M_1=1.0$, and argument of pericenter $\omega_{\rm{enc}}=90\degr$, while the encounter inclination $i_{\rm{enc}}$ and eccentricity $e_{\rm{enc}}$ vary.
$i_{\rm{enc}}$ is changing along the columns, while $e_{\rm{enc}}$ along the rows of the mosaic of the small plots.
The large plot shows the encounter with $i_{\rm{enc}}=30\degr$ and $e_{\rm{enc}}=2.0$ and is the same as the small plot highlighted with the thicker frame.
All the plots cover the same ranges for all the variables (that is $a_1=0$--1000\,au on the horizontal axis, $e_1=0$--1 on the vertical axis of the lower panel, $i_1=0$--$180\degr$ in the middle panel, and $\omega_1=-180$--$180\degr$ in the upper panel) and have the same scale.
The orbital elements are calculated in the coordinate system with the reference plane of the initial disc.}
\label{fig:encounter_orb_pars}
\end{figure*}

The transferred particles have specific distributions in the space of orbital elements and these are given by the parameters of the encounter (see also Sect.~\ref{sec:transferred_orbits}).
In Fig.~\ref{fig:encounter_orb_pars}, we show distributions of the orbital elements of the captured population for encounters with a range in encounter inclination $i_{\rm{enc}}$ and eccentricity $e_{\rm{enc}}$.
The transferred population clearly depend on both parameters.
For example, the last line of the mosaic shows encounters with $e_{\rm{enc}}=2.5$. 
Here, the distribution of semimajor axis $a_1$ and eccentricity $e_1$ are moving to higher values with increasing encounter inclination $i_{\rm{enc}}$ (see the bottom panels of the plots in the last line of the mosaic);
the orbits also have higher inclination $i_1$ (middle panels);
and the range of argument of pericenter $\omega_1$ of the transferred orbits is shrinking (top panels).
Each encounter parameter effects the captured population and its final distribution in the orbital elements space is given by a complex combination of the individual signatures.
The population of the transferred particles can therefore be used to constrain the encounter through which it was delivered \citep{sednitos2015}.

\subsection{Argument of periastron of the encounter}
\label{sec:omega_enc}

\begin{figure}
\center
\includegraphics[width=0.49\textwidth]{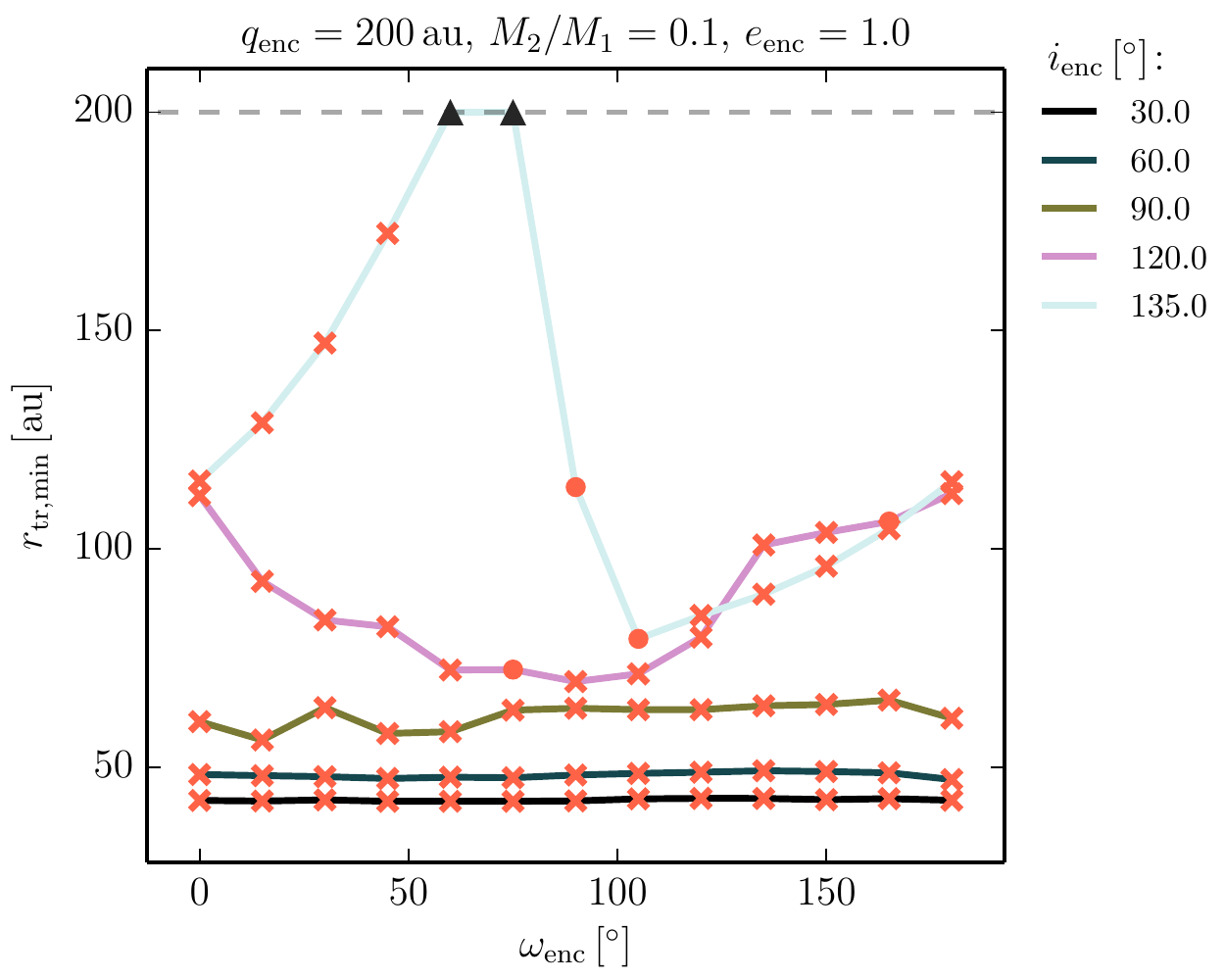}
\includegraphics[width=0.49\textwidth]{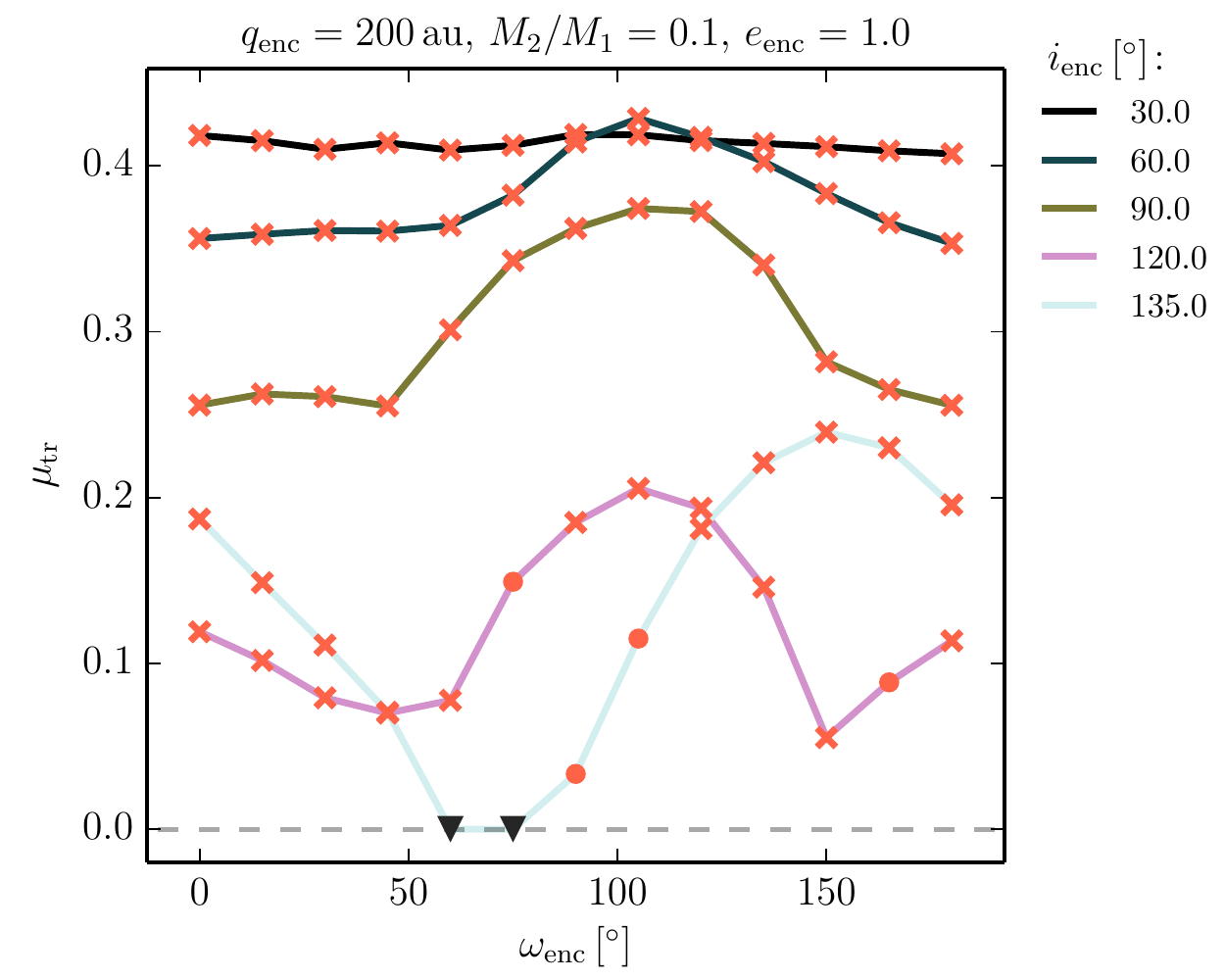}
\caption{Minimal disc radius $r_{\rm{tr,min}}$ ({\it Top}) and the transfer efficiency $\mu_{\rm{tr}}$ ({\it Bottom}) as a function of the argument of periastron of the encounter $\omega_{\rm{enc}}$.
The encounters have fixed pericentre $q_{\rm{enc}}=200$\,au, mass ratio $M_2/M_1=0.1$, and eccentricity $e_{\rm{enc}}=1.0$.
Lines of different color correspond to different inclinations of the encounter plane with respect to the disc $i_{\rm{enc}}$ as indicated on the right.
Bullets depict the encounters with completely covered transfer region, while crosses the encounters with $r_{\rm{tr,max}}>200$\,au.
Black triangles indicate the runs with $r_{\rm{tr,min}}>200$\,au and the dashed gray lines correspond to the upper and lower limit of the $r_{\rm{tr,min}}$ and $\mu_{\rm{tr}}$, in the {\it Top} and {\it Bottom} plot, respectively.
}
\label{fig:omega_rmin_mu}
\end{figure}

We investigated the role of the argument of periastron of the encounter $\omega_{\rm{enc}}$ on the transfer region and efficiency.
We carried out simulations sampling $\omega_{\rm{enc}}$ in the range of 0--180$\degr$ for different inclinations $i_{\rm{enc}}$ and fixed pericentre $q_{\rm{enc}}=200$\,au, mass ratio $M_2/M_1=0.1$, and eccentricity $e_{\rm{enc}}=1.0$ (see Table~\ref{tab:grid_par}, section {\it varying} $\omega_{\rm{enc}}$). 
Within the considered resolution, these encounter parameters result in mass transfer for inclinations $i_{\rm{enc}}\leq135\degr$ (Figs.~\ref{fig:rmin_incl} and \ref{fig:mu_tr_incl}).
In Fig.~\ref{fig:omega_rmin_mu}, we show the minimal disc radius of the transferred particles $r_{\rm{tr,min}}$ and the transfer efficiency $\mu_{\rm{tr}}$ as a function of $\omega_{\rm{enc}}$ ({\it Top} and {\it Bottom} plot, respectively).
The minimal radius $r_{\rm{tr,min}}$ is independent of $\omega_{\rm{enc}}$ for the prograde inclinations $i_{\rm{enc}}\leq90\degr$.
For the retrotrograde encounters, $r_{\rm{tr,min}}$ has a clear minimum for $\omega_{\rm{enc}}\approx105\degr$.
The transfer efficiency $\mu_{\rm{tr}}$ changes with $\omega_{\rm{enc}}$ for all the considered inclinations $i_{\rm{enc}}>30\degr$.
The higher $i_{\rm{enc}}$, the larger the variations of $\mu_{\rm{tr}}$.
Except for the encounter with the retrograde inclination $i_{\rm{enc}}=135\degr$ and regardless of $i_{\rm{enc}}$, the transfer efficiency is approximately constant for $\omega_{\rm{enc}}\aplt45\degr$ and $\apgt150\degr$ and maximal for $\omega_{\rm{enc}}\approx90\degr$.
This can be understood given the geometrical meaning of the argument of periastron $\omega_{\rm{enc}}$ (Sect.~\ref{sec:ic}, Fig.~\ref{fig:encounter_ic})\,---\,during the encounters with $\omega_{\rm{enc}}\approx90\degr$, the star $M_1$ passes at smaller distance to the disc particles for a longer time and captures more particles.

\subsection{Eccentricity of the planetesimal disc}
\label{sec:ecc_disc}

We studied the effect of the initial eccentricity of the disc $e_{\rm{disc}}$ on the transfer region and efficiency.
For the encounters specified in the section {\it varying} $e_{\rm{disc}}$ of Table~\ref{tab:grid_par}, we set up the initial eccentricities of the disc particles randomly with a uniform distribution within the ranges 0.0--0.05 and 0.0--0.1 and their orbital phase within 0 and $2\pi$.

For most of the studied encounters, there is no substantial change due to the eccentricity of the disc particles $e_{\rm{disc}}$. 
The minimal radius of the transferred particles $r_{\rm{tr,min}}$, typically decreases by $\sim$5\% and 10\% compared to the circular disc for the eccentricities of 0.05 and 0.1, respectively.
The change of the maximal radius $r_{\rm{tr,max}}$ is of similar scale.
The relative number of transferred particles $n_{\rm{tr}}$ changes by up to about 10\% for both considered disc eccentricities and is both higher and lower with respect to the circular disc.
For several encounters, the change of $n_{\rm{tr}}$ is up to about $\pm20$\%. 
In these cases however, the total number of transferred particles is small ($n_{\rm{tr}}<0.1$).

\subsection{Surface density of the planetesimal disc}
\label{sec:sigma}

As described in Sect.~\ref{sec:ic}, the initial distribution of the disc particles corresponds to the surface number density $\propto 1/r$.
To estimate the dependence of the transfer efficiency on the initial surface density profile, we weighted the particle counts by $1/\sqrt{r}$ to re-scale the surface density to $\propto 1/r^{1.5}$, which corresponds to the minimum mass solar nebula \citep{1981PThPS..70...35H}.
Such surface density profile changes the transfer efficiencies presented in Sect.~\ref{sec:transfer_eff} by less than 5\% (10\%) for 95\% (all) the coplanar encounters.

\section{Analytic investigation into the truncation radius}
\label{sect:analytic}
In this section, we derive analytic expressions for the truncation radius for which the test particles, originally bound to star $M_2$, can be stripped from their parent star because of the tidal force of the passing star $M_1$.
The truncation radius correspond to the minimal radius of the unbound particles $r_{\rm{un,min}}$ defined in Sect.~\ref{sec:transfer_rad}.
The stars are assumed to move in a hyperbolic orbit that is unaffected by the test particles.

\subsection{Equations of motion}
\label{sect:analytic:EOM}
To describe the motion of a test particle, we use a rotating and pulsating frame $(x,y,z)$ in which the origin corresponds to the centre of mass of the massive bodies, and in which the massive bodies are located at fixed positions along the $x$ axis (note that this is different from the reference frame used in our simulations that is defined in Sect.~\ref{sec:ic} and illustrated in Fig.~\ref{fig:encounter_ic}). 
In this frame, the unit of length is varying with time, and it is given by the instantaneous relative position between the two bodies,
\begin{align}
\label{eq:R12}
R_{12} = \frac{R_\mathrm{p}(1+e)}{1+e \cos(\nu)} \equiv \frac{R_\mathrm{p} (1+e) }{\lambda}.
\end{align} 
Here, $R_\mathrm{p}$, $e>1$ and $\nu$ are the pericentre distance, eccentricity and true anomaly of the hyperbolic orbit, respectively; the range of the true anomaly is $-\arccos(-1/e) \leq \nu < \arccos(-1/e)$. For notational convenience, we define $\lambda \equiv 1 + e \cos(\nu)$. The massive bodies are located at $x=-\mu$ (body $M_1$) and $x=1-\mu$ (body $M_2$), where
\begin{align}
\label{eq:mu_def}
\mu \equiv \frac{M_2}{M_1+M_2};
\end{align}
note that the coordinates $x$, $y$ and $z$ are in units of $R_{12}$. The equations of motion for the test particle in the rotating and pulsating frame are given by \citet{duboshin_64}; for completeness, we also include a self-contained derivation in Appendix~\ref{app:EOM_der})
\begin{align}
\left \{
\begin{array}{cc}
\label{eq:EOM_gen}
\textstyle x''(\nu) - 2 y'(\nu) &= \frac{\partial \Omega}{\partial x}; \\
\displaystyle y''(\nu) + 2 x'(\nu) &= \frac{\partial \Omega}{\partial y}; \\
\displaystyle z''(\nu) + z(\nu) &= \frac{\partial \Omega}{\partial z}, \\
\end{array} \right.
\end{align}
where primes denote derivatives with respect to $\nu$. The `effective potential' $\Omega = \Omega(x,y,z,\nu)$ is given by
\begin{align}
\Omega = \frac{1}{\lambda} \left [ \frac{1}{2} \left(x^2+y^2+z^2 \right ) + \frac{1-\mu}{r_1} + \frac{\mu}{r_2} \right ].
\end{align}
Here $r_1$ and $r_2$ are the (scaled) distances between the test particle and the massive bodies $M_1$ and $M_2$, respectively, and are given by
\begin{subequations}
\label{eq:r1r2_def}
\begin{align}
r_1^2 &= (x+\mu)^2 + y^2 + z^2; \\
r_2^2 &= (x+\mu-1)^2 + y^2 + z^2.
\end{align}
\end{subequations}
For the purpose of determining the truncation radius, it is useful to apply a linear transformation to the coordinates in Eqs.~(\ref{eq:EOM_gen}), such that the origin of the new coordinate system corresponds to body $M_2$. The new coordinates are denoted by $(\tilde{x},\tilde{y},\tilde{z})$, and are related to the old coordinates via 
\begin{align}
\label{eq:lin_trans}
(\tilde{x},\tilde{y},\tilde{z}) = (x + \mu - 1,y,z).
\end{align}
The equation of motion for $\tilde{x}''(\nu)$ then reads
\begin{align}
\label{eq:EOM_tilde_x}
\tilde{x}'' - 2 \tilde{y}' = \frac{1}{\lambda} \left [ \tilde{x} + 1 - \mu - \left( \frac{1-\mu}{r_1^3}(\tilde{x}+1) + \frac{\mu}{r_2^3} \tilde{x} \right ) \right ].
\end{align}
A useful relation between $r_1$ and $r_2$, used frequently below, is
\begin{align}
\label{eq:r1r2_rel}
r_1^2 = 1 + 2 \tilde{x} + r_2^2,
\end{align}
which follows directly from Eqs.~(\ref{eq:r1r2_def}) and (\ref{eq:lin_trans}).

\subsection{Initial radial orbit around body $M_2$}
\label{sect:analytic:radial}
A simple estimate can be obtained by assuming that the test particle is initially in a radial orbit around body $M_2$ along the $\tilde{x}$ axis, i.e. $\tilde{y}=\tilde{z}=0$ and $y'(\nu)=0$. For detachment from body $M_2$, we adopt the criterion $\tilde{x}''(\nu)=0$ and $\tilde{x} = -r_2$ (note that $r_2>0$, and $\tilde{x}=r_2$ corresponds to a larger distance of the particle to body $M_1$, compare with Eq.~\ref{eq:r1r2_rel}). This approach is analogous to the method used to derive the Hill radius in the case that the massive bodies are in circular orbits (see for example section~5.6 of \citealt{valtonen_karttunen_06}). Substituting these conditions into Eq.~(\ref{eq:EOM_tilde_x}) and using Eq.~(\ref{eq:r1r2_rel}), we find
\begin{align}
\label{eq:r2_hill_exact}
0 = (1-r_2)^2 \left[ \mu (1+ r_2) + r_2^2 \right ] \, |1-r_2| - (1-\mu) r_2^2.
\end{align}
Analytic solutions of Eq.~(\ref{eq:r2_hill_exact}) are not readily available. It is possible, however, to obtain analytic expressions for small $r_2$, i.e. $r_2 \ll 1$, by expanding the right-hand side of Eq.~(\ref{eq:r2_hill_exact}) in terms of $r_2$, giving the condition $\mu - (3 - 2\mu)r_2^3 \approx 0$,
with (real) solution
\begin{align}
r_2 \approx \left ( \frac{\mu}{3-2\mu} \right)^{1/3}, \quad \quad \mathrm{for}\,\,r_2 \ll 1.
\end{align}
In terms of physical units, this corresponds to a truncation radius of
\begin{align}
\label{eq:r2_hill_appr1}
R_\mathrm{t}(\nu) = R_{12} r_2 \approx \frac{R_\mathrm{p}(1+e)}{\lambda} \left ( \frac{\mu}{3-2\mu} \right)^{1/3}.
\end{align}
As expected, the truncation radius is smallest at pericentre, $\nu=0$, in which case $\lambda = 1+e$, and, therefore,
\begin{align}
R_\mathrm{t}(0) \approx R_\mathrm{p} \left ( \frac{\mu}{3-2\mu} \right )^{1/3}.
\end{align}
For $\mu \approx 0$, this reduces to
\begin{align}
\label{eq:r2_hill_appr2}
R_\mathrm{t}(0) \approx R_\mathrm{p} \left ( \frac{\mu}{3} \right )^{1/3}, \quad \quad (r_2 \ll 1; \, \mu \approx 0),
\end{align}
which is precisely the Hill radius with the radius of the outer (circular) orbit replaced by the pericentre distance of the hyperbolic orbit. Note that $\mu\approx 0$ is assumed in the (circular) Hill problem \citep{valtonen_karttunen_06}.

\subsection{Initial circular orbit around body $M_2$ (prograde and retrograde)}
\label{sect:analytic:circ}
A criterion that is more appropriate to the configuration studied in the numerical simulations in Sects.~\ref{sec:transfer_rad} and \ref{sec:transfer_eff}, is based on the assumption that the test particles are initially in circular orbits around body $M_2$. The orbits do not remain circular during the encounter, but useful criteria can nevertheless be obtained with this simple assumption. 
We furthermore assume that the test particle orbits are coplanar with the massive bodies, that is $\tilde{z}=0$.

\begin{figure}
\center
\includegraphics[scale = 0.7]{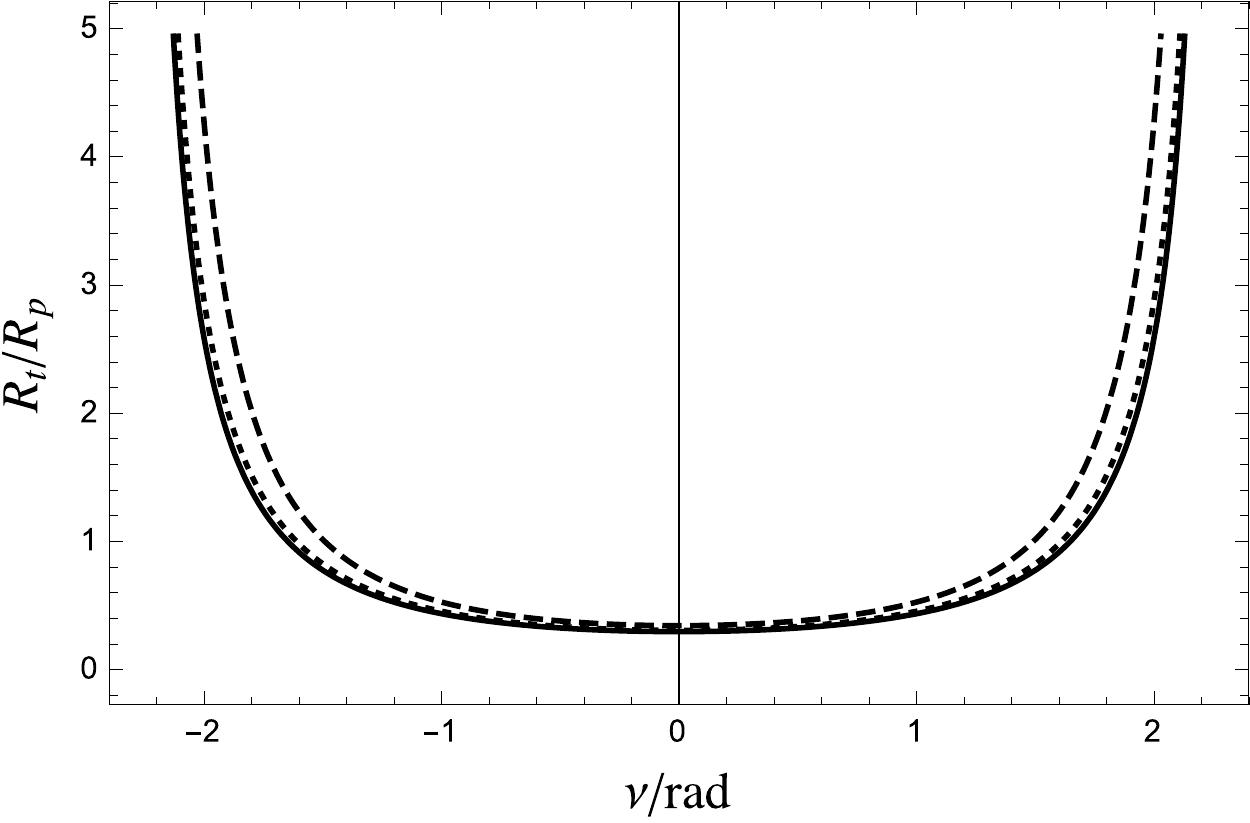}
\includegraphics[scale = 0.7]{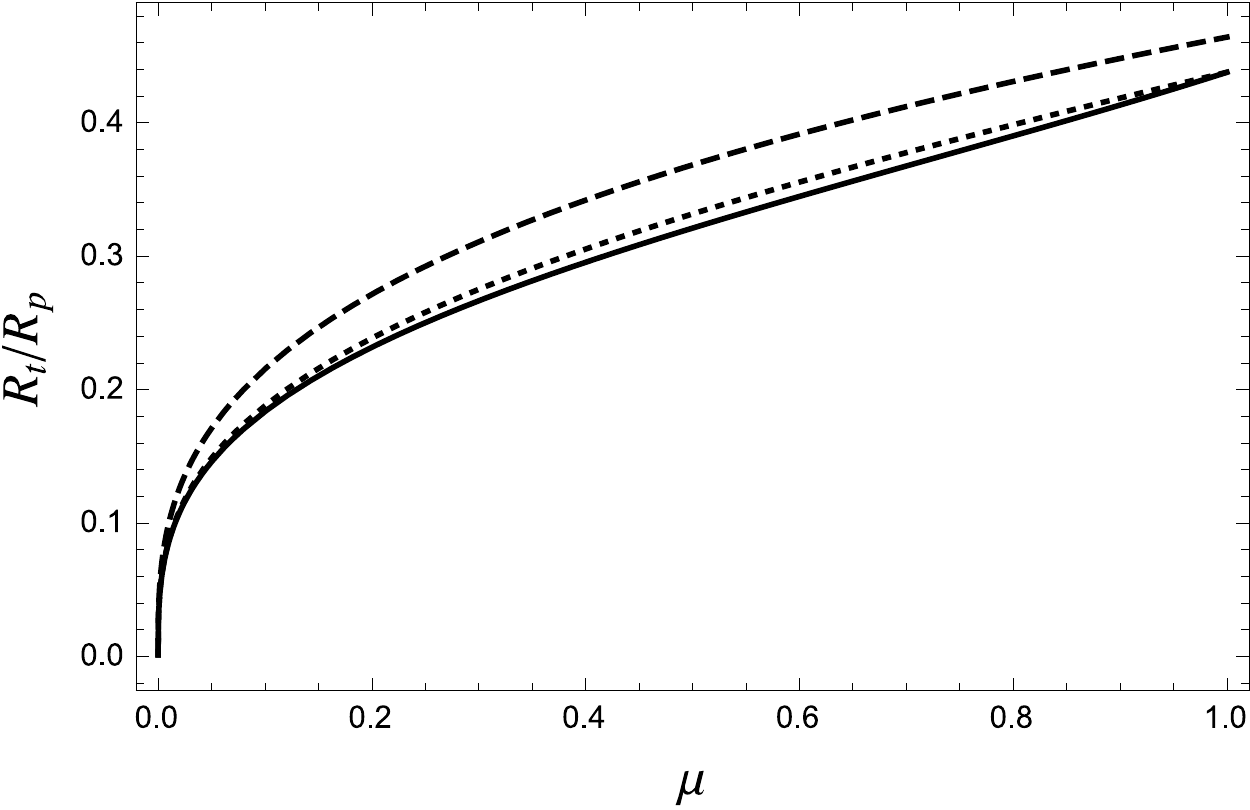}
\includegraphics[scale = 0.7]{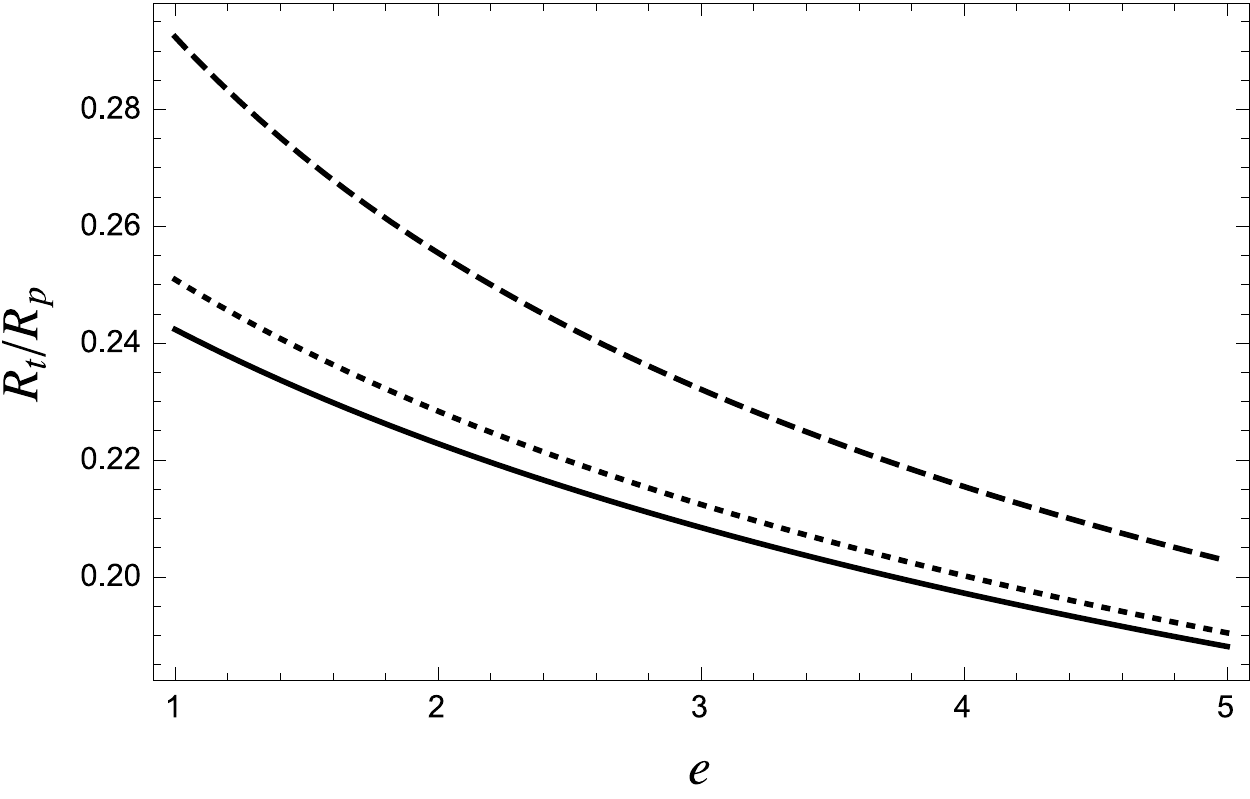}
\caption{Truncation radius for the case of initially circular test particle orbits (cf. Sect.~\ref{sect:analytic:circ}) and for prograde orbits, according to Eq.~(\ref{eq:r2_circ_appr1}) (dashed line), Eq.~\ref{eq:r2_circ_appr2}) (dotted line) and Eq.~(\ref{eq:r2_circ_exact}) (solid line). {\it Top} panel: as a function of $\nu$; {\it middle} panel: as a function of $\mu$ (setting $\nu=0$); {\it bottom} panel: as a function of $e$ (setting $\nu=0$). Where applicable, the fixed parameters are $\mu=0.2$ and $e=1.5$.}
\label{fig:comparison_circ}
\end{figure}

The closest point in the test particle orbit to body $M_1$ is when the test particle intersects the $\tilde{x}$ axis, that is $\tilde{y}=0$, and its position along the $\tilde{x}$ axis is $\tilde{x}=-r_2$. Furthermore, $\tilde{x}'(\nu)=0$, whereas $\tilde{y}'(\nu)$ is nonzero owing to the circular motion around body $M_2$. In particular, using the chain rule, $\tilde{y}'(\nu)$ can be written as
\begin{align}
\label{eq:y_prime_nu}
\tilde{y}'(\nu) = \frac{\partial \tilde{y}}{\partial \nu} = \frac{\partial \tilde{y}}{\partial t} \frac{\partial t}{\partial \nu} = \frac{\partial \tilde{y}}{\partial t} \frac{1}{\dot{\nu}} = \pm \frac{\tilde{x} \omega}{\dot{\nu}},
\end{align}
where $\dot{\nu}$ is given by e.g. Eq.~(\ref{eq:nu_dot}), and $\omega$ is the angular frequency of the test particle around body $M_2$, given by
\begin{align}
\label{eq:omega}
\omega = \sqrt{ \frac{G M_2}{(R_{12} r_2)^3}}
\end{align}
(note that $r_2$ is dimensionless, and it should therefore be multiplied by $R_{12}$ in Eq.~\ref{eq:omega}). 
The positive sign in Eq.~(\ref{eq:y_prime_nu}) corresponds to prograde motion of the disc test particles and the encounter of the massive bodies ($\tilde{x}<0$, therefore $\tilde{y}'<0$), whereas the negative sign corresponds to retrograde motion ($\tilde{y}'>0$). 

\begin{figure}
\center
\includegraphics[scale = 0.7]{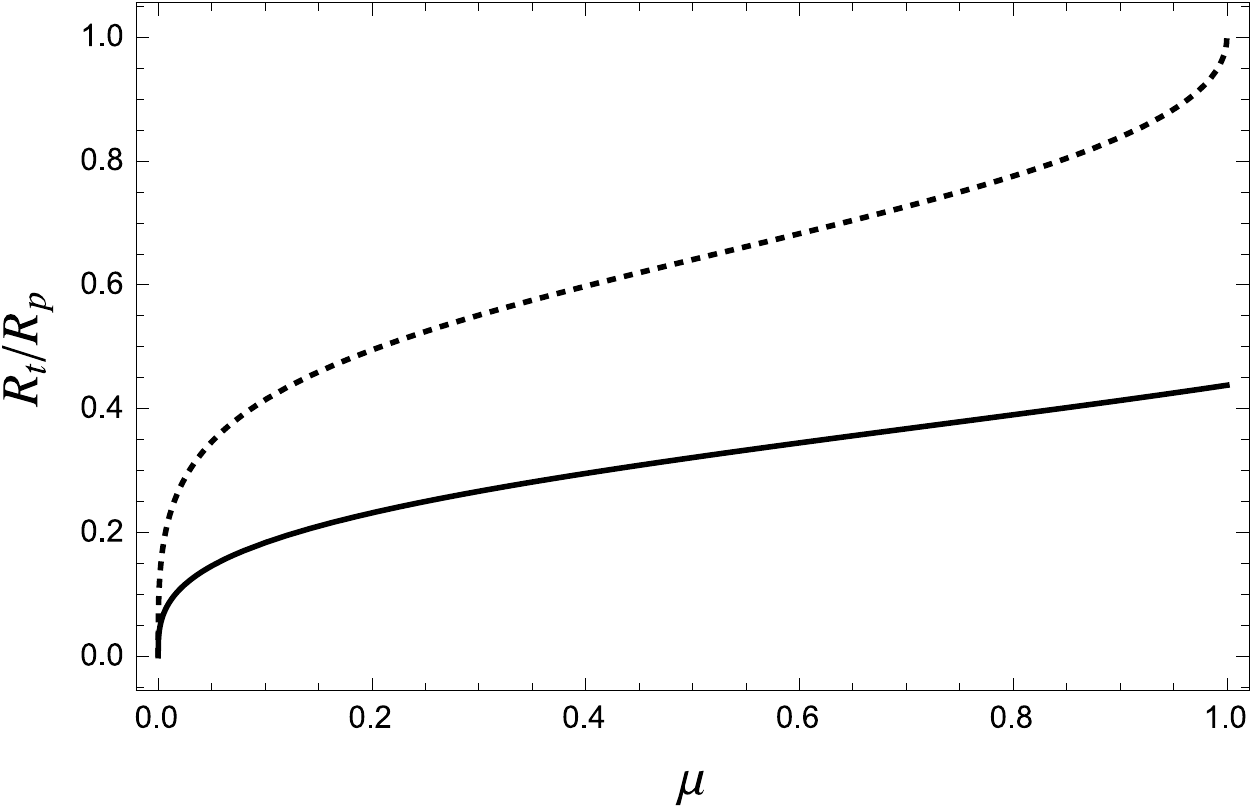}
\includegraphics[scale = 0.7]{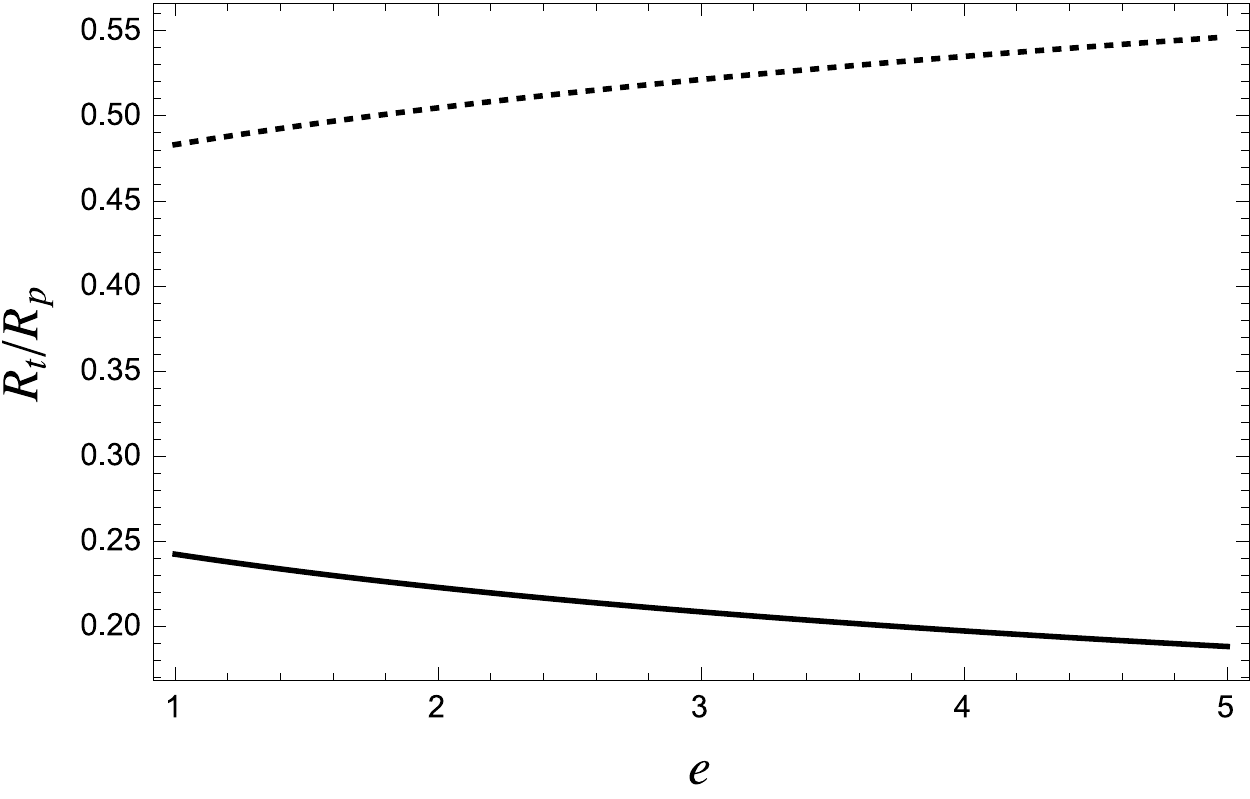}
\caption{Truncation radius for the case of initially circular test particle orbits, for the case of prograde (solid lines) and retrograde (dotted lines) orbits, according to Eq.~(\ref{eq:r2_circ_exact}). In the {\it top} and {\it bottom} panels, $R_\mathrm{t}$ is plotted as a function of $\mu$ and $e$, respectively, where the fixed parameters are $\nu=0$, $\mu=0.2$ and $e=1.5$. }
\label{fig:circ_pro_retro}
\end{figure}

Substituting these conditions into Eq.~(\ref{eq:EOM_tilde_x}), we find the following condition for detachment of the test particle from body $M_2$ (that is $\tilde{x}''(\nu)=0$),
\begin{align}
\label{eq:r2_circ_exact}
0 &= \pm \nonumber 2 \sqrt{\mu r_2 \lambda } \\
& - \frac{(1-r_2)^2 \left [ \mu(1+r_2) + r_2^2 \right ] \, |1 - r_2| - (1-\mu)r_2^2}{(1-r_2) \, |1-r_2| }.
\end{align}
Equation~(\ref{eq:r2_circ_exact}) is not amenable to analytical solutions. Nevertheless, in the case of prograde orbits, analytic solutions can be obtained by expanding it in terms of $r_2$, assuming $r_2 \ll 1$. To third order in $r_2$,
\begin{align}
\label{eq:r2_circ_cond_appr}
\mu - 2 \sqrt{r_2^3 \mu \lambda} - (3-2\mu) r_2^3 + \mathcal{O}\left(r_2^4 \right) = 0.
\end{align}
Including only terms of order $r_2^{3/2}$ in Eq.~(\ref{eq:r2_circ_cond_appr}), this gives
\begin{subequations}
\label{eq:r2_circ_appr1}
\begin{align}
R_\mathrm{t}(\nu) &\approx R_\mathrm{p} (1+e) \left (  \frac{\mu}{4 \lambda^4} \right )^{1/3}, \\
R_\mathrm{t}(0) &\approx R_\mathrm{p} \left [ \frac{\mu}{4(1+e)} \right ]^{1/3};
\end{align}
\end{subequations}
also including terms of order $r_2^3$, 
\begin{subequations}
\label{eq:r2_circ_appr2}
\begin{align}
\nonumber R_\mathrm{t}(\nu) &\approx \frac{R_\mathrm{p} (1+e)}{\lambda} \left [ \frac{\mu}{(3 - 2 \mu)^2} \right ]^{1/3} \\
&\quad \times \left [2\lambda + 3 - 2 \mu - 2 \sqrt{\lambda(\lambda + 3 -2\mu)} \right ]^{1/3}, \\
\nonumber R_\mathrm{t}(0) &\approx R_\mathrm{p} \left [ \frac{\mu}{(3-2\mu)^2} \right ]^{1/3} \\
&\quad \times \left [5+2e-2\mu -2\sqrt{(1+e)(4+e-2\mu)} \right ]^{1/3}.
\end{align}
\end{subequations}
Unfortunately, in the case of retrograde orbits (negative sign in Eq.~\ref{eq:r2_circ_exact}), we were unable to find useful analytic approximations to the solution. 

In Fig.~\ref{fig:comparison_circ}, we compare, for prograde orbits, the expressions in Eqs.~(\ref{eq:r2_circ_appr1}) (dashed lines) and (\ref{eq:r2_circ_appr2}) (dotted lines) to the unapproximated solution obtained by numerically solving Eq.~(\ref{eq:r2_circ_exact}) (solid lines). In the {\it top} panel, we show the dependence of $R_\mathrm{t}$ on the true anomaly; as expected, the minimum value occurs at pericentre. In the {\it middle} and {\it bottom} panels, we show the dependence on $\mu$ and $e$, respectively, setting $\nu=0$. For the parameters chosen in Fig.~\ref{fig:comparison_circ} ($\mu=0.2$ and $e=1.5$), Eq.~(\ref{eq:r2_circ_appr1}) gives a reasonable approximation (within a few tens of per cent) of the exact solution. Eq.~(\ref{eq:r2_circ_appr2}) is a better approximation, differing no more than a few per cent. 

In Fig.~\ref{fig:circ_pro_retro}, we show numerical solutions of Eq.~(\ref{eq:r2_circ_exact}) for the cases of prograde (solid lines) and retrograde (dotted lines) orbits. In the {\it top} and {\it bottom} panels, $R_\mathrm{t}$ is plotted as a function of $\mu$ and $e$, respectively, where the fixed parameters are $\nu=0$, $\mu=0.2$ and $e=1.5$. As expected, retrograde orbits are more stable, in the sense that $R_\mathrm{t}$ is always larger for retrograde orbits compared to prograde orbits (in Fig.~\ref{fig:circ_pro_retro}, typically by a factor of $\sim 2$). Interestingly, $R_\mathrm{t}$ decreases with $e$ for prograde orbits, whereas for retrograde orbits, $R_\mathrm{t}$ increases with $e$. The dependence on $e$ in both cases is weak, however.

The approximate solution for $R_\mathrm{t}$ given by Eq.~(\ref{eq:r2_circ_appr1}) has the same dependency on $\mu$ and $e$ as the expression for the minimal radius of the unbound particles $r_{\rm{un,min}}$ of \citet[][here Eq.~\ref{eq:r_truncation_Kob2001}]{2001Icar..153..416K}.
The two expressions differ only by a constant\,---\,\citet{{2001Icar..153..416K}} give $\approx0.34$, while here we derive $\approx0.63$.
Based on extensive simulations covering parameter space in the mass ratio and pericenter distance (larger than considered here), \citet{2014A&A...565A.130B} derived an empirical formula for the truncation of the disc size after a parabolic prograde encounter, given as $0.28\cdot R_\mathrm{p} \cdot (M_1/M_2)^{-0.32}$.
Their result has the same dependency on the encounter pericenter $R_\mathrm{p}$ but differs for the mass ratio\,---\,our expression~(\ref{eq:r2_circ_appr1}) $\propto(M_1/M_2 + 1)^{-1/3}$.
However, \citet{2014A&A...565A.130B} defined the size of the disc depending on the drop in the disc radial surface density profile after the encounter and their expression might therefore not be directly comparable with our result.

\subsection{Comparison with the simulations}
\label{sec:comparison}

\begin{figure}
\center
\includegraphics[width=0.49\textwidth]{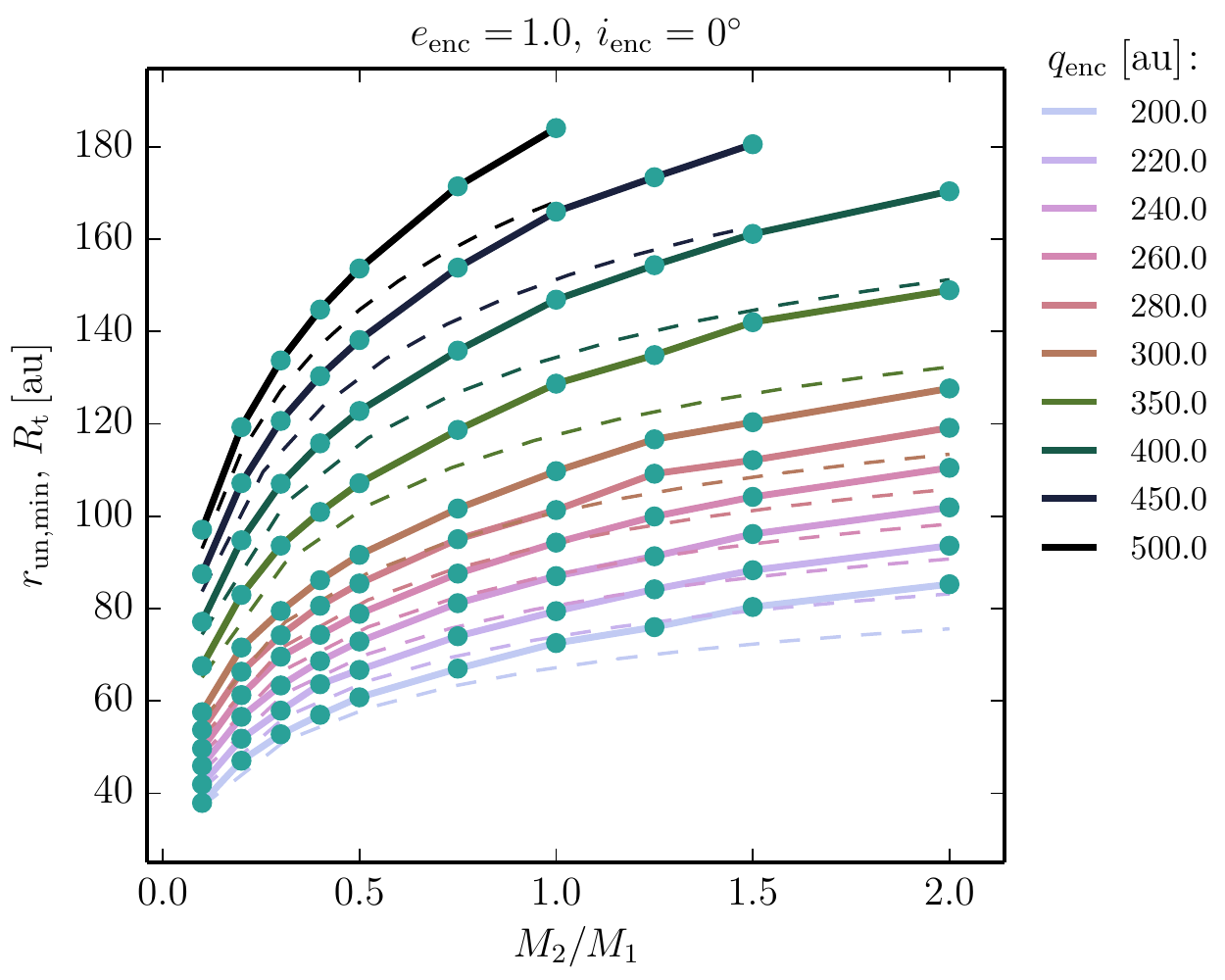} \\
\includegraphics[width=0.49\textwidth]{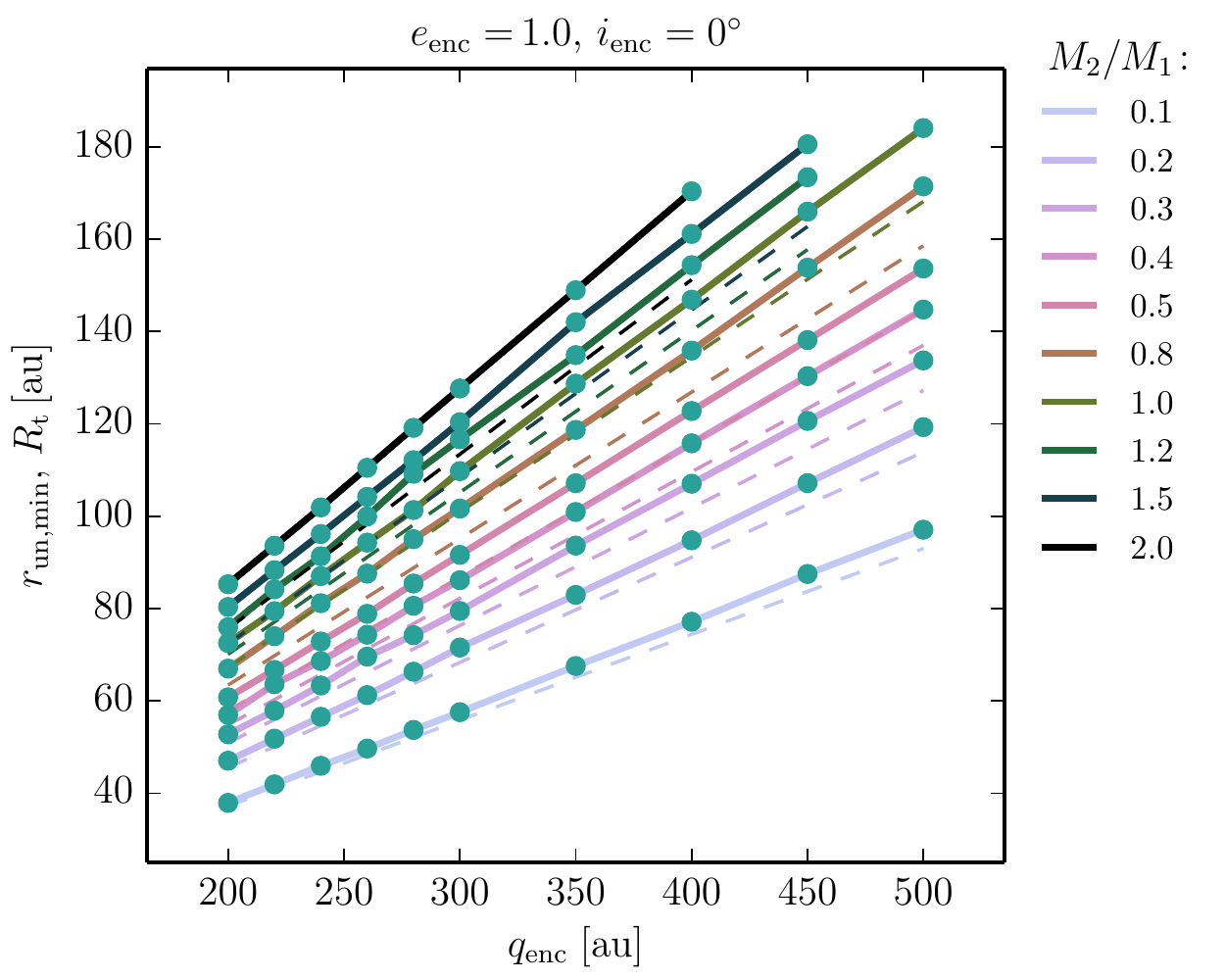} \\
\caption{Comparison of the minimal radius of the particles unbound from $M_2$ from our simulations, $r_{\rm{un,min}}$, and given by Eq.~(\ref{eq:r2_circ_exact}), $R_{\rm{t}}$.
{\it Top} panel shows $r_{\rm{un,min}}$ and $R_{\rm{t}}$ as a function of $M_2/M_1$ for different $q_{\rm{enc}}$,
{\it bottom} panel then $r_{\rm{un,min}}$ and $R_{\rm{t}}$ as a function of $q_{\rm{enc}}$ for different $M_2/M_1$.
The dots connected by full lines indicate the simulated results, $r_{\rm{un,min}}$, while the dashed lines correspond to the analytic approximation of Eq.~(\ref{eq:r2_circ_exact}), $R_{\rm{t}}$.}
\label{fig:com_e1}
\end{figure}

The comparison of the analytic estimation of the truncation radius $R_{\rm{t}}$ and the minimal radius of the unbound particles  derived from our simulated results $r_{\rm{un,min}}$ is presented in Fig.~\ref{fig:com_e1}.
We show $r_{\rm{un,min}}$ of the parabolic ($e_{\rm{enc}}=1.0$) coplanar ($i_{\rm{enc}}=0\degr$) encounters as a function of the mass ratio $M_2/M_1$ and pericentre $q_{\rm{enc}}$.
The analytic and simulated results are in a good agreement ($r_{\rm{un,min}}$ and $R_{\rm{t}}$ do not differ by more than about 11\% of $r_{\rm{un,min}}$).
The {\it bottom} plot clearly shows that $r_{\rm{un,min}}$ is a linear function of the pericentre $q_{\rm{enc}}$, in agreement with Eqs.~(\ref{eq:r2_circ_appr1}) and (\ref{eq:r2_circ_appr2}).
These expressions represent a good approximation to the solution of Eq.~(\ref{eq:r2_circ_exact}) (see Fig.~\ref{fig:comparison_circ}).

As we showed in Sect.~\ref{sec:transfer_rad}, the minimal radius of the unbound particles $r_{\rm{un,min}}$ is similar to the minimal transfer radius $r_{\rm{tr,min}}$ (see Fig.~\ref{fig:rmin_tr_unbound}) and the expressions~(\ref{eq:r2_circ_exact}) and its approximate solutions (\ref{eq:r2_circ_appr1}) and (\ref{eq:r2_circ_appr2}) present a good estimate the minimal transfer radius for coplanar prograde encounters.

The difference between $R_{\rm{t}}$ and $r_{\rm{un,min}}$ increases for faster encounters with higher eccentricities or nonzero inclinations $i_{\rm{enc}}$.
In the case of coplanar eccentric encounters, the analytic model describes the qualitative behavior of $r_{\rm{un,min}}$ well, but tends to underestimate $r_{\rm{un,min}}$ by up to about 30\% for the most eccentric case of $e_{\rm{enc}}=4.5$.
We also derived analytic expressions of $R_{\rm{t}}$ for the inclined (non-coplanar) encounters (not given here).
Their values of $R_{\rm{t}}$ are quantitatively in agreement with $r_{\rm{un,min}}$ presented in Sect.~\ref{sec:inc_enc} and Fig.~\ref{fig:rmin_incl}, but $R_{\rm{t}}$ tends to underestimate $r_{\rm{un,min}}$.
Our analytic approach is probably too simplistic to represent the faster and inclined encounters well, for example, due to the assumption that the orbits of the test particles stay circular up to the moment when they are unbound from the parent star.

\section{Summary and Conclusions}
\label{sec:conclusions}

We studied the mass transfer between debris discs during close stellar encounters that are expected to often happen in star clusters and stellar associations where most stars form \citep{2011A&A...532A.120L}.
We carried out simulations of an encounter of two stars on parabolic and hyperbolic orbit (eccentricity up to 4.5) where one of the stars has a debris disc represented by test particles.
The mass ratio of the star initially surrounded by the disc to the one initially without was varied in the range of 0.1--2.0.
The stars approached each other as close as 200--500\,au and the disc extent was fixed to 30--200\,au.
We also considered the effect of the geometry of the encounter (inclination of the disc with respect to the plane of the encounter and the argument of periastron of the encounter).

The disc particles are transferred to the other star during encounters from a substantial part of the covered parameter space.
We identify a restricted radial range in the initial disc, the so-called {\it transfer region} (Sect.~\ref{sec:transfer_rad}), from where the particles are transferred.
The limiting radii of the region depend on the parameters of the encounter.
We derive an analytic description of the minimal radius from where the particles can be unbound from their parent star (Sect.~\ref{sect:analytic}) and we show how this radius compares with the minimal radius from where the particles are transferred, that is from where they end up unbound from their parent star and bound to the other star.
The minimal radius of the transferred particles is typically up to 5\% larger than the one of the unbound particles.
The analytic description of the minimal radius of the unbound particles $r_{\mathrm{un,min}}$ can be approximated (within a few tens of per cent) as
\begin{align*}
r_{\mathrm{un,min}} \approx q \left[\frac{\mu}{4(1+e)}\right]^{1/3}, \quad \mathrm{where} \quad \mu=\frac{M_2}{M_1+M_2},
\end{align*}
and $q$ is the pericenter of the encounter, $e$ its eccentricity, $M_1$ and $M_2$ are the masses of the star initially without and with the disc.

The transfer efficiency\,---\,defined as the ratio of the number of transferred particles to the number of particles initially orbiting within the transfer region\,---\,depends on the encounter parameters.
The efficiency is generally higher for smaller mass of the star initially with the disc, for closer approach of the two stars, and for slower encounters (that is the ones with smaller the eccentricity).
As much as $\sim45$\% of the particles can be captured by the other star (Sect.~\ref{sec:transfer_eff}).

The geometry of the disc and the orbit of the encounter plays an important role for the transfer region and efficiency.
If the encounter orbit is inclined with respect to the disc, the minimal radius of the transfer particles is larger than for the coplanar case, and the transfer efficiency is lower.
Within our resolution, the geometry with a coplanar and counter-rotating disc and encounter orbit results in no transferred particles (for the outer disc edge of 200\,au considered in our simulations), regardless of other encounter parameters.
With the exception of a highly inclined retrograde orbit, the argument of pericenter of the encounter has a weaker effect.
Generally, the transfer efficiency is the highest for encounters with the pericentre direction perpendicular to the intersection of the two planes, but does not change by more than 10\% for different values of the argument of pericenter.

The transferred particles acquire specific orbits around their new host and the population occupies a specific space of orbital elements depending on the parameters of the encounter (Fig.~\ref{fig:encounter_orb_pars}).
The minimal semimajor axis of the transferred orbits is a linear function of the pericentre of the encounter, where the coefficient of the proportionality depends on the encounter eccentricity.
The transferred orbits tend to be highly eccentric, with most of the encounters resulting in an eccentricity distribution with a median value of about 0.8.
However, depending on the parameters of the encounter, the median eccentricity can be as low as 0.4 and particles can be transferred into almost circular orbits ($e\aplt0.1$).
If the planes of the disc and the encounter are inclined with respect to each other, the directions of the angular momenta of more than 70\% of the transferred particles are restricted to a cone with a typical opening angle of less than $15\degr$.

We find that if the disc is initially eccentric (ranges from 0.0 to 0.05 and 0.1 were considered), the transfer efficiency decreases by no more than 10\%.
Similarly, the initial surface density profile of the disc has a weak effect.

The eccentricities of the transferred orbits are sufficiently high ($e \apgt 0.05 \sqrt{(a/[35\,\mathrm{au}])}$) for the collisions between the planetesimals to be destructive and grind them into dust \citep{2002AJ....123.1757K}.
The relative mass of the transferred planetesimals depends on the transferred region and the transfer efficiency and for the can be as high as about 40\,per\,cent of the total initial mass of the planetesimal disc.
The observability of the dust produced from the transferred planetesimals is given by further properties such as the size and temperature of the grains or their orbits and distribution around the new host star, and the estimate is beyond the scope of this work.
However we expect that the transferred material can produce an observable IR excess if its mass is at least comparable to the observed lower limits ($\apgt 5 \times 10^{-4}$\,M$_{\earth}$, \citealp{2004MNRAS.351L..54G}, \citealp{2008ARA&A..46..339W}).

We expect that many stars could have experienced transfer among their debris discs and planetary systems.
Considering the orbital characteristics of the transferred orbits, capturing planets from foreign systems presents a formation channel for objects on wide orbits of arbitrary inclinations, typically having high eccentricity but possibly also close-to-circular (with eccentricities about 0.1).
The orbital elements of the transferred population (planetesimals or planets) can be distinct from the ones of the native objects formed around the star. 
This might help to identify the captured population which can be used to constrain the encounter.

%%%

\section*{Acknowledgments}

We thank the anonymous referee for comments that helped to put our work better in context.
We thank Susanne Pfalzner for valuable suggestions and comments on the manuscript.
We thank Alan Heays, Anton Walsh and Paola Pinilla for useful discussions.
We acknowledge the Leiden/ESA Astrophysics Program for Summer Students (LEAPS) for the support of MH.
This work was supported by the Interuniversity Attraction Poles Programme initiated by the Belgian Science Policy Office (IAP P7/08 CHARM) and by the Netherlands Research Council NWO (grants \#643.200.503, \#639.073.803 and \#614.061.608) and by the Netherlands Research School for Astronomy (NOVA).
Part of the numerical computations were carried out using the Little Green Machine at Leiden University.
We acknowledge the use of The Minor Planet Center (MPC, \url{http://www.minorplanetcenter.net/}) database.

%%%%%%%%%%%%%%%%%%%%%
%%%
\bibliographystyle{mnras}
\bibliography{capture}
%%%

\appendix
\section{Derivation of the equations of motion in a rotating and pulsating frame for the hyperbolic restricted three-body problem}
\label{app:EOM_der}
Here, we give a self-contained derivation of the equations of motion in a rotating and pulsating frame for the hyperbolic restricted three-body problem, which we used in Sect.~\ref{sect:analytic}. Similar equations have been given by \citet{duboshin_64,lukyanov_10}. Our derivation is an extension of the derivation presented in \citet{baoyin_ea_10}, where the equations of motion were derived for the elliptic restricted three-body problem. 

We assume that two massive bodies, bodies 1 and 2, are in a hyperbolic orbit, with separation $R_{12}$ given by Eq.~(\ref{eq:R12}). Their position vectors with respect to an inertial frame are denoted by $\ve{R}_i=\{X_i,Y_i,Z_i\}$, with $i \in \{1,2\}$. Without loss of generality, the orbital plane of the massive bodies is set to coincide with the $XY$ plane, i.e. $Z_1=Z_2=0$. The Newtonian equations of motion for the test particle, with position vector $\ve{R} = \{X,Y,Z\}$, are given by
\begin{align}
\label{eq:EOM_inertial}
\frac{\mathrm{d} ^2 \ve{R}}{\mathrm{d} t^2} = - G M_1 \frac{\ve{R}-\ve{R}_1}{||\ve{R}-\ve{R}_1||^3} - G M_2 \frac{\ve{R}-\ve{R}_2}{||\ve{R}-\ve{R}_2||^3}.
\end{align}

Next, we define a rotating and pulsating frame $\ve{r}=(x,y,z)$ in which the massive bodies are at fixed positions along the $x$ axis, $x_1=-\mu$ and $x_2=1-\mu$, where $\mu$ is defined in Eq.~(\ref{eq:mu_def}), and the unit of length is the instantaneous separation $R_{12}=R_{12}(\nu)$. The relation between the dimensionless vector $\ve{r}$ in the new frame, and the dimensional vector $\ve{R}$ in the old frame, is given by
\begin{align}
\label{eq:trans_Rr}
\ve{R} = R_{12} \bf{M} \cdot \ve{r}, \quad \quad 
\bf{M} = 
\begin{pmatrix}
\cos(\nu) & -\sin(\nu) & 0 \\
\sin(\nu) & \cos(\nu) & 0 \\
0 & 0 & 1 \\
\end{pmatrix}.
\end{align}

Our procedure is to express both sides of Eq.~(\ref{eq:EOM_inertial}) in terms of quantities pertaining only to the new frame and with the time transformed to the true anomaly $\nu$. This yields a set of second-order differential equations for $\ve{r}''(\nu)$, i.e. Eqs.~(\ref{eq:EOM_gen}).

First, we consider the left-hand side Eq.~(\ref{eq:EOM_inertial}). By applying the chain rule, the inertial acceleration can be written as
\begin{align}
\label{eq:d2t_to_d2nu}
\frac{\mathrm{d}^2 \ve{R}}{\mathrm{d} t^2} = \dot{\nu}^2 \frac{\mathrm{d} \ve{R}}{\mathrm{d} \nu^2} + \ddot{\nu} \frac{\mathrm{d} \ve{R}}{\mathrm{d} \nu}.
\end{align}
For hyperbolic orbits, $\dot{\nu}$ and $\ddot{\nu}$ are given by
\begin{subequations}
\label{eq:nu_dot_nu_ddot}
\begin{align}
\label{eq:nu_dot} \dot{\nu} &= \sqrt{ \frac{G(M_1+M_2)}{R_\mathrm{p}^3}} \frac{\lambda^2}{(1+e)^{3/2}}; \\
\ddot{\nu} &= - \frac{G(M_1+M_2)}{R_\mathrm{p}^3} \frac{2e \lambda^{3} \sin(\nu)}{(1+e)^3},
\end{align}
\end{subequations}
where $\lambda \equiv 1+e \cos(\nu)$. Applying Eq.~(\ref{eq:trans_Rr}) to Eq.~(\ref{eq:d2t_to_d2nu}) and substituting Eqs.~(\ref{eq:nu_dot_nu_ddot}), we find
\begin{subequations}
\label{eq:EOM_inertial_left}
\begin{align}
&\nonumber \frac{\mathrm{d}^2 X}{\mathrm{d} t^2} = - \alpha \left \{ \cos(\nu) x(\nu) - \sin(\nu) y(\nu) \right. \\
  &\quad \left. + \lambda \cos(\nu) \left [ 2y'(\nu) - x''(\nu) \right ] + \lambda \sin(\nu) \left [2x'(\nu) + y''(\nu) \right ] \right \}; \\
&\nonumber \frac{\mathrm{d}^2 Y}{\mathrm{d} t^2} = - \alpha \left \{ \sin(\nu) x(\nu) + \cos(\nu) y(\nu) \right. \\
  &\quad \left. + \lambda \sin(\nu) \left [ 2y'(\nu) - x''(\nu) \right ] - \lambda \cos(\nu) \left [2x'(\nu) + y''(\nu) \right ] \right \}; \\
& \frac{\mathrm{d}^2 Z}{\mathrm{d} t^2} = \alpha \left \{ \lambda z''(\nu) + (\lambda-1) z(\nu) \right \},
\end{align}
\end{subequations}
where
\begin{align}
\alpha \equiv \frac{G(M_1+M_2)}{R_\mathrm{p}^2} \frac{\lambda^2}{(1+e)^2}.
\end{align}

Next, we consider the right-hand side of Eq.~(\ref{eq:EOM_inertial}). By applying Eq.~(\ref{eq:trans_Rr}) to $||\ve{R}-\ve{R}_i||^2 = (X-X_i)^2 + (Y-Y_i)^2 + Z^2$, the latter can be written as
\begin{subequations}
\begin{align}
||\ve{R}-\ve{R}_1||^2 &= R_{12}^2 \left [ (x + \mu)^2 + y^2 + z^2 \right ] \equiv R_{12}^2 r_1^2; \\
||\ve{R}-\ve{R}_2||^2 &= R_{12}^2 \left [ (x + \mu - 1)^2 + y^2 + z^2 \right ] \equiv R_{12} r_2^2,
\end{align}
\end{subequations}
where we introduced the dimensionless distances $r_1$ and $r_2$ (cf. Eq.~\ref{eq:r1r2_def}). Also applying Eq.~(\ref{eq:trans_Rr}) to the terms $\ve{R}-\ve{R}_i$ in Eq.~(\ref{eq:EOM_inertial}), the components $A_X$, $A_Y$ and $A_Z$ of the right-hand of Eq.~(\ref{eq:EOM_inertial}), i.e. the inertial gravitational acceleration, can be written as
\begin{subequations}
\label{eq:EOM_inertial_right}
\begin{align}
\nonumber A_X &= - \alpha \left \{ \cos(\nu) \mu (1-\mu) \left (\frac{1}{r_1^3} - \frac{1}{r_2^3} \right ) \right . \\
&\quad \left. + \left ( \frac{1-\mu}{r_1^3} + \frac{\mu}{r_2^3} \right ) \left [ \cos(\nu) x(\nu) - \sin(\nu) y(\nu) \right ] \right \}; \\
\nonumber A_Y &= - \alpha \left \{ \sin(\nu) \mu (1-\mu) \left (\frac{1}{r_1^3} - \frac{1}{r_2^3} \right ) \right . \\
&\quad \left. + \left ( \frac{1-\mu}{r_1^3} + \frac{\mu}{r_2^3} \right ) \left [ \sin(\nu) x(\nu) + \cos(\nu) y(\nu) \right ] \right \}; \\
A_Z &= - \alpha \left ( \frac{1-\mu}{r_1^3} + \frac{\mu}{r_2^3} \right ) z(\nu).
\end{align}
\end{subequations}
Equating the three components in Eqs.~(\ref{eq:EOM_inertial_left}) and (\ref{eq:EOM_inertial_right}) and solving for $\{x''(\nu),y''(\nu),z''(\nu)\}$, we find the equations of motion given by Eqs.~(\ref{eq:EOM_gen}).

%%%
\label{lastpage}
\end{document}